\documentclass[aps,prx,twocolumn,superscriptaddress,notitlepage]{revtex4-1}

\usepackage{xcolor}

\usepackage{graphicx}
\usepackage{epsfig}
\usepackage{epstopdf}
\usepackage{float}
\usepackage{subfigure}
\usepackage{bbold}

\usepackage{dcolumn}
\usepackage{latexsym}
\usepackage{amsmath,amsfonts,amsthm,bm}
\usepackage{wasysym}
\usepackage{bm}

\usepackage[colorlinks,bookmarks=false,citecolor=blue,linkcolor=red,urlcolor=blue]{hyperref}
\usepackage{verbatim}

\usepackage{booktabs}

\usepackage{ulem} 

\usepackage{algorithm}
\usepackage{algorithmic}
\usepackage{indentfirst}
\usepackage{listings}

\newcommand{\RNum}[1]{\uppercase\expandafter{\romannumeral #1\relax}}

\def\be{\begin{equation}}
\def\ee{\end{equation}}
\def\bea{\begin{eqnarray}}
\def\eea{\end{eqnarray}}

\usepackage{array}
\newcommand{\PreserveBackslash}[1]{\let\temp=\\#1\let\\=\temp}
\newcolumntype{C}[1]{>{\PreserveBackslash\centering}p{#1}}
\newcolumntype{R}[1]{>{\PreserveBackslash\raggedleft}p{#1}}
\newcolumntype{L}[1]{>{\PreserveBackslash\raggedright}p{#1}}

\usepackage{color}

\definecolor{darkblue}{rgb}{0,0.02,0.45}
\definecolor{darkred}{rgb}{0.45,0.02,0} 

\makeatletter
\newcommand{\thickhline}{%
\noalign {\ifnum 0=`}\fi \hrule height 0.7pt
\futurelet \reserved@a \@xhline
}
\newcolumntype{"}{@{\hskip\tabcolsep\vrrule width 0.7pt\hskip\tabcolsep}}
\makeatother

\usepackage{times}

\begin{document}
\title{Phonon dynamics in the generalized Kitaev spin liquid}
\author{Susmita Singh}
\affiliation{School of Physics and Astronomy, University of Minnesota, Minneapolis, Minnesota 55455, USA}
\author{P. Peter Stavropoulos}
\affiliation{School of Physics and Astronomy, University of Minnesota, Minneapolis,  Minnesota 55455, USA}
\author{Natalia B. Perkins}
\affiliation{School of Physics and Astronomy, University of Minnesota, Minneapolis,  Minnesota 55455, USA}
\affiliation{Institute for Advanced Study, Technical University of Munich, Germany, D-85748 Garching, Germany}
\date{\today}

\begin{abstract}
   Candidate materials for the Kitaev spin liquid generically have residual interactions beyond the Kitaev coupling. It therefore becomes necessary to understand how signatures of the quantum spin liquid, e.g., fractionalization of the spin excitations, are affected by the presence of these interactions. Recently it was shown that phonon dynamics is an indirect but effective probe to study fractionalized excitations in the Kitaev spin liquid. Ultrasound experiments can measure sound attenuation, which  should show characteristic  temperature and angular dependence of the sound attenuation coefficient if the scattering of phonons happens predominantly on Majorana fermions.
So far the computation of the sound attenuation was only done in the pure spin-phonon coupled Kitaev model, without taking into account residual interactions. 
In order to understand experimental signatures, here we present a mean-field study of the  sound attenuation in the generalized $J$-$K$-$\Gamma$ model, which is relevant to candidate materials. Our findings show that as long as the system is in the spin liquid phase, characteristic features of the sound attenuation remain observable even in the presence of residual interactions.
 
\end{abstract}

\maketitle

\section{Introduction}
Quantum spin liquids (QSLs) are fascinating magnetic states characterized by exotic properties such as long-range entanglement, topological order, emergent gauge theories and spin fractionalization. Over the last decades QSLs have been the subject of intense research efforts from pure scientific curiosity of their exotic properties  \cite{Anderson1973, Wen2002,Kitaev2006,Balents2010,Savary2016,KnolleMoessner2019,Broholm2020} as well as from a technology viewpoint as potential platforms for topological quantum
computation \cite{Kitaev2003,Kitaev2006}. QSLs 
are usually ensured by frustration, either from a particular geometry of the lattice structure or from competing
spin interactions, making identification of a QSL a major challenge. 

A plethora of works in QSL research \cite{Knolle2017,Motome2019,Takagi2019,Trebst2022} was spurred by the exactly solvable Kitaev honeycomb model, which hosts a QSL ground state where the spin fractionalizes into itinerant Majorrana fermions and localized $Z_2$ fluxes \cite{Kitaev2006}.
Remarkably, it was proposed that the bond-dependent form of spin interactions in the Kitaev model can be realized in materials consisting of heavy transition metal ions with large spin-orbit coupling in  $4d$ and $5d$ groups \cite{Jackeli2009,Chaloupka2010,Kubota2015,Rau2016,Trebst2022,Takagi2019}. The candidate Kitaev materials include the honeycomb iridates $\text{Na}_2\text{IrO}_3$ \cite{Yogesh2010,Liu2011,Choi2012,Ye2012,Comin2012,Hwan2015}, $\alpha\text{-Li}_2\text{IrO}_3$ \cite{Yogesh2012,Williams2016}, $\text{H}_3\text{LiIr}_2\text{O}_6$ \cite{Kitagawa2018}, and the ruthenium compound $\alpha\text{-RuCl}_3$ \cite{Plumb2014,Sandilands2015,Sears2015,Majumder2015,Johnson2015,Sandilands2016,Banerjee2016,Banerjee2017,Do2017}. 
In these materials, a lot of effort was spent  to obtain 
 combined  evidence of fractionalization  from various types of dynamical probes, such as inelastic neutron scattering, Raman scattering, resonant inelastic x-ray scattering, scanning tunneling microscopy, ultrafast spectroscopy, terahertz non-linear coherent spectroscopy and phonon dynamics \cite{Sandilands2015,Sandilands2016,Banerjee2016,Banerjee2017,Wu2018,Dirk2020,Ruiz2021,Halloran2022,Yang2022,Miao2020,Mu2022}. The possibility to compute
the corresponding response functions analytically in the
Kitaev model provides a unique opportunity to explore
the characteristic fingerprints of the QSL physics in the
dynamical probes on a more quantitative level \cite{Knolle2014a,Knolle2014b,
Knolle2015,nasu2014vaporization,nasu2016fermionic,Perreault2015,Perreault2016,Gabor2016,Gabor2017,Gabor2019,Rousochatzakis2019,udagawa2021,Wan2019,Choi2020, Mengxing2020,Metavitsiadis2020,Kexin2021,Feng2022}.

 While the dominance of Kitaev interaction
is well established in most of the Kitaev materials \cite{Winter2017,Takagi2019,Trebst2022},
they generically have other symmetry allowed interactions beyond the Kitaev coupling, such as the nearest neighbor symmetric off-diagonal interaction $\Gamma$ and the Heisenberg interaction $J$ \cite{Rau2014,Sizyuk2014,Williams2016,Wang2017,Ran2017}.  Extensive studies of the $J$-$K$-$\Gamma$ model using a wide range of techniques, including exact diagonalization \cite{Chaloupka2010,Chaloupka2013,Rau2014,JRau2014,Gotfryd2017}, density matrix renormalization group \cite{Gohlke2017,Gohlke2018,Gordon2019}, tensor-network method \cite{Osorio2014,Lee2020}, parton mean field theories \cite{Burnell2011,Schaffer2012,Knolle2018} and variational Monte Carlo approaches \cite{Wang2019,Zhang2021} have shown that the QSL state has a finite region of stability even in the presence of finite  $J$ and $\Gamma$ interactions.  In real life, additional interactions
 compete with the Kitaev coupling  and often result in  a long-range order
below some temperature $T_{\rm N}$ which  has the same energy scale as these subdominant interactions (the magnon excitation frequencies  also have the same  energy scale) \cite{Rousochatzakis2019}.  Thus, 
the observation of the features of the QSL  is possible at temperatures and frequencies above the energy scale of the subdominant  interactions, and
  it is therefore important to understand the experimental signatures of fractionalization in the generalized $J$-$K$-$\Gamma$ QSL.

\begin{figure}
	\centering
	\includegraphics[width=1.0\columnwidth]{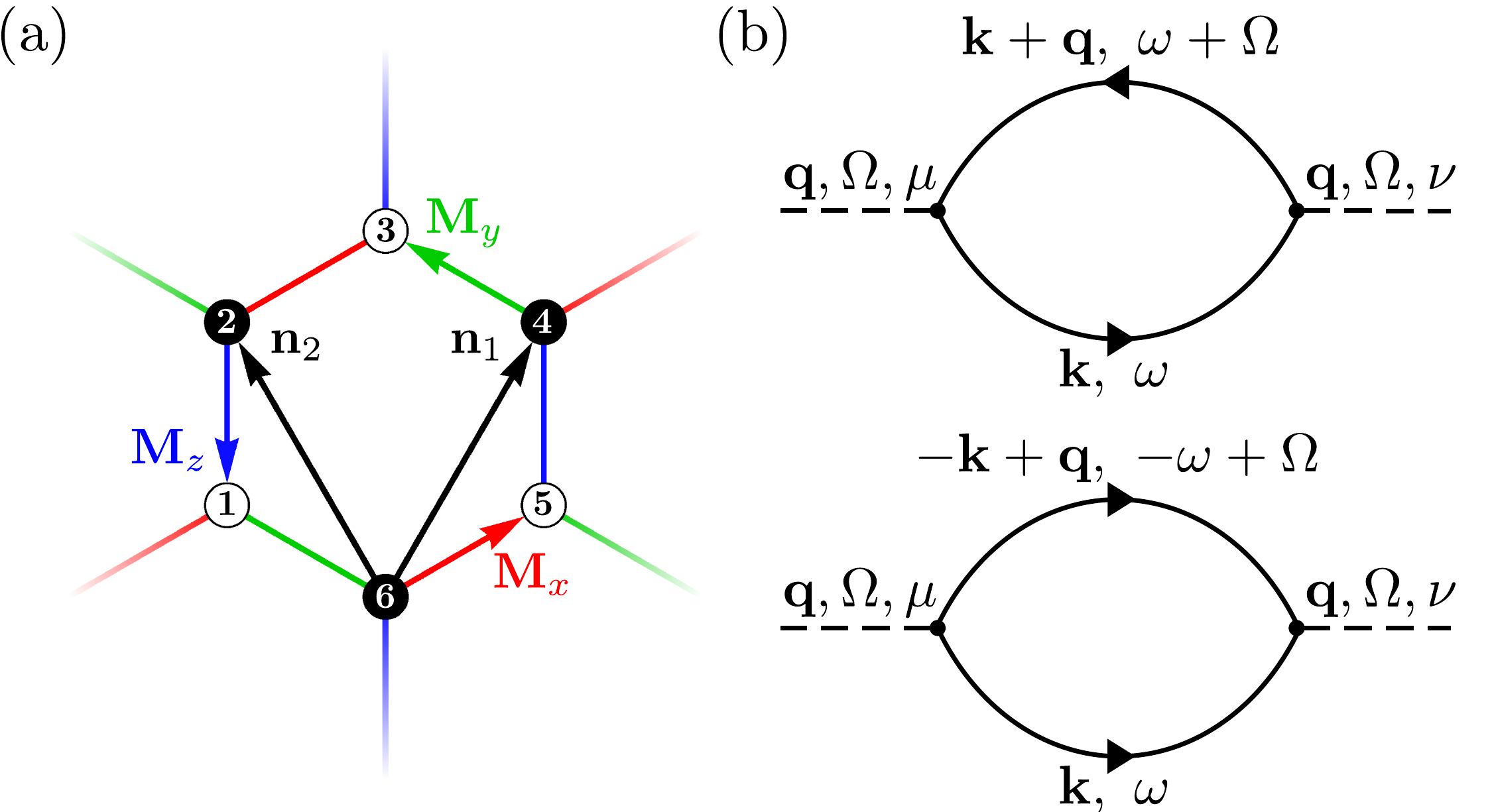}
     \caption{(a) The honeycomb structure, with black and white sites indicating the two sub-lattices. Red, green, and blue bonds show X-, Y-, and Z-bonds, respectively. The bond directions are $\mathbf{M}_x = (\frac{\sqrt{3}}{2},\frac{1}{2})$, $\mathbf{M}_y = (-\frac{\sqrt{3}}{2},\frac{1}{2})$, and $\mathbf{M}_z = (0,-1)$, and the unit cell is defined by $\mathbf{n}_1 = (\frac{\sqrt{3}}{2},\frac{3}{2})$, $\mathbf{n}_2 = (\frac{-\sqrt{3}}{2},\frac{3}{2})$. (b) The two processes that contribute to the phonon polarization bubble: particle-hole (ph) on the top, and particle-particle (pp) at the bottom.}
\label{Fig:baseset}
\end{figure}

 Here  we analyze the phonon dynamics as a dynamical probe in the generalized $J$-$K$-$\Gamma$ QSL.
  Our study is motivated by several experiments  in $\alpha\text{-RuCl}_3$  \cite{Hentrich2018,Kasahara2018,Pal2020,Miao2020,Haoxiang2021,Andreas2023}  and recent  theoretical studies \cite{Mengxing2020,Metavitsiadis2020,Kexin2021,Kexin2021-s,Metavitsiadis2022,Feng2022}
  that  indicated that  phonon dynamics  might be a useful probe for signatures of 
  fractionalization. In Kitaev materials, such as $\alpha\text{-RuCl}_3$ as well as others, the strong spin-orbit coupling ties the spin degrees of freedom to the lattice and subsequently to its vibrations,
thus the study of phonon dynamics presents an avenue for probing spin fractionalization in the QSL state. Specifically, the sound attenuation\textemdash how lattice vibrations diminish while traveling through a material\textemdash may be measured by ultrasound experiments~\cite{Pippard1955,Akhiezer1957,Blount1959,Tsuneto1961,Ott1985,Kazumi1994,Andreas2023}.


 So far, the studies of phonon dynamics have been limited to the pure Kitaev model, without taking into account residual interactions relevant to real materials.  They revealed that the Majorana fermion-phonon scattering  has a particular angular dependence and a linear in temperature dependence at low enough $T$ when the sound velocity  $v_s$ is smaller than the Fermi velocity $v_F$ characterizing the
 low-energy  Dirac-like spectrum
 of Majorana fermions \cite{Metavitsiadis2020,Mengxing2020,Haoxiang2021}. 
 The effect of the $Z_2$ flux excitations at finite temperatures was also calculated \cite{Kexin2021}.  It is important to answer whether these features of phonon dynamics found for the pure Kitaev model survive in the presence of the residual interactions. 
 
 In this work we answer this question by studying the  sound attenuation in the generic $J$-$K$-$\Gamma$ model using a self-consistent slave fermion
 mean-field (MF)  framework  \cite{Burnell2011,Wen2002,Schaffer2012,Knolle2018}.  The sound-attenuation coefficient $\alpha_s(\textbf{q})$ is then calculated from the imaginary part of the diagonal components of the phonon self-energy $\Pi(\textbf{q},\Omega)$  \cite{Mengxing2020}:
\begin{equation}
    \alpha_s^{\mu}(\textbf{q})  \propto -\frac{1}{v_s^2q}\text{Im}[\Pi^{\mu\mu}(\textbf{q},\Omega)]_{\Omega=v_sq},
\end{equation}
where $\mu$ is the phonon polarization component, and $\textbf{q}$ and $\Omega=v_s q$ are the phonon's momentum and frequency.
 We find that  in the $J$-$K$-$\Gamma$ spin liquid the sixfold angular symmetry of the sound attenuation survives   both  for $v_s<v_F$  and for $v_s>v_F$,  and its linear in temperature ($\sim T$) dependence still holds at small temperatures  for $v_s<v_F$.
  Moreover, as we move away from the pure Kitaev point, the static $Z_2$ fluxes start dispersing, providing more low-energy states for the phonon to scatter on and resulting in higher values of the sound-attenuation coefficient for nonzero $J$ and $\Gamma$. We show that this increase in intensity is intricately linked to the change in the fermionic spectrum. We also see that processes that are forbidden for the pure Kitaev model due to kinematic constraints become allowed in the generic $J$-$K$-$\Gamma$ model, which again happens due to the change in the fermionic spectrum.

The rest of the paper is organized as follows: In Sec.~\ref{sec:model} we describe the full model, with details of the MF spin Hamiltonian,  the phonon Hamiltonian, and  the spin-phonon coupling Hamiltonian presented in Secs.~\ref{sec:h_s},  \ref{sec_h_ph}, and \ref{sec:h_c}, respectively.  In Sec.~\ref{sec:bubble},  we present the calculation of the one-loop phonon  polarization bubble. Then, in Sec.~\ref{sec:kinematic} we describe the kinematic constraints for the phonon dynamics.  
In Sec.\ref{Conc}, we present a  short  summary and discuss the possibility for  the spin fractionalization   to be seen in the sound-attenuation  measurements by the ultrasound experiments. 
Some of the technical details and auxiliary information are relegated to Appendixes  \ref{app:mf_spin_ham}\textendash\ref{app:addnumres}.

\section{Spin-phonon $\bm{J}$-$\bm{K}$-$\bm{\Gamma}$ model}\label{sec:model}
The Hamiltonian is composed of the spin, phonon  and the spin-phonon interaction terms:
\begin{equation}\label{h_tot}
    \mathcal{H}=\mathcal{H}_{\text{s}}+\mathcal{H}_{\text{ph}}+\mathcal{H}_{\text{c}}.
\end{equation}
 The first term  $\mathcal{H}_{\text{s}}$ is  the generalized  $J$-$K$-$\Gamma$ model, the second term is  the phonon Hamiltonian $\mathcal{H}_{\text{ph}}$ leading to two-dimensional free phonons, and the third term is the magnetoelastic coupling $\mathcal{H}_{\text{c}}$. 

\subsection{Spin Hamiltonian fermionic mean field}\label{sec:h_s}
The  generic $J$-$K$-$\Gamma$ spin Hamiltonian on the honeycomb lattice is 
\begin{equation}\label{h_s}
    \mathcal{H}_{\text{s}}=\sum_{\substack{\langle ij \rangle_{\alpha}, \\ \beta\neq\gamma\neq\alpha}}K\sigma_i^{\alpha}\sigma_j^{\alpha}+J\bm{\sigma}_i\cdot\bm{\sigma}_j+\Gamma(\sigma_i^{\beta}\sigma_j^{\gamma}+\sigma_i^{\gamma}\sigma_j^{\beta}),
\end{equation}
where $\bm{\sigma}=(\sigma^x,\sigma^y,\sigma^z)$ are the Pauli matrices, $\alpha=x,y,z$ is the spin component that follows from the X-, Y-, or Z-bond type, respectively, and $\beta,\gamma$ are the two remaining spin components on the bond  [see Fig. \ref{Fig:baseset} (a)]. The spin model can be decomposed into Majorana fermions by
\begin{equation}\label{majo_rep}
    \sigma_i^{\alpha}=ib_i^{\alpha}c_i.
\end{equation}
Keeping all terms, and proceeding with a MF decomposition, we get
\begin{equation}\label{h_s_maj}
\begin{split}
    \mathcal{H}_{\text{s}}^{\text{MF}}=&-\sum_{\langle ij \rangle_{\alpha}}\biggl\{(J+K)\bigl[\kappa_{\alpha}^{\alpha\alpha}(ic_ic_j)+\kappa_{\alpha}^0(ib_i^{\alpha}b_j^{\alpha})-\kappa_{\alpha}^{\alpha\alpha}\kappa_{\alpha}^0\bigr]\\
    &+J\sum_{\beta\neq\alpha}\bigl[\kappa_{\alpha}^{\beta\beta}(ic_ic_j)+\kappa_{\alpha}^0(ib_i^{\beta}b_j^{\beta})-\kappa_{\alpha}^{\beta\beta}\kappa_{\alpha}^0\bigr]\\
    &+\Gamma \sum_{\beta\neq\gamma\neq\alpha}\bigl[\kappa_{\alpha}^{\beta\gamma}(ic_ic_j)+\kappa_{\alpha}^0(ib_i^{\beta}b_j^{\gamma})-\kappa_{\alpha}^{\beta\gamma}\kappa_{\alpha}^0\bigr]\biggr\},
\end{split}
\end{equation}
where the MF bond parameters are
\begin{equation}\label{mf1}
\kappa_{\alpha}^0=\langle ic_ic_j\rangle,\ 
\kappa_{\alpha}^{\beta\gamma}=\langle ib_i^{\beta}b_j^{\gamma}\rangle,
\end{equation}  
with $i\in\text{sublattice A}$ and  $j\in\text{sublattice B}$. In principle, the magnetic channels, $m_i^{\alpha}=\langle ib_i^{\alpha}c_i\rangle$, and cross Majorana bond terms $\langle ib_i^{\alpha}c_j\rangle$ are also present but are found to be zero in the spin liquid regime. So, they are omitted here for simplicity.

In the momentum space, the MF Hamiltonian reads
\begin{equation}\label{majo_ham1}
\mathcal{H}_{\text{s}}^{\text{MF}}=\sum_{\textbf{k}}{\mathcal A}_{-\textbf{k}}^TH_{\text{s}}^{\text{MF}}(\textbf{k}){\mathcal A}_{\textbf{k}},
\end{equation} 
with the Majorana fermion basis ordered as
\begin{equation}\label{majo_ham1_basis}
{\mathcal A}^T_{\textbf{k}}=\left[\begin{array}{cccccccc}
c_{\textbf{k},A}&b_{\textbf{k},A}^{x}&b_{\textbf{k},A}^{y}&b_{\textbf{k},A}^{z}&c_{\textbf{k},B}&b_{\textbf{k},B}^{x}&b_{\textbf{k},B}^{y}&b_{\textbf{k},B}^{z}
\end{array}\right].
\end{equation}
By changing into complex fermions
\begin{equation}\label{map_fc}
\begin{array}{cc}
 f_{\textbf{k},A(B),\uparrow}=\frac{1}{2}(c_{\textbf{k},A(B)}-ib_{\textbf{k},A(B)}^z),\\
 f_{\textbf{k},A(B),\downarrow}=\frac{1}{2}(b_{\textbf{k},A(B)}^y-ib_{\textbf{k},A(B)}^x),
\end{array}
\end{equation}
and diagonalizing in every momentum block we arrive at
\begin{equation}\label{diag_h_s}
\begin{split}
 \mathcal{H}_{s, {\rm diag}}^{\text{MF}}(\textbf{k})=&{\mathcal B}(\textbf{k})^{\dagger}U(\textbf{k})^{\dagger}\left[ V H_{\text{s}}^{\text{MF}}(\textbf{k})V^{\dagger}\right]U(\textbf{k}){\mathcal B}(\textbf{k})\\
 =&{\mathcal B}(\textbf{k})^{\dagger}D(\textbf{k}){\mathcal B}(\textbf{k}),
 \end{split}
\end{equation}
where the matrix $V$ is defined by the mapping in Eq.~\eqref{map_fc}, $U(\textbf{k})$ is the unitary diagonalizing matrix, $D(\textbf{k})$ is the diagonalized MF Hamiltonian at the momentum $\textbf{k}$, and
${\mathcal B}^T_{\textbf{k}}=\left[\begin{array}{cccccccc}
    \beta_{-\textbf{k},1}^{\dagger} &   \beta_{-\textbf{k},2}^{\dagger} &   \beta_{-\textbf{k},3}^{\dagger} &   \beta_{-\textbf{k},4}^{\dagger} &  \beta_{\textbf{k},1} &  \beta_{\textbf{k},2} &  \beta_{\textbf{k},3} &  \beta_{\textbf{k},4} 
\end{array}\right]$ represents 
the Bogoliubov fermion eigenstates $\beta_{\textbf{k},i}$, arranged in ascending order in eigenvalues. The explicit Hamiltonian matrix forms as well as details on the self-consistent procedure can be found in Appenddix~\ref{app:mf_spin_ham}.

\begin{figure}
	\centering
	\includegraphics[width=1.0\columnwidth]{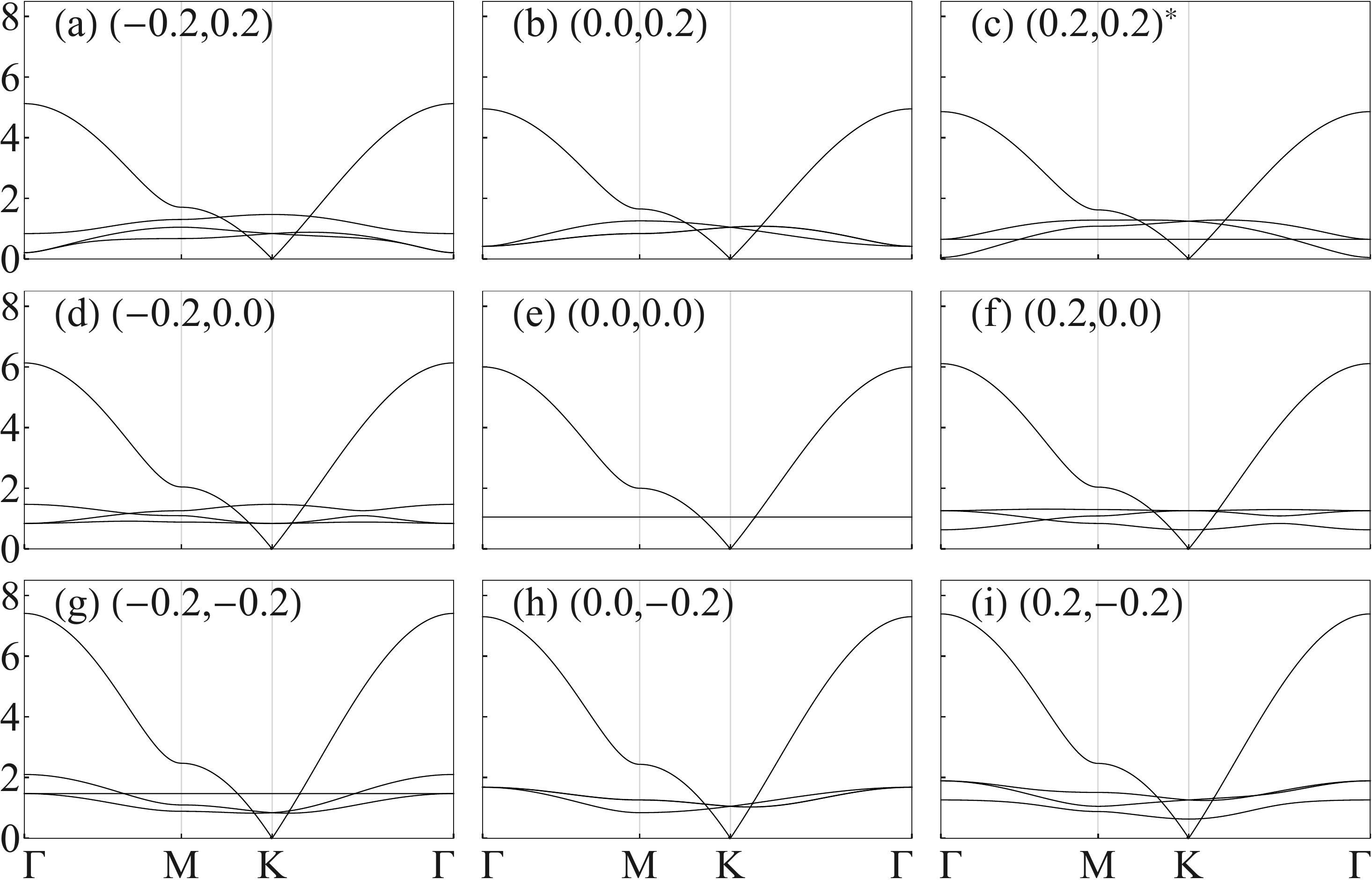}
     \caption{The MF fermionic spectrum for different values of ($\Gamma$, $J$) with $K=-1$, fixed. The central (e) plot corresponds to the pure Kitaev point, with $J=\Gamma=0$. $\Gamma$ increases in steps of 0.2 along the \textit{x} axis and $J$ increases in steps of 0.2 along the \textit{y} axis. The spectrum in panel (c) corresponds to $(\Gamma,J)=(0.19,0.19)$, which is right at the edge of gap closing for the flux bands.
}
\label{Fig:disp}
\end{figure}

In Fig.~\ref{Fig:disp} we present the MF  fermionic spectrum, which we obtain 
for various parameters $J$ and $\Gamma$, while keeping $K=-1$ fixed.
When $J=\Gamma=0$  [Fig. \ref{Fig:disp} (e)], the spin model reduces to the original Kitaev model \cite{Kitaev2006}, and we see the characteristic dispersing mode owing to the free Majaoranas hopping on the lattice, with Dirac cones at the $K$ points of the  Brillouin zone (BZ) and the flat bands corresponding to the static flux operators 
${W}_p = \sigma_1^x \sigma_2^y \sigma_3^z \sigma_4^x \sigma_5^y \sigma_6^z$ [see Fig. \ref{Fig:baseset} (a)].
Moving away from the the exactly solvable point but staying within the spin liquid state, the flux bands acquire a small dispersion. The original dispersing mode remains pronounced and the Dirac cones remain at the $K$ points however; the Fermi velocity on the cone changes. In addition, for the positive value of $J$ (upper  row in Fig. \ref{Fig:disp}), more low-energy states appear  near the $\Gamma$ point of the BZ. 
As we will see later, the modified structure of the fermionic spectrum will be essential for an understanding  of the sound attenuation in the  generalized  $J$-$K$-$\Gamma$  QSL. 

\subsection{The phonon Hamiltonian}\label{sec_h_ph}
The $\mathcal{H}_{\text{ph}}$ term in Eq.~\eqref{h_tot} is the Hamiltonian for acoustic phonons on the honeycomb lattice and is given by,
\begin{equation}
    \mathcal{H}_{\text{ph}}=\mathcal{H}_{\text{ph}}^{\text{kinetic}}+\mathcal{H}_{\text{ph}}^{\text{elastic}}.
\end{equation}
Here, $\mathcal{H}_{\text{ph}}^{\textmd{kinetic}}=\sum_{\textbf{q}}\frac{\textbf{p}_{-\textbf{q}}\textbf{p}_{\textbf{q}}}{2\rho \delta_V}$, with $\textbf{p}_{\textbf{q}}=\rho\delta_t \textbf{u}_{\textbf{q}}$, $\rho$ is the mass density of the lattice ion, $\textbf{u}_{\textbf{q}}=\{u_x,u_y\}$ is the lattice displacement vector, and $\delta_V$ is the area enclosed in one unit cell. The elastic energy is comprised of the strain tensor, $\epsilon_{\alpha\beta}=\frac{1}{2}(\partial_{\alpha}u_{\beta}+\partial_{\beta}u_{\alpha})$. The combinations $\epsilon_{xx}+\epsilon_{yy}$ and $\{\epsilon_{xx}-\epsilon_{yy},2 \epsilon_{xy}\}$ form the basis of the point group  $D_{3d}$  irreducible  representations (irreps) $A_{1g}^{\text{ph}}$ and $E_g^{\text{ph}}$, respectively. The longitudinal and transverse components of the acoustic phonon spectrum and the polarization vectors (defined through $\textbf{u}_\textbf{q}=\sum_{\nu}\hat{\textbf{e}}^{\nu}_{\textbf{q}}\tilde{u}^{\nu}_{\textbf{q}}$; $\nu=\parallel,\perp$ labels the polarization) are 
\begin{equation}
\begin{split}
    \Omega_{\textbf{q}}^{\parallel}=&v_{s}^{\parallel}q, \quad \hat{\textbf{e}}_{\textbf{q}}^{\parallel}=\{\cos\theta_{\textbf{q}},\sin\theta_{\textbf{q}}\},\\
    \Omega_{\textbf{q}}^{\perp}=&v_{s}^{\perp}q, \quad \hat{\textbf{e}}_{\textbf{q}}^{\perp}=\{-\sin\theta_{\textbf{q}},\cos\theta_{\textbf{q}}\},
    \end{split}
\end{equation}
with $q=\sqrt{q_x^2+q_y^2}$ and $\theta_{\textbf{q}}$ is the angle made by $\textbf{q}$ with the \textit{x} axis.

\subsection{The spin-phonon coupling Hamiltonian}\label{sec:h_c}
The coupling term in Eq.~\eqref{h_tot} is the magnetoelastic coupling, which arises from the change in the coupling constants $J$, $K$, $\Gamma$ due to the lattice vibrations. We assume that the coupling constants depend only on the distance between the atoms and that the positions of the spins deviate slightly from their equilibrium positions. The spin-phonon coupling Hamiltonian can then be decomposed into the $A_{1g}$ and $E_g$ symmetry channels as
\begin{equation}\label{hc_irr}
\mathcal{H}_{\text{c}}=\mathcal{H}_{\text{c}}^{A_{1g}}+\mathcal{H}_{\text{c}}^{E_g},
\end{equation}
with
\begin{equation}\label{ha1g}
\begin{split}
\mathcal{H}_{\text{c}}^{A_{1g}}=&\lambda_{A_{1g}}\sum_{\textbf{r}}(\epsilon_{xx}+\epsilon_{yy})\bigl\{(J+K)f_{K}^{A_{1g}}\\
&+Jf_{J}^{A_{1g}}+\Gamma f_{\Gamma}^{A_{1g}}\},
\end{split}
\end{equation}
\begin{equation}\label{heg}
\begin{split}
\mathcal{H}_{\text{c}}^{E_g}=&\lambda_{E_g}\sum_{\textbf{r}}\bigl\{(\epsilon_{xx}-\epsilon_{yy})\bigl((J+K)f_{K}^{E_{g,1}}\\
&+Jf_{J}^{E_{g,1}}+\Gamma f_{\Gamma}^{E_{g,1}}\bigr)+2\epsilon_{xy}\bigl((J+K)f_{K}^{E_{g,2}}\\
&+Jf_{J}^{E_{g,2}}+\Gamma f_{\Gamma}^{E_{g,2}}\bigr)\bigr\}.
\end{split}
\end{equation}
Here $\lambda_{A_{1g}}$, $\lambda_{E_g}$ are the two  unknown independent  constants that characterize the spin-phonon coupling that might be different but of similar strength.
The functions $f_{K}^{A_{1g}}$, $f_{K}^{E_{g,i}}$, etc., are the basis functions for the irreps of  the $D_{3d}$ point group in  the spin space given in the Tables~\ref{table:irrA1g} and \ref{table:irrEg} of Appendix~\ref{app:spin_phonon_coupling}.

Next, we rewrite the spin-phonon coupling Hamiltonian in the MF framework. We first use the Majorana representation of spin defined in \eqref{majo_rep}, and then Fourier transform the
coupling Hamiltonian. We  obtain
\begin{equation}\label{eq:kspaceHc}
\mathcal{H}_{\text{c}}=\mathcal{H}_{\text{c}}^{A_{1g}}+\mathcal{H}_{\text{c}}^{E_g},
\end{equation}
with
\begin{equation}
\begin{split}
\mathcal{H}_{\text{c}}^{A_{1g}}=&\frac{\lambda_{A_{1g}}}{2}\sum_{\textbf{q}}\sum_{\textbf{k}}(iq_xu_{x,\textbf{q}}+iq_yu_{y,\textbf{q}})
\\
&\times
\bigl[{\mathcal A}_{-\textbf{k}-\textbf{q}}^TH_{\text{c}}^{A_{1g}}(\textbf{q},\textbf{k}){\mathcal A}_{\textbf{k}}\bigr],
\end{split}
\end{equation}
\begin{equation}
\begin{split}
\mathcal{H}_{\text{c}}^{E_g}=&\frac{\lambda_{E_{g}}}{2}\sum_{\textbf{q}}\sum_{\textbf{k}}\bigl((iq_xu_{x,\textbf{q}}-iq_yu_{y,\textbf{q}})\\
&\times\bigl[{\mathcal A}_{-\textbf{k}-\textbf{q}}^TH_{\text{c}}^{E_{g,1}}(\textbf{q},\textbf{k}){\mathcal A}_{\textbf{k}}\bigr]\\
&+(iq_xu_{y,\textbf{q}}+iq_yu_{x,\textbf{q}})\bigl[{\mathcal A}_{-\textbf{k}-\textbf{q}}^T H_{\text{c}}^{E_{g,2}}(\textbf{q},\textbf{k}){\mathcal A}_{\textbf{k}}\bigr]\bigr).
\end{split}
\end{equation}
The matrix $H_{\text{c}}^{A_{1g}}(\textbf{q},\textbf{k})$ has the form
\begin{equation}
    H_{\text{c}}^{A_{1g}}(\textbf{q},\textbf{k})=\left[\begin{array}{cc}
0&iM_{A_{1g}}(\textbf{k)}\\
-i[M_{A_{1g}}(\textbf{q}+\textbf{k})]^{\dagger}&0
\end{array}\right],
\end{equation}
and matrices $H_{\text{c}}^{E_{g,1}}(\textbf{q},\textbf{k})$, $H_{\text{c}}^{E_{g,2}}(\textbf{q},\textbf{k})$ have the same structure as $H_{\text{c}}^{A_{1g}}(\textbf{q},\textbf{k})$ with $M_{A_{1g}}$ replaced by $M_{E_{g,1}}$, $M_{E_{g,2}}$, respectively. The detailed form of these matrices is given in Appendix~\ref{app:spin_phonon_coupling}.

To obtain the spin-phonon coupling vertices, we express the phonon modes in terms of the longitudinal and and transverse eigenmodes. This gives
\begin{equation}
\begin{split}
\mathcal{H}_{\textbf{q},\textbf{k}}^{\parallel}=&\tilde{\textbf{u}}_{\textbf{q},\parallel}{\mathcal A}_{-\textbf{k}-\textbf{q}}^T\hat{\lambda}_{\textbf{q},\textbf{k}}^{\parallel}{\mathcal A}_{\textbf{k}},\\
\mathcal{H}_{\textbf{q},\textbf{k}}^{\perp}=&\tilde{\textbf{u}}_{\textbf{q},\perp}{\mathcal A}_{-\textbf{k}-\textbf{q}}^T\hat{\lambda}_{\textbf{q},\textbf{k}}^{\perp}{\mathcal A}_{\textbf{k}},
\end{split}
\end{equation}
where the spin-phonon coupling vertices are
\begin{equation}\label{lambda_lon}
\begin{split}
\hat{\lambda}_{\textbf{q},\textbf{k}}^{\parallel}=&\frac{\lambda_{A_{1g}}}{2}q H_{\text{c}}^{A_{1g}}(\textbf{q},\textbf{k})\\
&+\frac{\lambda_{E_g}}{2}q\bigl[\cos2\theta_q H_{\text{c}}^{E_{g,1}}(\textbf{q},\textbf{k})+\sin2\theta_q H_{\text{c}}^{E_{g,2}}(\textbf{q},\textbf{k})\bigr],
\end{split}
\end{equation}
\begin{equation}\label{lambda_tr}
\hat{\lambda}_{\textbf{q},\textbf{k}}^{\perp}=\frac{\lambda_{E_g}}{2}q\bigl[-\sin2\theta_q H_{\text{c}}^{E_{g,1}}(\textbf{q},\textbf{k})+\cos2\theta_q H_{\text{c}}^{E_{g,2}}(\textbf{q},\textbf{k})\bigr].
\end{equation}
For the purpose of calculating the phonon polarization bubble, we write the coupling Hamiltonian in the basis of the  Bogoliubov quasiparticles of Eq.~\eqref{diag_h_s}:
\begin{equation}
\begin{array}{cc}
   \mathcal{H}_{\textbf{q},\textbf{k}}^{\mu}=\tilde{u}_{\textbf{q},\mu}{\mathcal B}^{\dagger}_{\textbf{k}+\textbf{q}}\tilde{\lambda}_{\textbf{q},\textbf{k}}^{\mu}{\mathcal B}_{\textbf{k}},  \\
    \tilde{\lambda}_{\textbf{q},\textbf{k}}^{\mu}=U(\textbf{k}+\textbf{q})^{\dagger}[V\lambda_{\textbf{q},\textbf{k}}^{\mu}V^{\dagger}]U(\textbf{k}). 
\end{array}
\end{equation}
where $\mu=\parallel,\ \perp$ and  the coupling vertices are divided into four blocks according to the division into creation and annihilation sectors:
\begin{equation}\label{lam_block}
\tilde{\lambda}_{\textbf{q},\textbf{k}}^{\mu}=\left[\begin{array}{cc}
\tilde{\lambda}_{\textbf{q},\textbf{k},11}^{\mu}&\tilde{\lambda}_{\textbf{q},\textbf{k},12}^{\mu}\\
\tilde{\lambda}_{\textbf{q},\textbf{k},21}^{\mu}&\tilde{\lambda}_{\textbf{q},\textbf{k},22}^{\mu}
\end{array}\right].
\end{equation}

\subsection{The phonon polarization bubble}\label{sec:bubble}
To see the effects of the spin-phonon coupling on phonon dynamics, we calculate the one-loop phonon self-energy, which in the leading order is given by the  polarization bubble $\Pi_{\text{ph}}^{\mu\nu}(\textbf{q},\Omega)$. In the Matsubara formalism, it can be written as \cite{Mengxing2020, Kexin2021}:
\begin{equation}
\begin{split}
\Pi^{\mu\nu}(\textbf{q},\tau)=&\langle T_{\tau}\bigl({\mathcal B}^{\dagger}_{\textbf{k}+\textbf{q},l}[\tilde{\lambda}_{\textbf{q},\textbf{k}}^{\mu}]_{lm}{\mathcal B}_{\textbf{k},m}\bigr)(\tau)\\
&\times\bigl({\mathcal B}_{\textbf{k},m}^{\dagger}[\tilde{\lambda}_{-\textbf{q},\textbf{k}+\textbf{q}}^{\nu}]_{ml}{\mathcal B}_{\textbf{k}+\textbf{q},l}\bigr)(0)\rangle.
\end{split}
\end{equation}
where $\langle \hat{O}\rangle_{\omega_n}=\int_0^{\beta}d\tau e^{i\omega_n \tau}\langle \hat{O}(\tau)\rangle$, and $T_{\tau}$ is the imaginary time ordering operator.
Using Wick's theorem, the polarization bubble can be written explicitly as
\begin{equation}
\begin{split}
\Pi^{\mu\nu}(\textbf{q},\tau)=&\langle T_{\tau}{\mathcal B}^{\dagger}_{\textbf{k}+\textbf{q},l}(\tau){\mathcal B}_{\textbf{k}+\textbf{q},l}(0)\rangle\langle{\mathcal B}_{\textbf{k},m}(\tau){\mathcal B}^{\dagger}_{\textbf{k},m}(0)\rangle\\
&\times[\tilde{\lambda}^{\mu}_{\textbf{q},\textbf{k}}]_{lm}[\tilde{\lambda}^{\nu}_{-\textbf{q},\textbf{k}+\textbf{q}}]_{ml}, \quad l,m=1,2,\ldots,8.
\end{split}
\end{equation}
Performing the Fourier transform, we can write
\begin{equation}\label{pi_omega}
\begin{split}
    \Pi^{\mu\nu}(\textbf{q},\Omega)=&\sum_{\textbf{k}}\sum_{{l,m}}\bigl\{P_{{lm}}^{{g}\bar{g}}(\textbf{k}+\textbf{q},\textbf{k})[\tilde{\lambda}^{\mu}_{\textbf{q},\textbf{k},11}]_{lm}[\tilde{\lambda}^{\nu}_{-\textbf{q},\textbf{k}+\textbf{q},11}]_{ml}\\
    &+P_{{lm}}^{\bar{g}g}(\textbf{k}+\textbf{q},\textbf{k})[\tilde{\lambda}^{\mu}_{\textbf{q},\textbf{k},22}]_{lm}[\tilde{\lambda}^{\nu}_{-\textbf{q},\textbf{k}+\textbf{q},22}]_{ml}\\
     &+P_{{lm}}^{gg}(\textbf{k}+\textbf{q},\textbf{k})[\tilde{\lambda}^{\mu}_{\textbf{q},\textbf{k},12}]_{lm}[\tilde{\lambda}^{\nu}_{-\textbf{q},\textbf{k}+\textbf{q},21}]_{ml}\\
     &+P_{{lm}}^{\bar{g}\bar{g}}(\textbf{k}+\textbf{q},\textbf{k})[\tilde{\lambda}^{\mu}_{\textbf{q},\textbf{k},21}]_{lm}[\tilde{\lambda}^{\nu}_{-\textbf{q},\textbf{k}+\textbf{q},12}]_{ml}\bigr\}.
\end{split}    
\end{equation}
Here  $P_{{lm}}^{g\bar{g}}(\textbf{k}+\textbf{q},\textbf{k})$, $P_{{lm}}^{\bar{g}g}(\textbf{k}+\textbf{q},\textbf{k})$, $P_{{lm}}^{gg}(\textbf{k}+\textbf{q},\textbf{k})$, $P_{{lm}}^{\bar{g}\bar{g}}(\textbf{k}+\textbf{q},\textbf{k})$ are convolutions of the Matsubara Green's functions of the free fermions $\beta_i$ of the diagonalized MF spin Hamiltonian, and $l,m$ refer to the fermion flavors corresponding to the four bands of the spectrum. The first two terms in Eq.~\eqref{pi_omega} contribute to the particle-hole (ph) channel, while the last two contribute to the particle-particle (pp) channel. Furthermore, it is worth noting that the intensity of the Majorana fermion-phonon scattering depends on the temperature and the fermionic spectrum, which are encoded in the terms $P_{{lm}}^{g\bar{g}}(\textbf{k}+\textbf{q},\textbf{k})$, $P_{{lm}}^{\bar{g}g}(\textbf{k}+\textbf{q},\textbf{k})$, $P_{{lm}}^{gg}(\textbf{k}+\textbf{q},\textbf{k})$, $P_{{lm}}^{\bar{g}\bar{g}}(\textbf{k}+\textbf{q},\textbf{k})$ through the Fermi distribution functions, while the angular dependence of the Majorana fermion-phonon scattering is contained in the matrix elements of the coupling vertices $[\lambda_{\textbf{q},\textbf{k}}^{\mu}]_{lm}$, describing the coupling between the fermion eigenmodes of flavor $l$, $m$ and momentum $\textbf{k}+\textbf{q}$  and $\textbf{k}$ to the acoustic phonon with 
momentum $\textbf{q}$ and polarization $\mu$. Further details of the terms in Eq.~\eqref{pi_omega} are relegated to Appendix~\ref{app:phonon_bubble}.


\onecolumngrid
\begin{center}
\begin{figure}
	\centering
	\includegraphics[width=1.0\columnwidth]{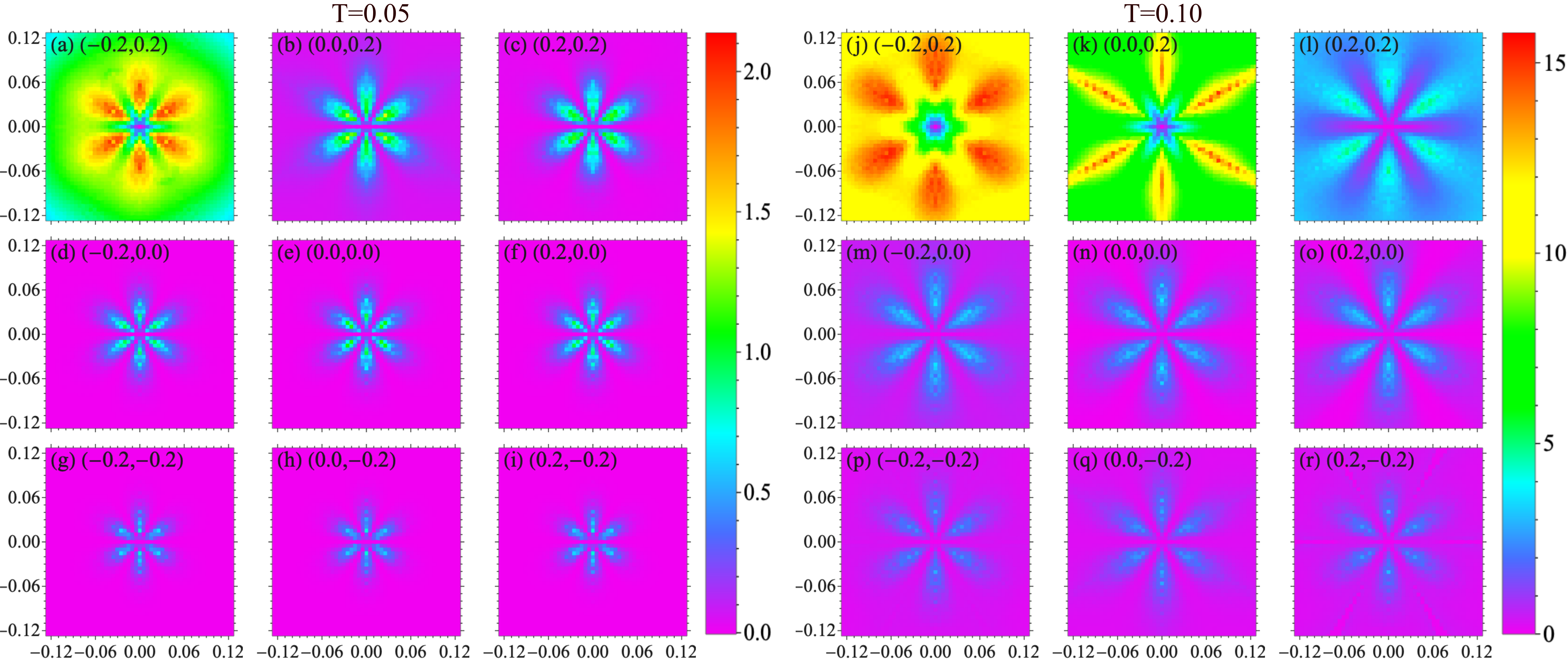}
     \caption{The longitudinal component of the sound-attenuation coefficient $\alpha_s^{\parallel}(\textbf{q})$ for $v_s<v_F$, for different values of ($\Gamma$,$J$) with $K=-1$, fixed. We show plots for $T=0.05$ on the left, and for $T=0.1$ on the right, with temperature, $T$, in units of the Kitaev interaction strength $K$. The pp contribution for $v_S<v_F$ is vanishingly small (except for $J=0.2$, $\Gamma=0.2$). Here we plot only the ph contribution coming from the $E_g$ irrep. The phonon momentum $\textbf{q}$ belongs to the region $(q_x,q_y)\in [-0.12\pi,0.12\pi]^2$. We used $v_s=0.3$ with $v_F=3$, determined from the slope of the Dirac cone of the Kitaev spectrum. The imaginary energy broadening was taken to be $\delta=0.2$.
}
\label{Fig:alon_vlf_grid}
\end{figure}
\end{center}

\twocolumngrid

\section{Results}\label{sec:results}

\subsection{The kinematic constraints in phonon dynamics}\label{sec:kinematic}
Before we present the numerical results of the sound-attenuation coefficient, we look at the kinematic constraints for the scattering of the phonons on Majorana fermions, which play a key role in the polarization bubble calculation. 

In our MF calculations, we consider a translationally invariant system, and the Majorana fermion-phonon scattering is constrained by both energy and momentum conservation, which makes the phonon scattering  on Majoarana fermions strongly velocity dependent. It can happen through two distinct channels\textemdash the ph and pp channels, seen in the one-loop bubbles in Fig.~\ref{Fig:baseset}(b).

In the ph-process, a phonon mode with momentum $\textbf{q}$ and energy $\Omega_{\textbf{q}}=v_s|\textbf{q}|$ scatters a fermion with momentum $\textbf{k}$ to a fermion with momentum $\textbf{k}+\textbf{q}$, with a corresponding energy constraint $\epsilon_{\textbf{k}+\textbf{q},m}-\epsilon_{\textbf{k},l}=\Omega_{\textbf{q}}$, with $m,l$ labeling fermionic flavors. The predominant way to satisfy these constraints in the ph-process is for both momenta $\textbf{k}$ and $\textbf{k}+\textbf{q}$ to  be in the vicinity of the same Dirac point $\textbf{K}$ (or $-\textbf{K}$) with energies on the same Dirac cone.  

For the pp process, a phonon with energy $\Omega_{\textbf{q}}$ decays into two fermions with positive energy, one with momentum $-\textbf{k}$ and another with momentum $\textbf{k}+\textbf{q}$, and  the energy constraint reads $\epsilon_{-\textbf{k},m}+\epsilon_{\textbf{k}+\textbf{q},l}=\Omega_{\textbf{q}}$. Opposite to the ph-process case, the predominant way to satisfy these constrains in the  pp-process involves two cones, with one fermion of momentum  $-\textbf{k}$ in the vicinity of the Dirac point $\textbf{K}$ and the other fermion with momentum $\textbf{k}+\textbf{q}$ in the vicinity the other Dirac point $-\textbf{K}$, and both Dirac cones are needed to satisfy these constraints. Beyond the Dirac-like dispersion at the $\textbf{K}$-points, there is another way to satisfy the pp-process, namely to keep both momenta $-\textbf{k}$ and $\textbf{k}+\textbf{q}$ near zero, in the vicinity of the BZ center. This becomes relevant with many states becoming significantly lowered in energy at the $\Gamma$ point.


From the above we can see that the energy spectra of the fermions and phonons form the phase space for these constraints and define the relative contribution of the ph- and pp-processes to the phonon scattering. In the earlier studies of the phonon dynamics in the pure Kitaev model (analytically   in the zero-flux sector \cite{Mengxing2020}  and numerically for the random fluxes  \cite{Metavitsiadis2020,Kexin2021}), 
 it was shown that  only the 
low-energy states on the Dirac cones, characterized by Fermi velocity $v_F$, are relevant for the scattering of acoustic phonons with small $q$. There, whether the ph or pp constraints were  satisfied, they could be interpreted geometrically by considering intersections between the fermionic Dirac cone and the phonon energy cone $\Omega=v_sq$. From this geometric intersection of cones, it followed that ph processes are possible only for $v_s<v_F$ and pp processes are possible only for $v_s\geq v_F$. Differently from \cite{Mengxing2020,Kexin2021}, here the flux degrees of freedom are captured in the flat bands, which acquire a small dispersion when moving away from the pure Kitaev point, as seen in Fig.~\ref{Fig:disp}. Also,  $J$ and $\Gamma$ interactions modify the Fermi velocity $v_F$. 
As a result we get more low-energy states in addition to those on the Dirac cones. 
Altogether, this leads to  the change in the contribution from the ph and pp processes to the phonon scattering, and makes it possible to get ph-processes even for $v_s>v_F$ and vice versa. We indeed see this from our numerical calculations. In general, the overall intensity of the phonon scattering will be increased due to the presence of  $J$ and $\Gamma$. 
\subsection{Numerical results for the sound attenuation}\label{sec:res}
The contribution to the phonon self-energy coming from the $A_{1g}$ irrep is small compared to the contribution from the $E_g$ symmetry channel. While we still show the numerical  results for the $A_{1g}$ channel  in Appendix~\ref{app:addnumres},  in the following we focus only on the results from the  $E_g$ contribution to the sound attenuation.

\subsubsection{Sound attenuation for $v_s<v_F$}\label{sec:res_num_1}

\begin{figure}
	\centering
	\includegraphics[width=1.0\columnwidth]{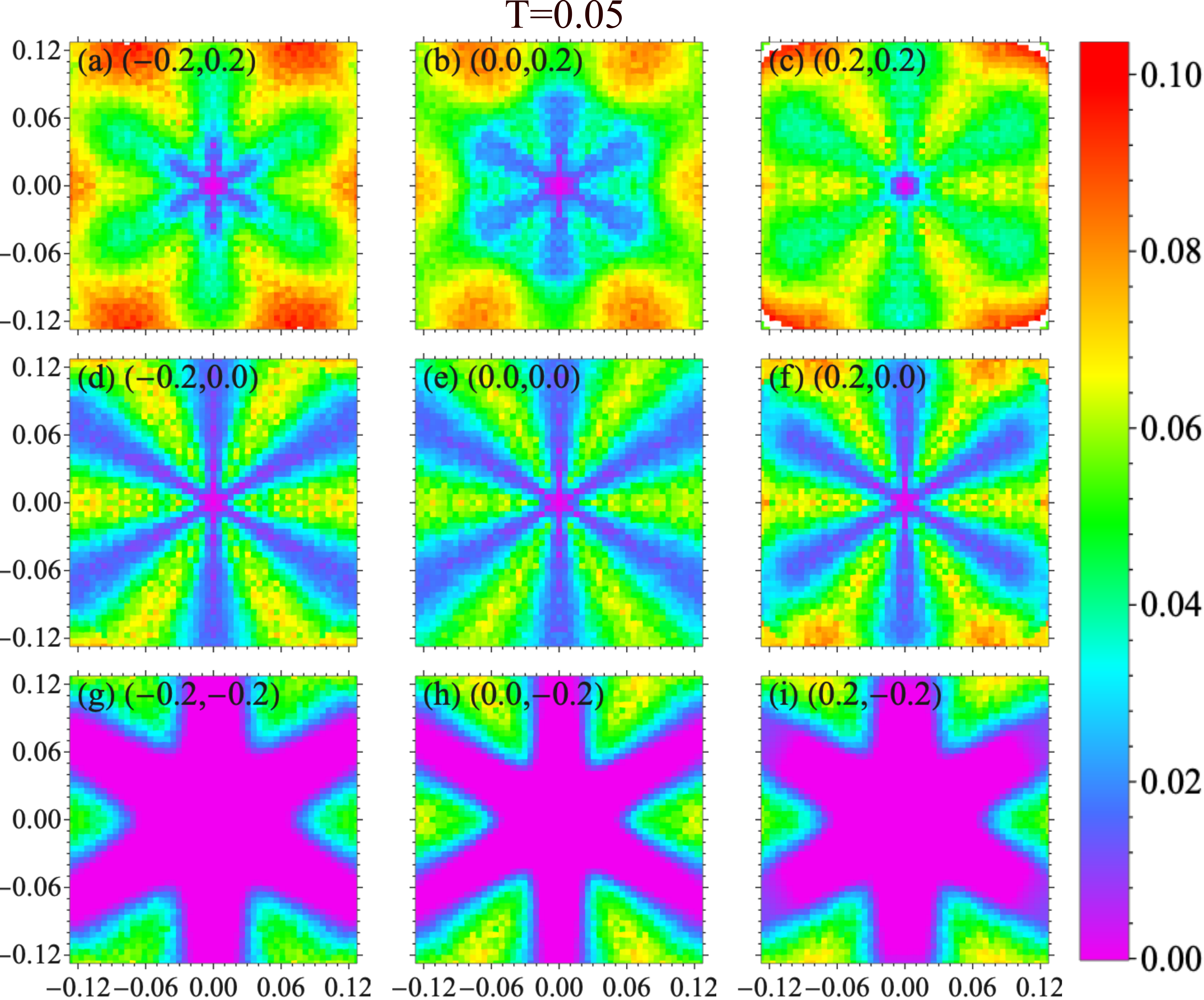}
     \caption{The longitudinal component of the sound-attenuation coefficient $\alpha_s^{\parallel}(\textbf{q})$ for $v_s>v_F$, for different values of ($\Gamma$,$J$) with $K=-1$, fixed. We show plots for $T=0.05$ in units of the Kitaev interaction $K$. The ph contribution for $v_s>v_F$ is vanishingly small. Therefore, we only plot the pp contribution coming from the $E_g$ irrep. The phonon momentum $\textbf{q}$ belongs to the region $(q_x,q_y)\in [-0.12\pi,0.12\pi]^2$. We used $v_s=3.5$ with $v_F=3$, determined from the slope of the Dirac cone of the Kitaev spectrum. The imaginary energy broadening was taken to be $\delta=0.2$.
}
\label{Fig:alon_pp_vgf_grid}
\end{figure}


When $v_s<v_F$, the major contribution to the sound attenuation comes from the ph processes. We find an enhanced magnitude of the sound attenuation for $J>0$ [see Figs.~\ref{Fig:alon_vlf_grid} (a)-(c) and (j)-(l)], and a slightly diminished magnitude for $J<0$ [see Figs.~\ref{Fig:alon_vlf_grid} (g)-(i) and (p)-(r)]. The $\Gamma$ term affects the magnitude of the sound attenuation only marginally at small temperature, with enhanced magnitude for $\Gamma<0$ and diminished magnitude for $\Gamma>0$. The combination of these two effects renders Fig.~\ref{Fig:alon_vlf_grid} (a), with $J>0, \Gamma<0$, the brightest. The six-fold symmetry of the sound attenuation is preserved for all values of $J$ and $\Gamma$. As the temperature increases we see the flower shape of the sound attenuation spreading out with increasing magnitude. At $T\sim 0.3$, we see essentially equal magnitudes of the sound attenuation for $J>0$ and $J<0$. The six-fold symmetry of the sound attenuation is still preserved; however, the pattern changes slightly from the flower shape. We also find pp contributions for the case of $J>0,\Gamma>0$, which has a sixfold symmetry and decreases in magnitude with increasing temperature.

\subsubsection{Sound attenuation for $v_s>v_F$}\label{sec:res_num_2}
For $v_s>v_F$, the major contribution comes from the pp processes.  The sound attenuation (shown in Fig.~\ref{Fig:alon_pp_vgf_grid}) again has the sixfold angular symmetry.
Its  magnitude is almost independent on the temperature, so we only show  our results  at $T=0.05$.
As we can see, 
the sound attenuation has the maximum magnitude for 
 $J>0$ [see Figs.~\ref{Fig:alon_pp_vgf_grid} (a)-(c)]  and the minimum for  $J<0$ [see Figs.~\ref{Fig:alon_pp_vgf_grid} (g)-(i)]. The $\Gamma$ term has no significant effect on the magnitude of the sound attenuation. As temperature increases the sound attenuation due to the pp-processes decreases  as $1/T$ with increasing temperature. 

\subsubsection{Temperature evolution of the sound attenuation}
Our numerical calculations show that the sound attenuation increases linearly  in temperature $T$ from the ph processes,  and decreases as  $1/T$  with increasing temperature from the pp processes. In  Fig.~\ref{Fig:alpha_temp} we see  this characteristic behavior  in the attenuation coefficient of the longitudinal acoustic  phonon computed
in the generic $J$-$K$-$\Gamma$ model  with subdominant $J$ and $\Gamma$.  The ph processes' linear behavior is in agreement with  recent studies in the pure Kitaev model \cite{Metavitsiadis2020,Mengxing2020,Haoxiang2021}, which would only allow for ph processes when $v_s<v_F$. Since  $J$ and 
 $\Gamma$ modify the Fermi velocity, and also lower the energy of the states near the center of the BZ, it allows for simultaneous  contributions from both the ph and the pp processes.   We see, however, that for this set of parameters, the pp contribution  remains subdominant. Note also that the phonon momentum for this calculation was taken to be $|\textbf{q}|=0.008\pi$, with $\theta_{\textbf{q}}=\pi/6$, which corresponds to the direction of the  maximum of  attenuation in the sixfold symmetric petal structure of the attenuation coefficient of the longitudinal mode \cite{Mengxing2020}.
 

\subsection{Discussion}\label{Disc}

As we briefly discussed above, the overall intensity of the sound attenuation is determined by the combined effect of the density of low-energy fermionic states, their population which follows the Fermi-Dirac statistics, and the strength of the  coupling vertices.   Subdominant Heisenberg interaction $J$ and off-diagonal symmetric coupling  $\Gamma$ modify both the energy and the structure of the low-energy states, and they do it  in a distinct way.
From Fig.~\ref{Fig:disp} we see that the  slope of the Dirac's cone, characterized by $v_F$, decreases for $J>0$ and increases for $J<0$ compared to the pure Kitaev case, thus changing the ratio between $v_F$ and $v_s$.  In principle, this means that subdominant interactions can change the nature of the phonon scattering processes (from pp to ph, or vice versa), but
for our choice of $v_s$ it does not happen.
 The $\Gamma$ interaction does not affect $v_F$ significantly but it is very effective in adding dispersion and fully splitting the degeneracy
 of the three ``flux" bands.  Moreover, $J>0$ leads to the lowering of states near the center of the BZ ($\Gamma$ point)  since the system is approaching the transition to the stripy phase
 \cite{Chaloupka2010,Schaffer2012} and $\Gamma<0$ leads to the lowering of states near the $M$ point of the BZ, as the system approaches the zigzag ordered state \cite{JRau2014}.
  These lowered states will contribute to the sound attenuation at low temperatures, possibly through both pp and ph processes, as seen in Fig.~\ref{Fig:alon_vlf_grid}.

\begin{figure}
	\centering
	\includegraphics[width=0.8\columnwidth]{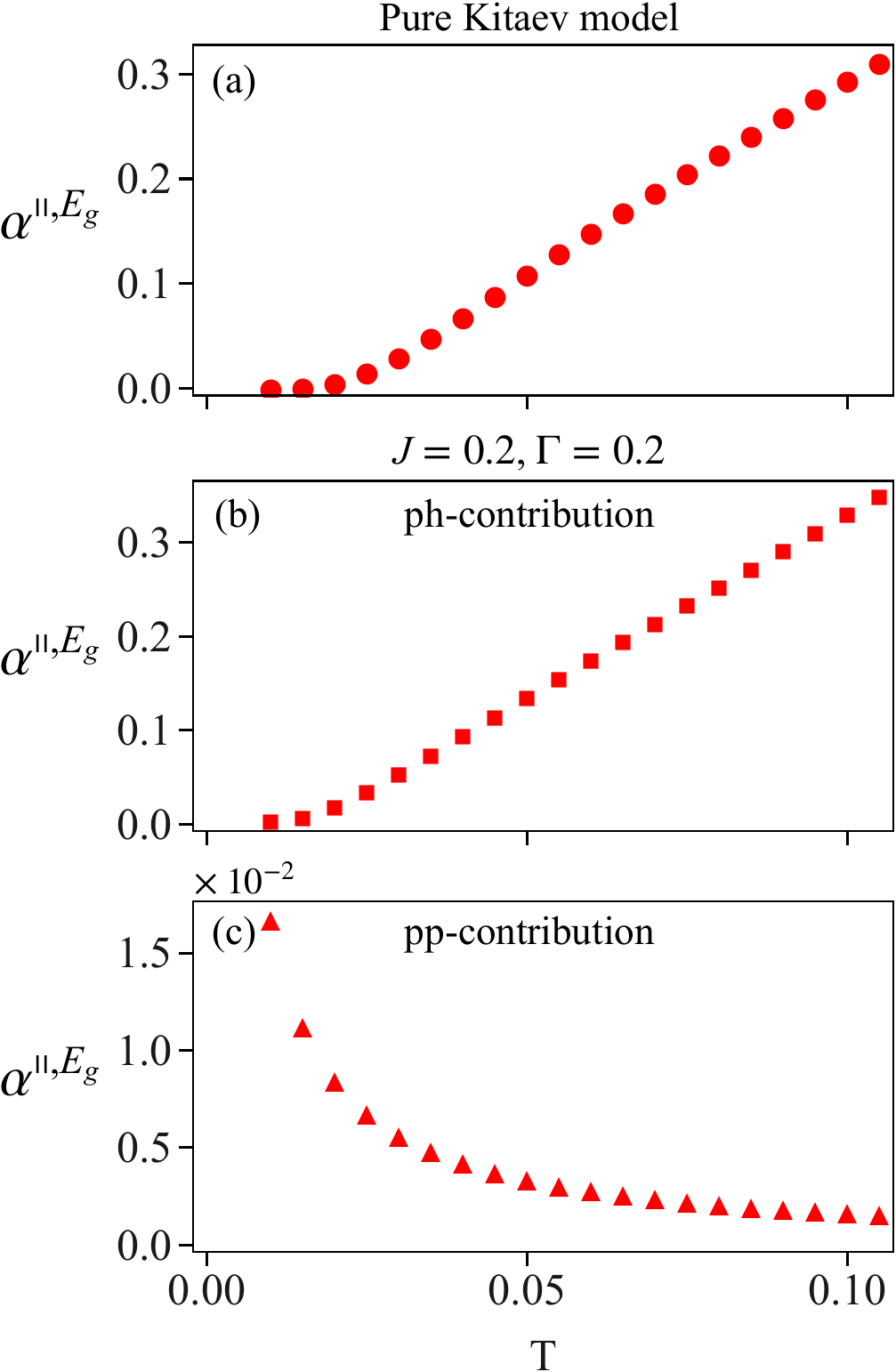}
     \caption{The temperature evolution of the sound-attenuation coefficients $\alpha^{\parallel,E_g}$ from the ph-processes for the pure Kitaev model (a), from the ph processes for $J=0.2, K=-1, \Gamma=0.2$ (b), and from the pp processes for $J=0.2, K=-1, \Gamma=0.2$, (c). The phonon momentum for this calculation was taken to be $|\textbf{q}|=0.008\pi$, with $\theta_{\textbf{q}}=\pi/6$. The imaginary energy broadening was taken to be $\delta=0.02.$ Note that, for both ph and pp plots, a maxima for $\alpha^{\parallel, E_g}$ coincides with a minima for $\alpha^{\perp,E_g}$ at $\theta_{\textbf{q}}=\pi/6$, \cite{Mengxing2020}. Here we omit the plots for $\alpha^{\perp,E_g}$ for which we get the same temperature dependence along one of its maxima directions (e.g., for $\theta_{\textbf{q}}=0$).
}
\label{Fig:alpha_temp}
\end{figure}

The contribution from the  pp processes  is maximum at $T=0$, when all low-energy states are unoccupied. We can see that the sound attenuation  is the ``brightest"  for $J>0$ 
(see Fig.~\ref{Fig:alon_pp_vgf_grid}). As temperature increases, the low-energy  states start getting
populated,
and the phonons can no longer decay into these filled states. We therefore see a decrease in the sound attenuation with increasing temperature from the pp processes in Fig.~\ref{Fig:alpha_temp} (c).
 Eventually,  at higher temperatures, ph contribution also decreases with increasing temperature, since  unoccupied states become unreachable due to the kinematic constraints.

{\color{black}

Finally, let us  compare our findings with recent ultrasound measurements in $\alpha-$RuCl$_3$ ~\cite{Andreas2023}. The reported velocity of 
the in-plane longitudinal acoustic phonons $v^{T}_{s,\text{expt}} \sim 16\, \text{ meV\AA}$, 
and for the in-plane transverse phonons $v^{L}_{s,\text{expt}} \sim 20 \,\text{ meV\AA}$. 
The estimate of the Fermi velocity of the Majorana fermions reported in the literature is about  $v_{F,\text{expt}} \sim 18\, \text{ meV\AA}$ \cite{Miao2020,LebertPRB2022,Haoxiang2021}. 
This puts $\alpha-$RuCl$_3$ in the regime $v_s^{T} < v_F < v_s^{L}$, and magically allows one to  study both  sound-attenuation channels  in the same compound. Indeed, it was found that phonon scattering  depends sensitively on the value of the phonon velocity:  while in-plane transverse modes show linear in $T$  behavior of the sound attenuation consistent with the character of the dominant  ph process, in-plane longitudinal mode shows almost  $T$ independent  decay for not too high temperatures. This behavior is consistent with the idea that the phonon attenuation is mostly occurring from scattering of the low-energy fermionic excitations, which describe the spin dynamics of the underlying Kitaev magnet.

Although the actual value of $K$ in $\alpha-$RuCl$_3$ has varying predictions, with a recent table summary found in Ref.~\cite{MaksimovPRR2020}, assuming an approximate value of $K\sim-6.0$ $\text{meV}$, would make our $v_F=3$ in units of $K$ corresponding to $18$ $\text{meV \AA}$. While our choice for the larger $v_s=3.5\rightarrow 21$ $\text{meV \AA}$, where sound attenuation is dominated by the pp process,
   is very  close to the actual value of $v^{L}_{s,\text{expt}}$, our choice for the smaller $v_s=0.3\rightarrow 1.8$ $\text{meV \AA}$, where sound attenuation is dominated by the ph process, is significantly smaller than the actual value of $v^{T}_{s,\text{expt}}$. To have a better comparison with  $\alpha-$RuCl$_3$, in Appendix~\ref{app:addnumres} we provide additional plots of sound attenuation with $v_s=$2, 2.25, and 2.5 (12, 13.5, and 15 meV \AA,  respectively).  There, we can see some modifications of the  pattern of the sound attenuation in the momentum space. This happens because  for  larger sound velocity the scattering processes are not confined anymore within the low-energy  cone's states  but also involve the states in the BZ originated primarily  from ``vison" branches. These modifications are particularly visible for enhanced temperatures.
}

\section{Conclusion}\label{Conc}

In this work, we studied the phonon dynamics in the QSL phase of the extended $J$-$K$-$\Gamma$ model and  showed that  the signatures  of the spin fractionalization in the Kitaev candidate materials can be seen in the sound-attenuation  of acoustic phonons measured by
 ultrasound experiments \cite{Andreas2023}.   We computed the sound attenuation coefficient by relating it to the imaginary part of the
phonon polarization bubble
and  explored how it changes in the presence of residual interactions. Since  generic $J$-$K$-$\Gamma$  QSL   is not exactly solvable, we utilized
a conventional mean-field fermionic parton approach. 
We find that similarly to the   phonon attenuation in the pure Kitaev spin liquid, the low-temperature scattering between acoustic phonons of velocity $v_s$ and Majorana fermions of
velocity $v_F$  is controlled by the relative magnitude of these velocities, which at low energies defines the kinematic constraints.
When the sound velocity $v_s$ is smaller than the Fermi velocity $v_F$,
the sound attenuation at low temperatures  is
dominated by the microscopic processes in which a fermion is excited to a higher energy  state (dubbed
as ph processes), with the attenuation
rate linear in temperature.
When the sound velocity  $v_s$ is larger than the Fermi velocity $v_F$, the phonon attenuation happens mostly through the microscopic processes
when a phonon decays into two fermions (dubbed as pp processes), which  has the maximum intensity at $T=0$ and then decreases with temperature, which can be interpreted as a consequence of the Pauli exclusion principle. 
However, contrary to the pure Kitaev model,  in the generic $J$-$K$-$\Gamma$ model both ph and pp processes can contribute simultaneously.
 We also find that  a distinct sixfold  symmetry in the sound attenuation is  still preserved even in the presence of residual interactions.
 

 As a final remark, we would like to emphasize that in our  computation we  made use of the $D_{3d}$ point-group symmetry of
the generalized  $J$-$K$-$\Gamma$ model. In this group, the magnetoelastic coupling is  reduced to two contributions, the $A_{1g}$ and the $E_g$. We find the $E_g$ contributions to the sound attenuation to be much larger than $A_{1g}$, assuming that  the couplings $\lambda_{A_{1g}}$ and  $\lambda_{E_{g}}$ are of the same order.
Although this is a reasonable assumption for a qualitative understanding,  a quantitative comparison   with experimental data requires one to study
the magnetoelastic couplings and elastic modulus tensor of Kitaev materials more carefully, e.g., by
 first-principles calculations. Moreover, to better understand the experimental signatures  in the candidate materials which are  magnetically ordered and only proximate to the QSL state, such as $\alpha$-$\text{RuCl}_3$ which is in the zigzag magnetic phase,
an analysis of sound attenuation in the magnetically ordered states neighboring to the quantum spin liquid state would also need to be examined.

\section{Acknowledgements}
The authors thank  Wolfram Brenig, Kexin Feng,
Rafael Fernandes, Andreas Hauspurg, Mengxing Ye, and Sergei Zherlitsyn for valuable discussions.
This work was supported by the U.S. Department of Energy, Office of Science, Basic Energy Sciences under Award No. DE-SC-0018056.
 N.B.P.  also acknowledges the hospitality and partial support  of the Technical University of Munich\textendash Institute for Advanced
Study.

\appendix

\section{Details of the  mean field spin Hamiltonian}\label{app:mf_spin_ham}
\subsection{The structure of the MF spin Hamiltonian}

The Majorana MF matrix $H_{\text{s}}^{\text{MF}}(\textbf{k})$ in Eq.~\eqref{majo_ham1}, which is written in the Majorana fermion basis $\mathcal{A}_{\textbf{k}}$ in Eq.~\eqref{majo_ham1_basis}, is found by performing a Fourier transform of Eq.~\eqref{h_s_maj} and has the matrix form  
\begin{equation}\label{majo_ham2}
H_{\text{s}}^{\text{MF}}(\textbf{k})=\left[\begin{array}{cc}
0&iM_{A_{1g}}(\textbf{k})\\
-i[M_{A_{1g}}(\textbf{k})]^{\dagger}&0
\end{array}\right],
\end{equation}
where $M_{A_{1g}}(\textbf{k})$ reads
\begin{equation}\label{m_mat_a1g}
M_{A_1g}(\textbf{k})=\left[\begin{array}{cccc}
f_{cc}(\textbf{k})&0&0&0\\
0&f_{xx}(\textbf{k})&f_{xy}(\textbf{k})&f_{xz}(\textbf{k})\\
0&f_{xy}(\textbf{k})&f_{yy}(\textbf{k})&f_{yz}(\textbf{k})\\
0&f_{xz}(\textbf{k})&f_{yz}(\textbf{k})&f_{zz}(\textbf{k})\\
\end{array}\right].
\end{equation}
The entries of $M_{A_{1g}}(\textbf{k})$ are
\begin{equation}\label{fa1g_1}
\begin{split}
f_{cc}(\textbf{k})=&-(J+K)\bigl\{\kappa_x^{xx}e^{i\textbf{k}\cdot\textbf{n}_1}+\kappa_y^{yy}e^{i\textbf{k}\cdot\textbf{n}_2}+\kappa_z^{zz}\bigr\}\\
&-J\bigl\{(\kappa_x^{yy}+\kappa_x^{zz})e^{i\textbf{k}\cdot\textbf{n}_1}
+(\kappa_y^{xx}+\kappa_y^{zz})e^{i\textbf{k}\cdot\textbf{n}_2}\\
&+(\kappa_z^{xx}+\kappa_z^{yy})\bigr\}
-\Gamma\bigl\{(\kappa_x^{yz}+\kappa_x^{zy})e^{i\textbf{k}\cdot\textbf{n}_1}\\
&+(\kappa_y^{xz}+\kappa_y^{zx})e^{i\textbf{k}\cdot\textbf{n}_2}
+(\kappa_z^{xy}+\kappa_z^{yx})\bigr\},
\end{split}
\end{equation}
\begin{equation}\label{fa1g_2}
\begin{split}
f_{xx}(\textbf{k})=&-(J+K)\kappa_x^0e^{i\textbf{k}\cdot\textbf{n}_1}-J(\kappa_y^0e^{i\textbf{k}\cdot\textbf{n}_2}+\kappa_z^0),\\
f_{yy}(\textbf{k})=&-(J+K)\kappa_y^0e^{i\textbf{k}\cdot\textbf{n}_2}-J(\kappa_x^0e^{i\textbf{k}\cdot\textbf{n}_1}+\kappa_z^0),\\
f_{zz}(\textbf{k})=&-(J+K)\kappa_z^0-J(\kappa_x^0e^{i\textbf{k}\cdot\textbf{n}_1}+\kappa_y^0e^{i\textbf{k}\cdot\textbf{n}_2}),
\end{split}
\end{equation}
\begin{equation}\label{fa1g_3}
\begin{split}
f_{yz}(\textbf{k})=&-\Gamma\kappa_x^0e^{i\textbf{k}\cdot\textbf{n}_1},\\
f_{xz}(\textbf{k})=&-\Gamma\kappa_y^0e^{i\textbf{k}\cdot\textbf{n}_2},\\
f_{xy}(\textbf{k})=&-\Gamma\kappa_z^0,
\end{split}
\end{equation}
 where $\textbf{n}_1$ and $\textbf{n}_2$ are  unit vectors defined in Fig. \ref{Fig:baseset}.



The Majorana fermion basis $\mathcal{A}_{\textbf{k}}$ can be rewritten in the complex fermion basis $\mathcal{F}_{\textbf{k}}$ as  
\begin{equation}\label{comp_basis}
\begin{array}{ccc}
     \left[\begin{array}{c}
c_{\textbf{k},A}\\
b_{\textbf{k},A}^{x}\\
b_{\textbf{k},A}^{y}\\
b_{\textbf{k},A}^{z}\\
c_{\textbf{k},B}\\
b_{\textbf{k},B}^{x}\\
b_{\textbf{k},B}^{y}\\
b_{\textbf{k},B}^{z}
\end{array}\right] = \mathcal{A}^T_{\textbf{k}}& \xrightarrow[]{V} & \mathcal{F}^T_{\textbf{k}}=\left[\begin{array}{c}
    f_{\textbf{k},A,\uparrow}\\
    f_{\textbf{k},A,\downarrow}\\
    f_{-\textbf{k},A,\uparrow}^{\dagger}\\
    f_{-\textbf{k},A,\downarrow}^{\dagger}\\
    f_{\textbf{k},B,\uparrow},\\
    f_{\textbf{k},B,\downarrow}\\
    f_{-\textbf{k},B,\uparrow}^{\dagger}\\
    f_{-\textbf{k},B,\downarrow}^{\dagger}
    \end{array}\right]. \end{array}
\end{equation}
The transformation matrix $V$ follows from Eq.~\eqref{map_fc} and reads
\begin{equation}\label{comp_basisV}    
    V=\left[
    \begin{array}{cc}
v&0\\
0&v\\
\end{array}\right],\quad v=\left[\begin{array}{cccc}
 1 &  0 &  0 & -i\\
 0 & -i &  1 &  0\\
 1 &  0 &  0 &  i\\
 0 &  i &  1 &  0\\
\end{array}\right]
\end{equation}
This allows us to transform the MF Hamiltonian from the Majorana fermions to the complex fermions:
\begin{equation}
\begin{array}{c}
\mathcal{A}_{-\textbf{k}}^TH_{\text{s}}^{\text{MF}}(\textbf{k})\mathcal{A}_{\textbf{k}}= \mathcal{F}^{\dagger}_{\textbf{k}}H_{\text{f}}^{\text{MF}}(\textbf{k})\mathcal{F}_{\textbf{k}}.
\end{array}
\end{equation}
The matrix $H_{\text{f}}^{\text{MF}}(\textbf{k})$ (written in the basis of the complex fermions) can be diagonalized by a unitary transformation for a given set of the MF parameters, resulting
in the spectrum, presented in Fig.~\ref{Fig:disp}.





\subsection{The self-consistent MF solution}
The solution of the MF spin Hamiltonian is calculated using the iterative self-consistent method. This consists of three steps. Step 1: for a set of mean-field values, evaluate eigenvalues and eigenvectors of $H_{\text{f}}^{\text{MF}}(\textbf{k})$ at every momentum point. Step 2: recalculate the MF values from the self-consistent equations. Step 3: check the fermionic constraints and substitute the MF values back into the MF Hamiltonian. Repeat steps 1\textendash 3 until convergence.

We start with step 1. Step 1 is performed by substituting a good guess for the values of the MF parameters in the complex fermion Hamiltonian $H_{\text{s,f}}^{\text{MF}}(\textbf{k})$. This Hamiltonian is then diagonalized at every momentum point
\begin{equation}
    \begin{array}{c}
\mathcal{F}^{\dagger}_{\textbf{k}}H_{\text{f}}^{\text{MF}}(\textbf{k})\mathcal{F}_{\textbf{k}} = \mathcal{B}^{\dagger}_{\textbf{k}}D(\textbf{k})\mathcal{B}_{\textbf{k}}
    \end{array}
\end{equation}
with $\mathcal{F}_{\textbf{k}}=U(\textbf{k})\mathcal{B}(\textbf{k})$, $U(\textbf{k})$ being the diagonalizing matrix and $\mathcal{B}(\textbf{k})$ being the Bogoliubov eigenstates $\mathcal{B}^T_{\textbf{k}}=\left[\begin{array}{cccccccc}
    \beta_{-\textbf{k},1}^{\dagger} &   \beta_{-\textbf{k},2}^{\dagger} &   \beta_{-\textbf{k},3}^{\dagger} &   \beta_{-\textbf{k},4}^{\dagger} &  \beta_{\textbf{k},1} &  \beta_{\textbf{k},2} &  \beta_{\textbf{k},3} &  \beta_{\textbf{k},4} 
\end{array}\right]$. The columns of the $U(\textbf{k})$ matrix consist of eigenvectors of the Hamiltonian at momentum point $\textbf{k}$.

To proceed with step 2, we assume that the MF parameters have a uniform value on the lattice, and calculate them on average. For example, the bond mean field parameter $\langle ic_ic_j\rangle_x$
 would be evaluated from the self-consistency equation



\begin{equation}\label{sel__cons1}
\begin{split}
    \langle ic_ic_j\rangle_x=&\frac{i}{N}\sum_{\langle ij\rangle_x,i\in A}\langle c_ic_j\rangle\\
    =&\frac{i}{N}\sum_{\textbf{k}}\langle c_{\textbf{k},A}c_{-\textbf{k},B}\rangle e^{-i\textbf{k}\cdot\textbf{n}_1}\\
    =&\frac{i}{N}\sum_{\textbf{k}}\langle(f_{\textbf{k},A,\uparrow}f_{-\textbf{k},B,\uparrow}+f_{\textbf{k},A,\uparrow}f_{\textbf{k},B,\uparrow}^{\dagger})\\
    &\times(f_{-\textbf{k},A,\uparrow}^{\dagger}f_{-\textbf{k},B,\uparrow+}+f_{-\textbf{k},A,\uparrow}^{\dagger}f_{\textbf{k},B,\uparrow+}^{\dagger})\rangle e^{-i\textbf{k}\cdot\textbf{n}_1}.
\end{split}    
\end{equation}
We can evaluate the variours terms in \eqref{sel__cons1} by using the diagonalizng matrix
\begin{equation}
\begin{split}
    \frac{i}{N}&\sum_{\textbf{k}}\langle f_{\textbf{k},A,\uparrow}f_{\textbf{k},B,\uparrow}^{\dagger}\rangle e^{-i\textbf{k}\cdot\textbf{n}_1}\\
    =&\frac{i}{N}\sum_{\textbf{k}}\sum_{l,l'}\langle U(\textbf{k})_{m,l}{\mathcal B}(\textbf{k})_lU(\textbf{k})_{n,l'}^*{\mathcal B}(\textbf{k})^{\dagger}_{l'}\rangle e^{-i\textbf{k}\cdot\textbf{n}_1}\\
    =&\frac{i}{N}\sum_{\textbf{k}}\sum_{l=5}^8U(\textbf{k})_{m,l}U(\textbf{k})_{n,l}^* e^{-i\textbf{k}\cdot\textbf{n}_1}.
    \end{split}
\end{equation}
Here $m,n$ correspond to the fermion flavors ($\textbf{k},A,\uparrow$) and ($\textbf{k},B,\uparrow$), respectively, and we have summed over the unoccupied states corresponding to the part of the eigenvector with positive eigenvalues. Similarly, we can evaluate other bilinear combinations of the complex fermions to get the updated values of the MF parameters.

Arriving now at step 3, we substitute these updated values of the MF parameters back into $H_{\text{f}}^{\text{MF}}(\textbf{k})$. We check that the fermionic single occupancy constraints are still preserved, by checking that $\langle f_{i,A(B),\uparrow}^{\dagger}f_{i,A(B),\uparrow}+f_{i,A(B),\downarrow}^{\dagger}f_{i,A(B),\downarrow}\rangle=1$. This ensures that we are working in the physical space of spin operators on the MF level. We iterate steps 1\textendash3 until we get a convergence of the MF parameter values.


\begin{table}[t]
\begin{center}
\begin{tabular}{| c| c|  }
\hline
irrep & $A_{1g}$ \\ 
\hline
$K$ &$f_{K}^{A_1g}=\sigma_{\textbf{r}}^x\sigma_{\textbf{r}+\textbf{M}_x}^x+\sigma_{\textbf{r}}^y\sigma_{\textbf{r}+\textbf{M}_y}^y+\sigma_{\textbf{r}}^z\sigma_{\textbf{r}+\textbf{M}_z}^z$ \\  
\hline
$J$&$f_{J}^{A_1g}=\sigma_{\textbf{r}}^x\sigma_{\textbf{r}+\textbf{M}_{y,z}}^x+\sigma_{\textbf{r}}^y\sigma_{\textbf{r}+\textbf{M}_{x,z}}^y+\sigma_{\textbf{r}}^z\sigma_{\textbf{r}+\textbf{M}_{x,y}}^z$ \\
\hline  
$\Gamma$&$f_{\Gamma}^{A_1g}=\sigma_{\textbf{r}}^y\sigma_{\textbf{r}+\textbf{M}_{x}}^z+\sigma_{\textbf{r}}^z\sigma_{\textbf{r}+\textbf{M}_{x}}^y+\sigma_{\textbf{r}}^x\sigma_{\textbf{r}+\textbf{M}_{y}}^z$ \\
&$+\sigma_{\textbf{r}}^z\sigma_{\textbf{r}+\textbf{M}_{y}}^x+\sigma_{\textbf{r}}^x\sigma_{\textbf{r}+\textbf{M}_{z}}^y+\sigma_{\textbf{r}}^y\sigma_{\textbf{r}+\textbf{M}_{z}}^x$ \\  
\hline
Phonon&$\epsilon_{xx}+\epsilon_{yy}$\\
\hline  
\end{tabular}
\caption{Basis functions of spins and phonons in the $A_{1g}$ irrep of the $D_{3d}$ point group. In the expressions for the $f_{J}$ functions, the summation over sub-indices of $\textbf{M}$ is assumed. }
\label{table:irrA1g}
\end{center}
\end{table}

 \begin{table}
\begin{center}
\begin{tabular}{| c| c| c|  }
\hline
irrep & $E_g$\\ 
\hline
$K$  & $(f_{K}^{E_{g,1}},f_{K}^{E_{g,2}})=(\sigma_{\textbf{r}}^x\sigma_{\textbf{r}+\textbf{M}_x}^x+\sigma_{\textbf{r}}^y\sigma_{\textbf{r}+\textbf{M}_y}^y-2\sigma_{\textbf{r}}^z\sigma_{\textbf{r}+\textbf{M}_z}^z,$\\
&$\sqrt{3}(\sigma_{\textbf{r}}^x\sigma_{\textbf{r}+\textbf{M}_x}^x-\sigma_{\textbf{r}}^y\sigma_{\textbf{r}+\textbf{M}_y}^y))$ \\  
\hline
$J$ & $(f_{J}^{E_{g,1}},f_{J}^{E_{g,2}})=(\sigma_{\textbf{r}}^x\sigma_{\textbf{r}+\textbf{M}_{y,z}}^x+\sigma_{\textbf{r}}^y\sigma_{\textbf{r}+\textbf{M}_{x,z}}^y-2\sigma_{\textbf{r}}^z\sigma_{\textbf{r}+\textbf{M}_{x,y}}^z,$\\
&$\sqrt{3}(\sigma_{\textbf{r}}^x\sigma_{\textbf{r}+\textbf{M}_{y,z}}^x-\sigma_{\textbf{r}}^y\sigma_{\textbf{r}+\textbf{M}_{x,z}}^y))$ \\  
\hline  
$\Gamma$ & $(f_{\Gamma}^{E_{g,1}},f_{\Gamma}^{E_{g,2}})=(\sigma_{\textbf{r}}^y\sigma_{\textbf{r}+\textbf{M}_{x}}^z+\sigma_{\textbf{r}}^z\sigma_{\textbf{r}+\textbf{M}_{x}}^y+\sigma_{\textbf{r}}^x\sigma_{\textbf{r}+\textbf{M}_{y}}^z$ \\
&$+\sigma_{\textbf{r}}^z\sigma_{\textbf{r}+\textbf{M}_{y}}^x-2(\sigma_{\textbf{r}}^x\sigma_{\textbf{r}+\textbf{M}_{z}}^y+\sigma_{\textbf{r}}^y\sigma_{\textbf{r}+\textbf{M}_{z}}^x),$ \\  
&$\sqrt{3}\left[\sigma_{\textbf{r}}^y\sigma_{\textbf{r}+\textbf{M}_{x}}^z+\sigma_{\textbf{r}}^z\sigma_{\textbf{r}+\textbf{M}_{x}}^y-(\sigma_{\textbf{r}}^x\sigma_{\textbf{r}+\textbf{M}_{y}}^z+\sigma_{\textbf{r}}^z\sigma_{\textbf{r}+\textbf{M}_{y}}^x)\right])$ \\  
\hline
Phonon&$\bigl(\epsilon_{xx}-\epsilon_{yy},2\epsilon_{xy}\bigr)$\\
\hline  
\end{tabular}
\caption{Basis functions of spins and phonons in the $E_{g}$ irrep of the $D_{3d}$ point group. In the expressions for the $f_{J}$ functions, the summation over subindices of $\textbf{M}$ is assumed.}
\label{table:irrEg}
\end{center}
\end{table}

\begin{figure*}
	\centering
\includegraphics[width=1.0\textwidth]{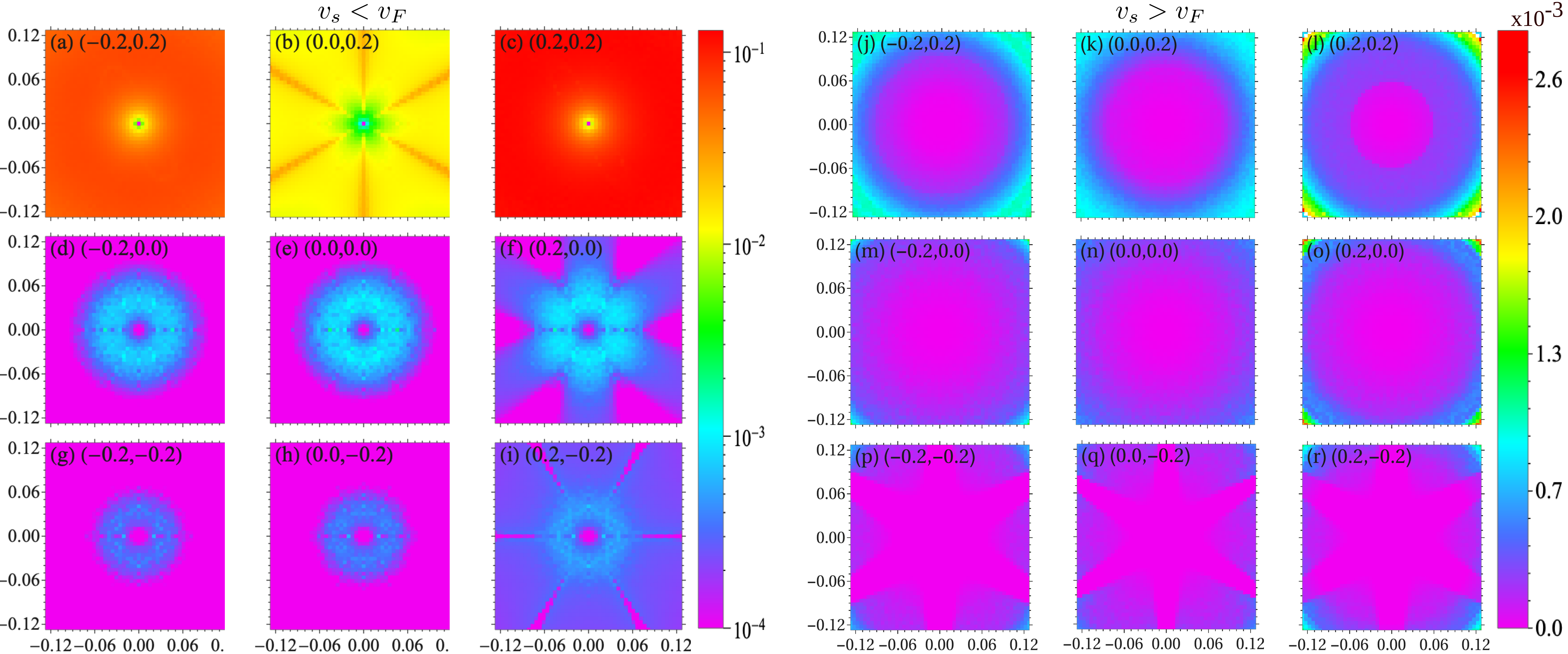}
   \caption{The $A_{1g}$ contribution to the longitudinal component of the sound attenuation coefficient $\alpha_s^{\parallel}(\textbf{q})$ for different values of ($\Gamma$,$J$), and $K=-1$ fixed, at $T=0.05$ in units of the Kitaev interaction strength $K$. The ph contribution coming from the $A_{1g}$ irrep is shown on the left, where $v_s=0.3<v_F$ on panels (a)\textendash(i), and the pp contribution coming from the $A_{1g}$ irrep is shown on the right, where $v_s=3.5>v_F$ on panels  (j)\textendash(r). The phonon momentum $\textbf{q}$ belongs to the region $(q_x,q_y)\in [-0.12\pi,0.12\pi]^2$, and $v_F=3$ determined from the slope of the Dirac cone of the Kitaev spectrum. The imaginary energy broadening was taken to be $\delta=0.2$.
}
\label{Fig:alon_ag_ph_and_pp}
\end{figure*}

\section{Details of the spin-phonon coupling Hamiltonian}\label{app:spin_phonon_coupling}
The basis functions for the $A_{1g}$ and $E_g$ irreps for the spins and phonon are given in Tables~\ref{table:irrA1g} and ~\ref{table:irrEg}. Using these  functions, we can explicitly write the symmetry allowed coupling Hamiltonian in Eq.~\eqref{hc_irr}. Through Fourier transforming and some algebra we arrive at Eq.~\eqref{eq:kspaceHc}, where we introduced the compact notation of the $H_{\text{c}}^{A_{1g}}$, $H_{\text{c}}^{E_{g,1}}$, and $H_{\text{c}}^{E_{g,2}}$ matrices. For  the $A_{1g}$ irrep, we get
\begin{equation}
    H_{\text{c}}^{A_{1g}}(\textbf{q},\textbf{k})=\left[\begin{array}{cc}
0&iM_{A_{1g}}(\textbf{k)}\\
-i[M_{A_{1g}}(\textbf{q}+\textbf{k})]^{\dagger}&0
\end{array}\right].
\end{equation}
Matrices $H_{\text{c}}^{E_{g,1}}$ and $H_{\text{c}}^{E_{g,2}}$ have the same structure as $H_{\text{c}}^{A_{1g}}$ with $M_{A_{1g}}$ replaced by $M_{E_{g,1}}$ and  $M_{E_{g,2}}$, respectively. The form of the $M$ matrices is
\begin{equation}\label{m_mat_eg}
M_{\rm{ir}}(\textbf{k})=\left[\begin{array}{cccc}
f_{cc,\rm{ir}}(\textbf{k})&0&0&0\\
0&f_{xx,\rm{ir}}(\textbf{k})&f_{xy,\rm{ir}}(\textbf{k})&f_{xz,\rm{ir}}(\textbf{k})\\
0&f_{xy,\rm{ir}}(\textbf{k})&f_{yy,\rm{ir}}(\textbf{k})&f_{yz,\rm{ir}}(\textbf{k})\\
0&f_{xz,\rm{ir}}(\textbf{k})&f_{yz,\rm{ir}}(\textbf{k})&f_{zz,\rm{ir}}(\textbf{k})\\
\end{array}\right].
\end{equation}
where index $\rm{ir}$=$A_{1g},E_{g,1},E_{g,2}$. The $M_{A_{1g}}(\textbf{k})$ and $M_{E_{g,(1,2)}}(\textbf{k})$ matrices enter the definition of the coupling vertices in Eqs.~\eqref{lambda_lon} and \eqref{lambda_tr}. The definitions of all the $f$ functions  are given below:
\begin{equation}
\begin{split}
f_{cc,E_{g,1}}(\textbf{k})=&-(J+K)\bigl\{\kappa_x^{xx}e^{i\textbf{k}\cdot\textbf{n}_1}+\kappa_y^{yy}e^{i\textbf{k}\cdot\textbf{n}_2}-2\kappa_z^{zz}\bigr\}\\
&-J\bigl\{(\kappa_x^{yy}+\kappa_x^{zz})e^{i\textbf{k}\cdot\textbf{n}_1}
+(\kappa_y^{xx}+\kappa_y^{zz})e^{i\textbf{k}\cdot\textbf{n}_2}\\
&-2(\kappa_z^{xx}+\kappa_z^{yy})\bigr\}
-\Gamma\bigl\{(\kappa_x^{yz}+\kappa_x^{zy})e^{i\textbf{k}\cdot\textbf{n}_1}\\
&+(\kappa_y^{xz}+\kappa_y^{zx})e^{i\textbf{k}\cdot\textbf{n}_2}-2(\kappa_z^{xy}+\kappa_z^{yx})\bigr\},
\end{split}
\end{equation}

\begin{equation}
\begin{split}
f_{xx,E_{g,1}}(\textbf{k})=&-(J+K)\kappa_x^0e^{i\textbf{k}\cdot\textbf{n}_1}-J(\kappa_y^0e^{i\textbf{k}\cdot\textbf{n}_2}-2\kappa_z^0),\\
f_{yy,E_{g,1}}(\textbf{k})=&-(J+K)\kappa_y^0e^{i\textbf{k}\cdot\textbf{n}_2}-J(\kappa_x^0e^{i\textbf{k}\cdot\textbf{n}_1}-2\kappa_z^0),\\
f_{zz,E_{g,1}}(\textbf{k})=&2(J+K)\kappa_z^0-J(\kappa_x^0e^{i\textbf{k}\cdot\textbf{n}_1}+\kappa_y^0e^{i\textbf{k}\cdot\textbf{n}_2}),
\end{split}
\end{equation}

\begin{equation}
\begin{split}
f_{yz,E_{g,1}}(\textbf{k})=-&\Gamma\kappa_x^0e^{i\textbf{k}\cdot\textbf{n}_1},\\
f_{xz,E_{g,1}}(\textbf{k})=&-\Gamma\kappa_y^0e^{i\textbf{k}\cdot\textbf{n}_2},\\
f_{xy,E_{g,1}}(\textbf{k})=&2\Gamma\kappa_z^0,
\end{split}
\end{equation}
\begin{equation}
\begin{split}
f_{cc,E_{g,2}}(\textbf{k})=&\sqrt{3}\biggl\{-(J+K)\bigl\{\kappa_x^{xx}e^{i\textbf{k}\cdot\textbf{n}_1}-\kappa_y^{yy}e^{i\textbf{k}\cdot\textbf{n}_2}\bigr\}\\
&-J\bigl\{(\kappa_x^{yy}+\kappa_x^{zz})e^{i\textbf{k}\cdot\textbf{n}_1}
-(\kappa_y^{xx}+\kappa_y^{zz})e^{i\textbf{k}\cdot\textbf{n}_2}\bigr\}\\
&-\Gamma\bigl\{(\kappa_x^{yz}+\kappa_x^{zy})e^{i\textbf{k}\cdot\textbf{n}_1}
-(\kappa_y^{xz}+\kappa_y^{zx})e^{i\textbf{k}\cdot\textbf{n}_2}\bigr\}\biggr\}
\end{split}
\end{equation}
\begin{equation}
\begin{split}
f_{xx,E_{g,2}}(\textbf{k})=&\sqrt{3}\biggl\{-(J+K)\kappa_x^0e^{i\textbf{k}\cdot\textbf{n}_1}+J\kappa_y^0e^{i\textbf{k}\cdot\textbf{n}_2}\biggr\},\\
f_{yy,E_{g,2}}(\textbf{k})=&\sqrt{3}\biggl\{(J+K)\kappa_y^0e^{i\textbf{k}\cdot\textbf{n}_2}-J\kappa_x^0e^{i\textbf{k}\cdot\textbf{n}_1}\biggr\},\\
f_{zz,E_{g,2}}(\textbf{k})=&\sqrt{3}\biggl\{-J(\kappa_x^0e^{i\textbf{k}\cdot\textbf{n}_1}-\kappa_y^0e^{i\textbf{k}\cdot\textbf{n}_2})\biggr\},
\end{split}
\end{equation}
\begin{equation}
\begin{split}
f_{yz,E_{g,2}}(\textbf{k})=&-\sqrt{3}\Gamma\kappa_x^0e^{i\textbf{k}\cdot\textbf{n}_1},\\
f_{xz,E_{g,2}}(\textbf{k})=&\sqrt{3}\Gamma\kappa_y^0e^{i\textbf{k}\cdot\textbf{n}_2},\\
f_{xy,E_{g,2}}(\textbf{k})=&0.
\end{split}
\end{equation}

\section{Matsubara frequency sums}\label{app:phonon_bubble}
The Matsubara Green's functions involved in Eq.~\eqref{pi_omega} are given below (we also explicitly show the intermediate step of the Matsubara summation using the residue method \cite{AltlandBook}):
\begin{equation}\label{matsubara_sum_ggbar}
    \begin{array}{rl}
    P_{{l,m}}^{g\bar{g}}= & T\displaystyle\sum\limits_{i\omega_n}g_l(\textbf{k}+\textbf{q},i\omega_n)\bar{g}_m(\textbf{k},i\Omega-i\omega_n) \\
    = & T\displaystyle\sum\limits_{i\omega_n}\dfrac{1}{i\omega_n-\epsilon_{\textbf{k}+\textbf{q},l}}\dfrac{1}{(i\Omega-i\omega_n)+\epsilon_{\textbf{k},m}} \\
    = &\dfrac{n_F(\epsilon_{\textbf{k}+\textbf{q},l})-n_F(\epsilon_{\textbf{k},m})}{i\Omega-\epsilon_{\textbf{k}+\textbf{q},l}+\epsilon_{\textbf{k},m}},
    \end{array}
\end{equation} 
\begin{equation}\label{matsubara_sum_gbarg}
    \begin{array}{rl}  
    P_{{l,m}}^{\bar{g}g}= & T\displaystyle\sum\limits_{i\omega_n}\bar{g}_l(\textbf{k}+\textbf{q},i\omega_n)g_m(\textbf{k},i\Omega-i\omega_n)\\
    = & T\displaystyle\sum\limits_{i\omega_n}\dfrac{1}{i\omega_n+\epsilon_{\textbf{k}+\textbf{q},l}}\dfrac{1}{(i\Omega-i\omega_n)-\epsilon_{\textbf{k},m}}\\
    = & \dfrac{n_F(-\epsilon_{\textbf{k}+\textbf{q},l})-n_F(-\epsilon_{\textbf{k},m})}{i\Omega+\epsilon_{\textbf{k}+\textbf{q},l}-\epsilon_{\textbf{k},m}},
    \end{array}
\end{equation} 
\begin{equation}\label{matsubara_sum_gg}
    \begin{array}{rl}
    P_{{l,m}}^{gg}= & T\displaystyle\sum\limits_{i\omega_n}g_l(\textbf{k}+\textbf{q},i\omega_n)g_m(\textbf{k},i\Omega-i\omega_n)\\
    = & T\displaystyle\sum\limits_{i\omega_n}\dfrac{1}{i\omega_n-\epsilon_{\textbf{k}+\textbf{q},l}}\dfrac{1}{(i\Omega-i\omega_n)-\epsilon_{\textbf{k},m}}\\
    = & \dfrac{n_F(\epsilon_{\textbf{k}+\textbf{q},l})-n_F(-\epsilon_{\textbf{k},m})}{i\Omega-\epsilon_{\textbf{k}+\textbf{q},l}-\epsilon_{\textbf{k},m}},\\
    \end{array}
\end{equation} 
\begin{equation}\label{matsubara_sum_gbargbar}
    \begin{array}{rl}
    P_{{l,m}}^{\bar{g}\bar{g}}= & T\displaystyle\sum\limits_{i\omega_n}\bar{g}_l(\textbf{k}+\textbf{q},i\omega_n)\bar{g}_m(\textbf{k},i\Omega-i\omega_n)\\
    = & T\displaystyle\sum\limits_{i\omega_n}\dfrac{1}{i\omega_n+\epsilon_{\textbf{k}+\textbf{q},l}}\dfrac{1}{(i\Omega-i\omega_n)+\epsilon_{\textbf{k},m}}\\
    = & \dfrac{n_F(-\epsilon_{\textbf{k}+\textbf{q},l})-n_F(\epsilon_{\textbf{k},m})}{i\Omega+\epsilon_{\textbf{k}+\textbf{q},l}+\epsilon_{\textbf{k},m}}.\\
    \end{array}
\end{equation} 
Here we have used the imaginary time free fermion propagators of the spin Hamiltonian, defined as
\begin{equation}
\begin{split}
    g_i(\textbf{k},i\omega_n)=&-\langle T_{\tau}\beta_{\textbf{k},i}(\tau)\beta_{\textbf{k},i}^{\dagger}(0)\rangle_{\omega_n}=\frac{1}{i\omega_n-\epsilon_{\textbf{k},i}},\\
    \bar{g}_i(\textbf{k},i\omega_n)=&-\langle T_{\tau}\beta_{\textbf{k},i}^{\dagger}(\tau)\beta_{\textbf{k},i}(0)\rangle_{\omega_n}=\frac{1}{i\omega_n+\epsilon_{\textbf{k},i}},
\end{split}    
\end{equation}
and $n_F$ is the Fermi-Dirac distribution function.

\begin{figure}[t]
	\centering
	\includegraphics[width=1.0\columnwidth]{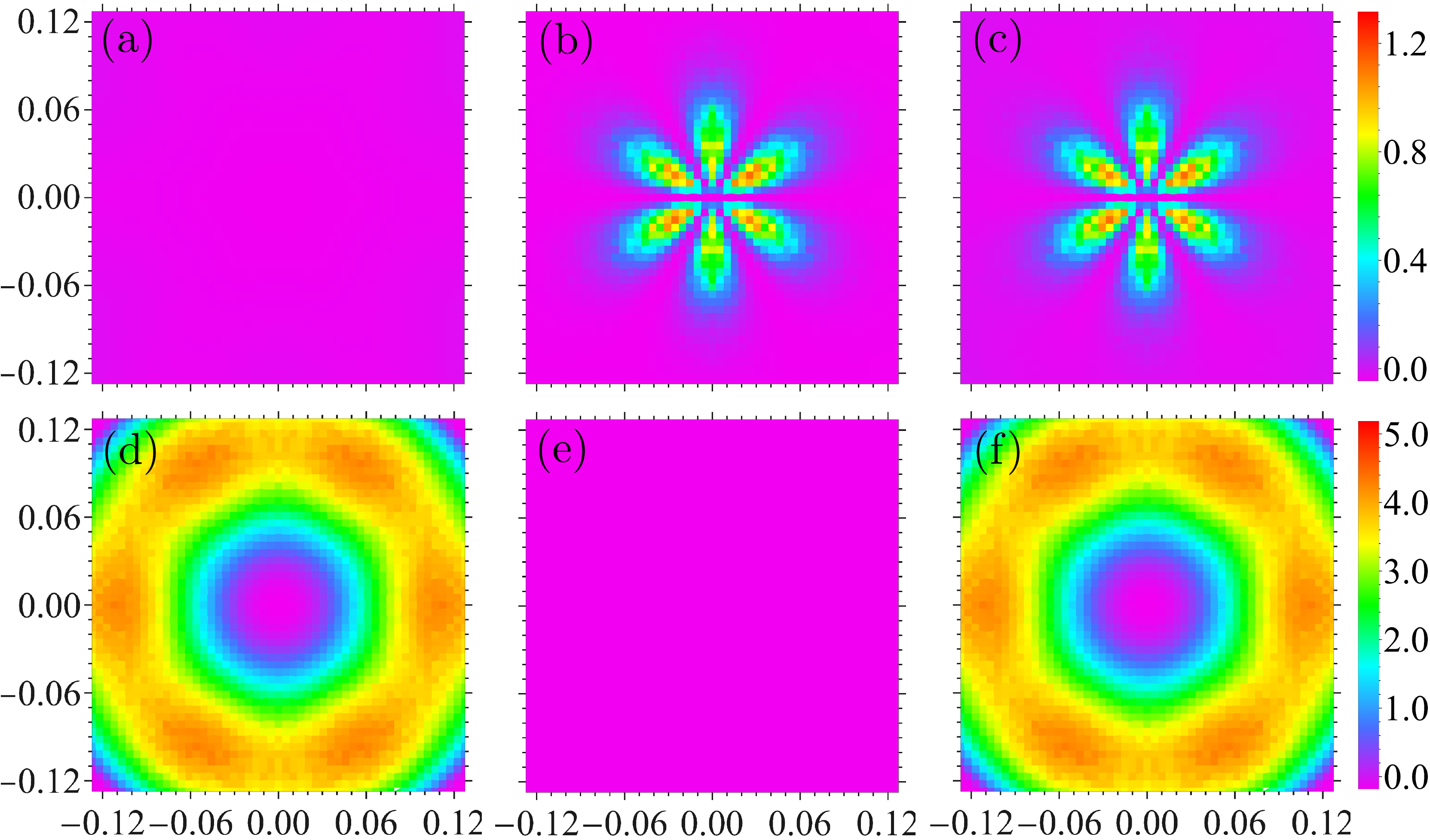}
     \caption{\color{black} Partitioned contributions to the sound attenuation for fixed parameters $(\Gamma,J,K)$=(0.2,0.2,-1), $v_s=0.3$, and $T=0.05$:
     Panels (a) and (d)  show the contribution from  momenta centered around the $\Gamma$-point, 
     panels (b) and (e) from momenta centered around the Dirac $K$-points, 
     and  panels (c) and (f) is the full summation from the entire BZ.
     The BZ sectors involve restricting momenta to a disk centered around the desired $\Gamma$-, K-, and K'-point,
     the radius of each disk being 1/2 the distance from $\Gamma\text{-point}$ to $\mathrm{K}\text{-point}$, 
     i.e., the maximum disk radius such that disk sectors do not overlap. 
     The top row (a)\textendash(c) shows the sound attenuation arising from the ph-channel processes, and the bottom row (d)\textendash(f)  from the pp-channel processes. The phonon momentum $\mathbf{q}$ belongs to the region $(q_x,q_y)\in [-0.12\pi,0.12\pi]^2$. }
\label{Fig:gamma_vs_k}
\end{figure}

\begin{figure}[t]
	\centering
	\includegraphics[width=1\columnwidth]{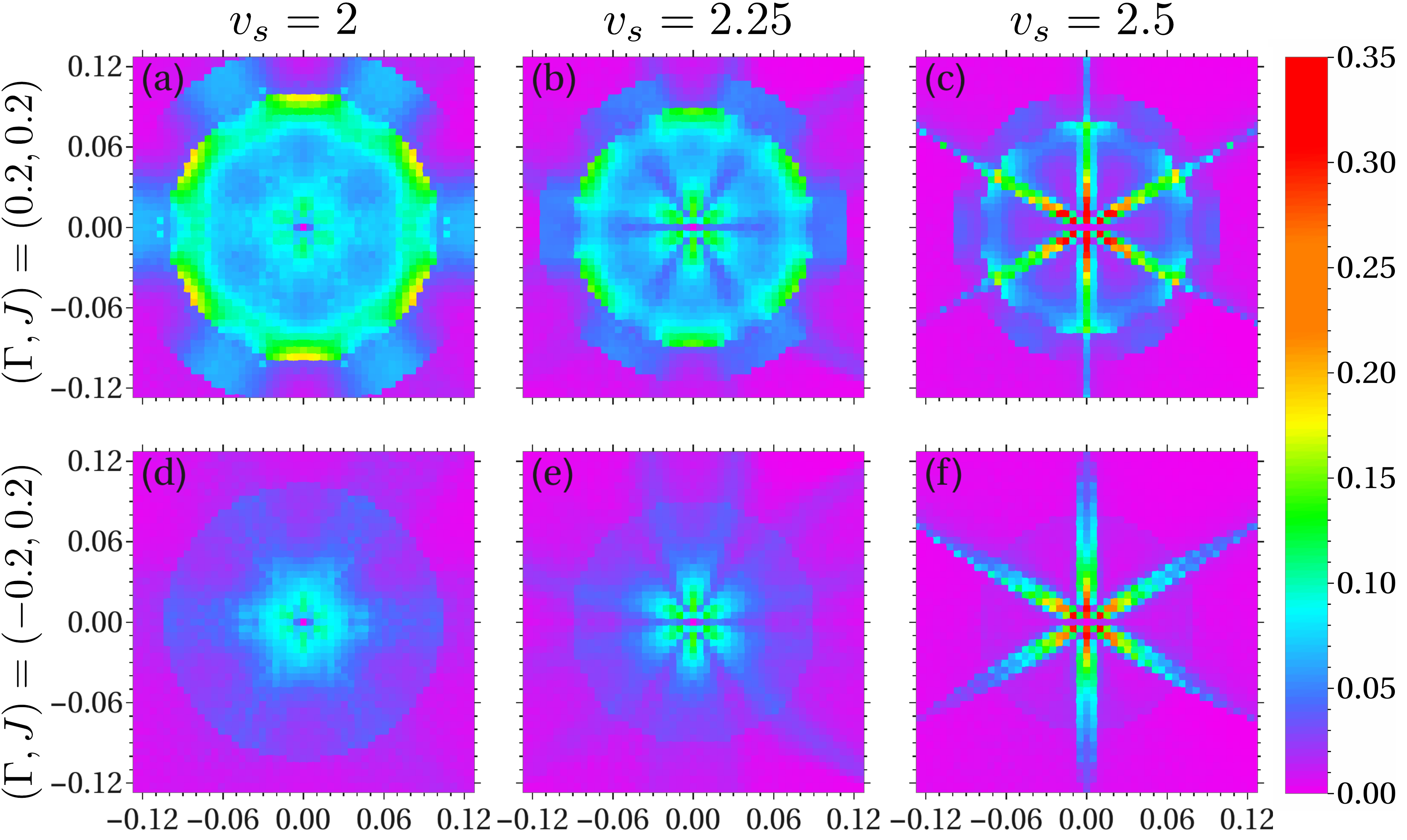}
   \caption{\color{black} The $E_{g}$ contribution from  the ph processes, to the longitudinal sound-attenuation coefficient $\alpha_s^{\parallel}(\textbf{q})$.  We set $T=0.1$ and $(\Gamma,J, K)=(0.2,0.2,-1)$ for the top row and
  $(\Gamma,J, K)=(-0.2,0.2,-1)$ for the bottom row.  
   Panels (a),(d), (b),(e), and (c),(f)  show  $\alpha_s^{\parallel}(\textbf{q})$ for  $v_s=2$, 2.25, and 2.5 respectively. The phonon momentum $\mathbf{q}$ belongs to the region $(q_x,q_y)\in [-0.12\pi,0.12\pi]^2$. }
\label{fig:sa_increasing_vs}
\end{figure}

\section{Additional numerical results for the sound attenuation}\label{app:addnumres}
 In Fig.~\ref{Fig:alon_ag_ph_and_pp} we show the $A_{1g}$ contribution to the sound-attenuation coefficient for the longitudinal phonon mode, $\alpha_s^{\parallel}(\textbf{q})$,  for both $v_s<v_F$ and  $v_s>v_F$,  computed for different values of ($\Gamma$, $J$)  and $K=-1$.  We clearly see that the intensity of the phonon attenuation in the $A_{1g}$ symmetry
 channel is smaller than that in the $E_g$ symmetry channel (see Figs. 3 and 4). While in the Kitaev limit ($J=\Gamma=0$), the contribution from the $A_{1g}$ symmetry channel seems to be symmetric,
  it still possesses weak sixfold angular dependence, which becomes more apparent in the presence of finite  $J$ and $\Gamma$.

{\color{black}  Further, as  we see a significant lowering    of the energy states near the center of the BZ for  the  parameters $(\Gamma,J,K)=(-0.2,0.2,-1)$  and $(\Gamma,J,K)=(0.2,0.2,-1)$ [see Figs.\ref{Fig:disp}(a) and \ref{Fig:disp}(c), respectively], it is reasonable to assume that these states give non-negligible contributions into the scattering processes leading to the sound-attenuation and, perhaps, are responsible for qualitative changes in the sound attenuation patterns and intensities (compare the different panels of  Figs.~\ref{Fig:alon_vlf_grid} and \ref{Fig:alon_pp_vgf_grid} of the main text).
To see this explicitly for one set of parameters $(\Gamma,J,K)=(0.2,0.2,-1)$, $v_s=0.3$, and $T=0.05$, in Fig. \ref{Fig:gamma_vs_k}  we plot separately
 the contributions to the sound attenuation from the states located near the  center of the BZ [Figs.\ref{Fig:gamma_vs_k}(a) and \ref{Fig:gamma_vs_k}(d)], from the states near the   Dirac $K$-points  [Figs.~\ref{Fig:gamma_vs_k}(b) and \ref{Fig:gamma_vs_k}(e)], and  from the entire BZ [Figs.~\ref{Fig:gamma_vs_k}(c) and \ref{Fig:gamma_vs_k}(f)].  The top row (a)\textendash(c) shows the sound attenuation arising from  the ph process, and the  bottom row (d)\textendash(f)  from the pp process. 
      While conventionally the pp process is not allowed for $v_s<v_F$ from the $K$-point centered Dirac cones, leading to (e) being identically zero, 
     the lowering of the vison bands at the $\Gamma$-point [see Fig.\ref{Fig:disp}(c)] leads to the significant pp process contributions  [see the very large value for the attenuation in panel (d)] for this parameter case. 
      Moreover, the almost circular like shape of the sound-attenuation pattern is because the vison bands have very little dispersion, i.e., are still almost flat.
      Also note that unlike (e),  the ph processes near the $\Gamma$-point  shown in panel (a) are not forbidden, and have a very small but  nonzero  contribution to the sound attenuation (we can see exceedingly faint blue colored intensity on the edges of the plot).}


{\color{black}
In Fig.~\ref{fig:sa_increasing_vs}, we  plot the sound-attenuation coefficient computed for  $(\Gamma,J,K)=(0.2,0.2,-1)$ [top row, panels (a)\textendash(c)] and  $(\Gamma,J,K)=(-0.2,0.2,-1)$ [bottom row, panels (d)\textendash(f)] for 
 progressively increased sound velocity $v_s$ from $2$ to $2.5$.
  While the features seen in the main text still generally persist, new behavior appears as a result of new accessible states and kinematic constraints. First note that $v_F=3$ holds true only in the case $J=\Gamma=0$.  As we can see from  the spectra in Fig~\ref{Fig:disp},  for the top row with $J=0.2$ the cones at the BZ $K$-points are more obtuse, leading to an effectively smaller $v_F$, while for the bottom row, with $J=-0.2$, the cones are more acute, leading to an effectively larger $v_F$.   When $J=0.2$, and  as $v_s$ increases, the character of the attenuation processes changes from the ph-like to the pp-like, and this is reflected in the change of the sound  attenuation
pattern.
}

\bibliography{ref.bib}

\begin{thebibliography}{93}%
\makeatletter
\providecommand \@ifxundefined [1]{%
 \@ifx{#1\undefined}
}%
\providecommand \@ifnum [1]{%
 \ifnum #1\expandafter \@firstoftwo
 \else \expandafter \@secondoftwo
 \fi
}%
\providecommand \@ifx [1]{%
 \ifx #1\expandafter \@firstoftwo
 \else \expandafter \@secondoftwo
 \fi
}%
\providecommand \natexlab [1]{#1}%
\providecommand \enquote  [1]{``#1''}%
\providecommand \bibnamefont  [1]{#1}%
\providecommand \bibfnamefont [1]{#1}%
\providecommand \citenamefont [1]{#1}%
\providecommand \href@noop [0]{\@secondoftwo}%
\providecommand \href [0]{\begingroup \@sanitize@url \@href}%
\providecommand \@href[1]{\@@startlink{#1}\@@href}%
\providecommand \@@href[1]{\endgroup#1\@@endlink}%
\providecommand \@sanitize@url [0]{\catcode `\\12\catcode `\$12\catcode
  `\&12\catcode `\#12\catcode `\^12\catcode `\_12\catcode `\%12\relax}%
\providecommand \@@startlink[1]{}%
\providecommand \@@endlink[0]{}%
\providecommand \url  [0]{\begingroup\@sanitize@url \@url }%
\providecommand \@url [1]{\endgroup\@href {#1}{\urlprefix }}%
\providecommand \urlprefix  [0]{URL }%
\providecommand \Eprint [0]{\href }%
\providecommand \doibase [0]{http://dx.doi.org/}%
\providecommand \selectlanguage [0]{\@gobble}%
\providecommand \bibinfo  [0]{\@secondoftwo}%
\providecommand \bibfield  [0]{\@secondoftwo}%
\providecommand \translation [1]{[#1]}%
\providecommand \BibitemOpen [0]{}%
\providecommand \bibitemStop [0]{}%
\providecommand \bibitemNoStop [0]{.\EOS\space}%
\providecommand \EOS [0]{\spacefactor3000\relax}%
\providecommand \BibitemShut  [1]{\csname bibitem#1\endcsname}%
\let\auto@bib@innerbib\@empty
\bibitem [{\citenamefont {Anderson}(1973)}]{Anderson1973}%
  \BibitemOpen
  \bibfield  {author} {\bibinfo {author} {\bibfnamefont {P.~W.}\ \bibnamefont
  {Anderson}},\ }\href@noop {} {\bibfield  {journal} {\bibinfo  {journal}
  {Materials Research Bulletin}\ }\textbf {\bibinfo {volume} {8}},\ \bibinfo
  {pages} {153} (\bibinfo {year} {1973})}\BibitemShut {NoStop}%
\bibitem [{\citenamefont {Wen}(2002)}]{Wen2002}%
  \BibitemOpen
  \bibfield  {author} {\bibinfo {author} {\bibfnamefont {X.-G.}\ \bibnamefont
  {Wen}},\ }\href {\doibase 10.1103/PhysRevB.65.165113} {\bibfield  {journal}
  {\bibinfo  {journal} {Phys. Rev. B}\ }\textbf {\bibinfo {volume} {65}},\
  \bibinfo {pages} {165113} (\bibinfo {year} {2002})}\BibitemShut {NoStop}%
\bibitem [{\citenamefont {Kitaev}(2006)}]{Kitaev2006}%
  \BibitemOpen
  \bibfield  {author} {\bibinfo {author} {\bibfnamefont {A.}~\bibnamefont
  {Kitaev}},\ }\href {https://doi.org/10.1016/j.aop.2005.10.005} {\bibfield
  {journal} {\bibinfo  {journal} {Annals of Physics}\ }\textbf {\bibinfo
  {volume} {321}},\ \bibinfo {pages} {2} (\bibinfo {year} {2006})}\BibitemShut
  {NoStop}%
\bibitem [{\citenamefont {Balents}(2010)}]{Balents2010}%
  \BibitemOpen
  \bibfield  {author} {\bibinfo {author} {\bibfnamefont {L.}~\bibnamefont
  {Balents}},\ }\href@noop {} {\bibfield  {journal} {\bibinfo  {journal}
  {Nature}\ }\textbf {\bibinfo {volume} {464}},\ \bibinfo {pages} {199}
  (\bibinfo {year} {2010})}\BibitemShut {NoStop}%
\bibitem [{\citenamefont {Savary}\ and\ \citenamefont
  {Balents}(2017)}]{Savary2016}%
  \BibitemOpen
  \bibfield  {author} {\bibinfo {author} {\bibfnamefont {L.}~\bibnamefont
  {Savary}}\ and\ \bibinfo {author} {\bibfnamefont {L.}~\bibnamefont
  {Balents}},\ }\href {http://stacks.iop.org/0034-4885/80/i=1/a=016502}
  {\bibfield  {journal} {\bibinfo  {journal} {Rep. Prog. Phys.}\ }\textbf
  {\bibinfo {volume} {80}},\ \bibinfo {pages} {016502} (\bibinfo {year}
  {2017})}\BibitemShut {NoStop}%
\bibitem [{\citenamefont {Knolle}\ and\ \citenamefont
  {Moessner}(2019)}]{KnolleMoessner2019}%
  \BibitemOpen
  \bibfield  {author} {\bibinfo {author} {\bibfnamefont {J.}~\bibnamefont
  {Knolle}}\ and\ \bibinfo {author} {\bibfnamefont {R.}~\bibnamefont
  {Moessner}},\ }\href {\doibase 10.1146/annurev-conmatphys-031218-013401}
  {\bibfield  {journal} {\bibinfo  {journal} {Annual Review of Condensed Matter
  Physics}\ }\textbf {\bibinfo {volume} {10}},\ \bibinfo {pages} {451}
  (\bibinfo {year} {2019})}\BibitemShut {NoStop}%
\bibitem [{\citenamefont {Broholm}\ \emph {et~al.}(2020)\citenamefont
  {Broholm}, \citenamefont {Cava}, \citenamefont {Kivelson}, \citenamefont
  {Nocera}, \citenamefont {Norman},\ and\ \citenamefont
  {Senthil}}]{Broholm2020}%
  \BibitemOpen
  \bibfield  {author} {\bibinfo {author} {\bibfnamefont {C.}~\bibnamefont
  {Broholm}}, \bibinfo {author} {\bibfnamefont {R.~J.}\ \bibnamefont {Cava}},
  \bibinfo {author} {\bibfnamefont {S.~A.}\ \bibnamefont {Kivelson}}, \bibinfo
  {author} {\bibfnamefont {D.~G.}\ \bibnamefont {Nocera}}, \bibinfo {author}
  {\bibfnamefont {M.~R.}\ \bibnamefont {Norman}}, \ and\ \bibinfo {author}
  {\bibfnamefont {T.}~\bibnamefont {Senthil}},\ }\href {\doibase
  10.1126/science.aay0668} {\bibfield  {journal} {\bibinfo  {journal}
  {Science}\ }\textbf {\bibinfo {volume} {367}} (\bibinfo {year} {2020}),\
  10.1126/science.aay0668}\BibitemShut {NoStop}%
\bibitem [{\citenamefont {Kitaev}(2003)}]{Kitaev2003}%
  \BibitemOpen
  \bibfield  {author} {\bibinfo {author} {\bibfnamefont {A.}~\bibnamefont
  {Kitaev}},\ }\href {\doibase http://dx.doi.org/10.1016/S0003-4916(02)00018-0}
  {\bibfield  {journal} {\bibinfo  {journal} {Annals of Physics}\ }\textbf
  {\bibinfo {volume} {303}},\ \bibinfo {pages} {2 } (\bibinfo {year}
  {2003})}\BibitemShut {NoStop}%
\bibitem [{\citenamefont {Hermanns}\ \emph {et~al.}(2018)\citenamefont
  {Hermanns}, \citenamefont {Kimchi},\ and\ \citenamefont
  {Knolle}}]{Knolle2017}%
  \BibitemOpen
  \bibfield  {author} {\bibinfo {author} {\bibfnamefont {M.}~\bibnamefont
  {Hermanns}}, \bibinfo {author} {\bibfnamefont {I.}~\bibnamefont {Kimchi}}, \
  and\ \bibinfo {author} {\bibfnamefont {J.}~\bibnamefont {Knolle}},\ }\href
  {\doibase 10.1146/annurev-conmatphys-033117-053934} {\bibfield  {journal}
  {\bibinfo  {journal} {Annual Review of Condensed Matter Physics}\ }\textbf
  {\bibinfo {volume} {9}},\ \bibinfo {pages} {17} (\bibinfo {year}
  {2018})}\BibitemShut {NoStop}%
\bibitem [{\citenamefont {Motome}\ and\ \citenamefont
  {Nasu}(2020)}]{Motome2019}%
  \BibitemOpen
  \bibfield  {author} {\bibinfo {author} {\bibfnamefont {Y.}~\bibnamefont
  {Motome}}\ and\ \bibinfo {author} {\bibfnamefont {J.}~\bibnamefont {Nasu}},\
  }\href {\doibase 10.7566/JPSJ.89.012002} {\bibfield  {journal} {\bibinfo
  {journal} {J. Phys. Soc. Jpn}\ }\textbf {\bibinfo {volume} {89}},\ \bibinfo
  {pages} {012002} (\bibinfo {year} {2020})}\BibitemShut {NoStop}%
\bibitem [{\citenamefont {Takagi}\ \emph {et~al.}(2019)\citenamefont {Takagi},
  \citenamefont {Takayama}, \citenamefont {Jackeli}, \citenamefont
  {Khaliullin},\ and\ \citenamefont {Nagler}}]{Takagi2019}%
  \BibitemOpen
  \bibfield  {author} {\bibinfo {author} {\bibfnamefont {H.}~\bibnamefont
  {Takagi}}, \bibinfo {author} {\bibfnamefont {T.}~\bibnamefont {Takayama}},
  \bibinfo {author} {\bibfnamefont {G.}~\bibnamefont {Jackeli}}, \bibinfo
  {author} {\bibfnamefont {G.}~\bibnamefont {Khaliullin}}, \ and\ \bibinfo
  {author} {\bibfnamefont {S.~E.}\ \bibnamefont {Nagler}},\ }\href {\doibase
  10.1038/s42254-019-0038-2} {\bibfield  {journal} {\bibinfo  {journal} {Nat.
  Rev. Phys.}\ }\textbf {\bibinfo {volume} {1}},\ \bibinfo {pages} {264}
  (\bibinfo {year} {2019})}\BibitemShut {NoStop}%
\bibitem [{\citenamefont {Trebst}\ and\ \citenamefont
  {Hickey}(2022)}]{Trebst2022}%
  \BibitemOpen
  \bibfield  {author} {\bibinfo {author} {\bibfnamefont {S.}~\bibnamefont
  {Trebst}}\ and\ \bibinfo {author} {\bibfnamefont {C.}~\bibnamefont
  {Hickey}},\ }\href {\doibase https://doi.org/10.1016/j.physrep.2021.11.003}
  {\bibfield  {journal} {\bibinfo  {journal} {Physics Reports}\ }\textbf
  {\bibinfo {volume} {950}},\ \bibinfo {pages} {1} (\bibinfo {year}
  {2022})}\BibitemShut {NoStop}%
\bibitem [{\citenamefont {Jackeli}\ and\ \citenamefont
  {Khaliullin}(2009)}]{Jackeli2009}%
  \BibitemOpen
  \bibfield  {author} {\bibinfo {author} {\bibfnamefont {G.}~\bibnamefont
  {Jackeli}}\ and\ \bibinfo {author} {\bibfnamefont {G.}~\bibnamefont
  {Khaliullin}},\ }\href {\doibase 10.1103/PhysRevLett.102.017205} {\bibfield
  {journal} {\bibinfo  {journal} {Phys. Rev. Lett.}\ }\textbf {\bibinfo
  {volume} {102}},\ \bibinfo {pages} {017205} (\bibinfo {year}
  {2009})}\BibitemShut {NoStop}%
\bibitem [{\citenamefont {Chaloupka}\ \emph {et~al.}(2010)\citenamefont
  {Chaloupka}, \citenamefont {Jackeli},\ and\ \citenamefont
  {Khaliullin}}]{Chaloupka2010}%
  \BibitemOpen
  \bibfield  {author} {\bibinfo {author} {\bibfnamefont {J.~c.~v.}\
  \bibnamefont {Chaloupka}}, \bibinfo {author} {\bibfnamefont {G.}~\bibnamefont
  {Jackeli}}, \ and\ \bibinfo {author} {\bibfnamefont {G.}~\bibnamefont
  {Khaliullin}},\ }\href {\doibase 10.1103/PhysRevLett.105.027204} {\bibfield
  {journal} {\bibinfo  {journal} {Phys. Rev. Lett.}\ }\textbf {\bibinfo
  {volume} {105}},\ \bibinfo {pages} {027204} (\bibinfo {year}
  {2010})}\BibitemShut {NoStop}%
\bibitem [{\citenamefont {Kubota}\ \emph {et~al.}(2015)\citenamefont {Kubota},
  \citenamefont {Tanaka}, \citenamefont {Ono}, \citenamefont {Narumi},\ and\
  \citenamefont {Kindo}}]{Kubota2015}%
  \BibitemOpen
  \bibfield  {author} {\bibinfo {author} {\bibfnamefont {Y.}~\bibnamefont
  {Kubota}}, \bibinfo {author} {\bibfnamefont {H.}~\bibnamefont {Tanaka}},
  \bibinfo {author} {\bibfnamefont {T.}~\bibnamefont {Ono}}, \bibinfo {author}
  {\bibfnamefont {Y.}~\bibnamefont {Narumi}}, \ and\ \bibinfo {author}
  {\bibfnamefont {K.}~\bibnamefont {Kindo}},\ }\href {\doibase
  10.1103/PhysRevB.91.094422} {\bibfield  {journal} {\bibinfo  {journal} {Phys.
  Rev. B}\ }\textbf {\bibinfo {volume} {91}},\ \bibinfo {pages} {094422}
  (\bibinfo {year} {2015})}\BibitemShut {NoStop}%
\bibitem [{\citenamefont {Rau}\ \emph {et~al.}(2016)\citenamefont {Rau},
  \citenamefont {Lee},\ and\ \citenamefont {Kee}}]{Rau2016}%
  \BibitemOpen
  \bibfield  {author} {\bibinfo {author} {\bibfnamefont {J.~G.}\ \bibnamefont
  {Rau}}, \bibinfo {author} {\bibfnamefont {E.~K.-H.}\ \bibnamefont {Lee}}, \
  and\ \bibinfo {author} {\bibfnamefont {H.-Y.}\ \bibnamefont {Kee}},\ }\href
  {\doibase 10.1146/annurev-conmatphys-031115-011319} {\bibfield  {journal}
  {\bibinfo  {journal} {Annual Review of Condensed Matter Physics}\ }\textbf
  {\bibinfo {volume} {7}},\ \bibinfo {pages} {195} (\bibinfo {year}
  {2016})}\BibitemShut {NoStop}%
\bibitem [{\citenamefont {Singh}\ and\ \citenamefont
  {Gegenwart}(2010)}]{Yogesh2010}%
  \BibitemOpen
  \bibfield  {author} {\bibinfo {author} {\bibfnamefont {Y.}~\bibnamefont
  {Singh}}\ and\ \bibinfo {author} {\bibfnamefont {P.}~\bibnamefont
  {Gegenwart}},\ }\href {\doibase 10.1103/PhysRevB.82.064412} {\bibfield
  {journal} {\bibinfo  {journal} {Phys. Rev. B}\ }\textbf {\bibinfo {volume}
  {82}},\ \bibinfo {pages} {064412} (\bibinfo {year} {2010})}\BibitemShut
  {NoStop}%
\bibitem [{\citenamefont {Liu}\ \emph {et~al.}(2011)\citenamefont {Liu},
  \citenamefont {Berlijn}, \citenamefont {Yin}, \citenamefont {Ku},
  \citenamefont {Tsvelik}, \citenamefont {Kim}, \citenamefont {Gretarsson},
  \citenamefont {Singh}, \citenamefont {Gegenwart},\ and\ \citenamefont
  {Hill}}]{Liu2011}%
  \BibitemOpen
  \bibfield  {author} {\bibinfo {author} {\bibfnamefont {X.}~\bibnamefont
  {Liu}}, \bibinfo {author} {\bibfnamefont {T.}~\bibnamefont {Berlijn}},
  \bibinfo {author} {\bibfnamefont {W.-G.}\ \bibnamefont {Yin}}, \bibinfo
  {author} {\bibfnamefont {W.}~\bibnamefont {Ku}}, \bibinfo {author}
  {\bibfnamefont {A.}~\bibnamefont {Tsvelik}}, \bibinfo {author} {\bibfnamefont
  {Y.-J.}\ \bibnamefont {Kim}}, \bibinfo {author} {\bibfnamefont
  {H.}~\bibnamefont {Gretarsson}}, \bibinfo {author} {\bibfnamefont
  {Y.}~\bibnamefont {Singh}}, \bibinfo {author} {\bibfnamefont
  {P.}~\bibnamefont {Gegenwart}}, \ and\ \bibinfo {author} {\bibfnamefont
  {J.~P.}\ \bibnamefont {Hill}},\ }\href {\doibase 10.1103/PhysRevB.83.220403}
  {\bibfield  {journal} {\bibinfo  {journal} {Phys. Rev. B}\ }\textbf {\bibinfo
  {volume} {83}},\ \bibinfo {pages} {220403} (\bibinfo {year}
  {2011})}\BibitemShut {NoStop}%
\bibitem [{\citenamefont {Choi}\ \emph {et~al.}(2012)\citenamefont {Choi},
  \citenamefont {Coldea}, \citenamefont {Kolmogorov}, \citenamefont
  {Lancaster}, \citenamefont {Mazin}, \citenamefont {Blundell}, \citenamefont
  {Radaelli}, \citenamefont {Singh}, \citenamefont {Gegenwart}, \citenamefont
  {Choi}, \citenamefont {Cheong}, \citenamefont {Baker}, \citenamefont
  {Stock},\ and\ \citenamefont {Taylor}}]{Choi2012}%
  \BibitemOpen
  \bibfield  {author} {\bibinfo {author} {\bibfnamefont {S.~K.}\ \bibnamefont
  {Choi}}, \bibinfo {author} {\bibfnamefont {R.}~\bibnamefont {Coldea}},
  \bibinfo {author} {\bibfnamefont {A.~N.}\ \bibnamefont {Kolmogorov}},
  \bibinfo {author} {\bibfnamefont {T.}~\bibnamefont {Lancaster}}, \bibinfo
  {author} {\bibfnamefont {I.~I.}\ \bibnamefont {Mazin}}, \bibinfo {author}
  {\bibfnamefont {S.~J.}\ \bibnamefont {Blundell}}, \bibinfo {author}
  {\bibfnamefont {P.~G.}\ \bibnamefont {Radaelli}}, \bibinfo {author}
  {\bibfnamefont {Y.}~\bibnamefont {Singh}}, \bibinfo {author} {\bibfnamefont
  {P.}~\bibnamefont {Gegenwart}}, \bibinfo {author} {\bibfnamefont {K.~R.}\
  \bibnamefont {Choi}}, \bibinfo {author} {\bibfnamefont {S.-W.}\ \bibnamefont
  {Cheong}}, \bibinfo {author} {\bibfnamefont {P.~J.}\ \bibnamefont {Baker}},
  \bibinfo {author} {\bibfnamefont {C.}~\bibnamefont {Stock}}, \ and\ \bibinfo
  {author} {\bibfnamefont {J.}~\bibnamefont {Taylor}},\ }\href {\doibase
  10.1103/PhysRevLett.108.127204} {\bibfield  {journal} {\bibinfo  {journal}
  {Phys. Rev. Lett.}\ }\textbf {\bibinfo {volume} {108}},\ \bibinfo {pages}
  {127204} (\bibinfo {year} {2012})}\BibitemShut {NoStop}%
\bibitem [{\citenamefont {Ye}\ \emph {et~al.}(2012)\citenamefont {Ye},
  \citenamefont {Chi}, \citenamefont {Cao}, \citenamefont {Chakoumakos},
  \citenamefont {Fernandez-Baca}, \citenamefont {Custelcean}, \citenamefont
  {Qi}, \citenamefont {Korneta},\ and\ \citenamefont {Cao}}]{Ye2012}%
  \BibitemOpen
  \bibfield  {author} {\bibinfo {author} {\bibfnamefont {F.}~\bibnamefont
  {Ye}}, \bibinfo {author} {\bibfnamefont {S.}~\bibnamefont {Chi}}, \bibinfo
  {author} {\bibfnamefont {H.}~\bibnamefont {Cao}}, \bibinfo {author}
  {\bibfnamefont {B.~C.}\ \bibnamefont {Chakoumakos}}, \bibinfo {author}
  {\bibfnamefont {J.~A.}\ \bibnamefont {Fernandez-Baca}}, \bibinfo {author}
  {\bibfnamefont {R.}~\bibnamefont {Custelcean}}, \bibinfo {author}
  {\bibfnamefont {T.~F.}\ \bibnamefont {Qi}}, \bibinfo {author} {\bibfnamefont
  {O.~B.}\ \bibnamefont {Korneta}}, \ and\ \bibinfo {author} {\bibfnamefont
  {G.}~\bibnamefont {Cao}},\ }\href {\doibase 10.1103/PhysRevB.85.180403}
  {\bibfield  {journal} {\bibinfo  {journal} {Phys. Rev. B}\ }\textbf {\bibinfo
  {volume} {85}},\ \bibinfo {pages} {180403} (\bibinfo {year}
  {2012})}\BibitemShut {NoStop}%
\bibitem [{\citenamefont {Comin}\ \emph {et~al.}(2012)\citenamefont {Comin},
  \citenamefont {Levy}, \citenamefont {Ludbrook}, \citenamefont {Zhu},
  \citenamefont {Veenstra}, \citenamefont {Rosen}, \citenamefont {Singh},
  \citenamefont {Gegenwart}, \citenamefont {Stricker}, \citenamefont {Hancock},
  \citenamefont {van~der Marel}, \citenamefont {Elfimov},\ and\ \citenamefont
  {Damascelli}}]{Comin2012}%
  \BibitemOpen
  \bibfield  {author} {\bibinfo {author} {\bibfnamefont {R.}~\bibnamefont
  {Comin}}, \bibinfo {author} {\bibfnamefont {G.}~\bibnamefont {Levy}},
  \bibinfo {author} {\bibfnamefont {B.}~\bibnamefont {Ludbrook}}, \bibinfo
  {author} {\bibfnamefont {Z.-H.}\ \bibnamefont {Zhu}}, \bibinfo {author}
  {\bibfnamefont {C.~N.}\ \bibnamefont {Veenstra}}, \bibinfo {author}
  {\bibfnamefont {J.~A.}\ \bibnamefont {Rosen}}, \bibinfo {author}
  {\bibfnamefont {Y.}~\bibnamefont {Singh}}, \bibinfo {author} {\bibfnamefont
  {P.}~\bibnamefont {Gegenwart}}, \bibinfo {author} {\bibfnamefont
  {D.}~\bibnamefont {Stricker}}, \bibinfo {author} {\bibfnamefont {J.~N.}\
  \bibnamefont {Hancock}}, \bibinfo {author} {\bibfnamefont {D.}~\bibnamefont
  {van~der Marel}}, \bibinfo {author} {\bibfnamefont {I.~S.}\ \bibnamefont
  {Elfimov}}, \ and\ \bibinfo {author} {\bibfnamefont {A.}~\bibnamefont
  {Damascelli}},\ }\href {\doibase 10.1103/PhysRevLett.109.266406} {\bibfield
  {journal} {\bibinfo  {journal} {Phys. Rev. Lett.}\ }\textbf {\bibinfo
  {volume} {109}},\ \bibinfo {pages} {266406} (\bibinfo {year}
  {2012})}\BibitemShut {NoStop}%
\bibitem [{\citenamefont {Hwan~Chun}\ \emph {et~al.}(2015)\citenamefont
  {Hwan~Chun}, \citenamefont {Kim}, \citenamefont {Kim}, \citenamefont {Zheng},
  \citenamefont {Stoumpos}, \citenamefont {Malliakas}, \citenamefont
  {Mitchell}, \citenamefont {Mehlawat}, \citenamefont {Singh}, \citenamefont
  {Choi}, \citenamefont {Gog}, \citenamefont {Al-Zein}, \citenamefont {Sala},
  \citenamefont {Krisch}, \citenamefont {Chaloupka}, \citenamefont {Jackeli},
  \citenamefont {Khaliullin},\ and\ \citenamefont {Kim}}]{Hwan2015}%
  \BibitemOpen
  \bibfield  {author} {\bibinfo {author} {\bibfnamefont {S.}~\bibnamefont
  {Hwan~Chun}}, \bibinfo {author} {\bibfnamefont {J.-W.}\ \bibnamefont {Kim}},
  \bibinfo {author} {\bibfnamefont {J.}~\bibnamefont {Kim}}, \bibinfo {author}
  {\bibfnamefont {H.}~\bibnamefont {Zheng}}, \bibinfo {author} {\bibfnamefont
  {C.~C.}\ \bibnamefont {Stoumpos}}, \bibinfo {author} {\bibfnamefont {C.~D.}\
  \bibnamefont {Malliakas}}, \bibinfo {author} {\bibfnamefont {J.~F.}\
  \bibnamefont {Mitchell}}, \bibinfo {author} {\bibfnamefont {K.}~\bibnamefont
  {Mehlawat}}, \bibinfo {author} {\bibfnamefont {Y.}~\bibnamefont {Singh}},
  \bibinfo {author} {\bibfnamefont {Y.}~\bibnamefont {Choi}}, \bibinfo {author}
  {\bibfnamefont {T.}~\bibnamefont {Gog}}, \bibinfo {author} {\bibfnamefont
  {A.}~\bibnamefont {Al-Zein}}, \bibinfo {author} {\bibfnamefont {M.~M.}\
  \bibnamefont {Sala}}, \bibinfo {author} {\bibfnamefont {M.}~\bibnamefont
  {Krisch}}, \bibinfo {author} {\bibfnamefont {J.}~\bibnamefont {Chaloupka}},
  \bibinfo {author} {\bibfnamefont {G.}~\bibnamefont {Jackeli}}, \bibinfo
  {author} {\bibfnamefont {G.}~\bibnamefont {Khaliullin}}, \ and\ \bibinfo
  {author} {\bibfnamefont {B.~J.}\ \bibnamefont {Kim}},\ }\href {\doibase
  10.1038/nphys3322} {\bibfield  {journal} {\bibinfo  {journal} {Nature
  Physics}\ }\textbf {\bibinfo {volume} {11}},\ \bibinfo {pages} {462}
  (\bibinfo {year} {2015})}\BibitemShut {NoStop}%
\bibitem [{\citenamefont {Singh}\ \emph {et~al.}(2012)\citenamefont {Singh},
  \citenamefont {Manni}, \citenamefont {Reuther}, \citenamefont {Berlijn},
  \citenamefont {Thomale}, \citenamefont {Ku}, \citenamefont {Trebst},\ and\
  \citenamefont {Gegenwart}}]{Yogesh2012}%
  \BibitemOpen
  \bibfield  {author} {\bibinfo {author} {\bibfnamefont {Y.}~\bibnamefont
  {Singh}}, \bibinfo {author} {\bibfnamefont {S.}~\bibnamefont {Manni}},
  \bibinfo {author} {\bibfnamefont {J.}~\bibnamefont {Reuther}}, \bibinfo
  {author} {\bibfnamefont {T.}~\bibnamefont {Berlijn}}, \bibinfo {author}
  {\bibfnamefont {R.}~\bibnamefont {Thomale}}, \bibinfo {author} {\bibfnamefont
  {W.}~\bibnamefont {Ku}}, \bibinfo {author} {\bibfnamefont {S.}~\bibnamefont
  {Trebst}}, \ and\ \bibinfo {author} {\bibfnamefont {P.}~\bibnamefont
  {Gegenwart}},\ }\href {\doibase 10.1103/PhysRevLett.108.127203} {\bibfield
  {journal} {\bibinfo  {journal} {Phys. Rev. Lett.}\ }\textbf {\bibinfo
  {volume} {108}},\ \bibinfo {pages} {127203} (\bibinfo {year}
  {2012})}\BibitemShut {NoStop}%
\bibitem [{\citenamefont {Williams}\ \emph {et~al.}(2016)\citenamefont
  {Williams}, \citenamefont {Johnson}, \citenamefont {Freund}, \citenamefont
  {Choi}, \citenamefont {Jesche}, \citenamefont {Kimchi}, \citenamefont
  {Manni}, \citenamefont {Bombardi}, \citenamefont {Manuel}, \citenamefont
  {Gegenwart},\ and\ \citenamefont {Coldea}}]{Williams2016}%
  \BibitemOpen
  \bibfield  {author} {\bibinfo {author} {\bibfnamefont {S.~C.}\ \bibnamefont
  {Williams}}, \bibinfo {author} {\bibfnamefont {R.~D.}\ \bibnamefont
  {Johnson}}, \bibinfo {author} {\bibfnamefont {F.}~\bibnamefont {Freund}},
  \bibinfo {author} {\bibfnamefont {S.}~\bibnamefont {Choi}}, \bibinfo {author}
  {\bibfnamefont {A.}~\bibnamefont {Jesche}}, \bibinfo {author} {\bibfnamefont
  {I.}~\bibnamefont {Kimchi}}, \bibinfo {author} {\bibfnamefont
  {S.}~\bibnamefont {Manni}}, \bibinfo {author} {\bibfnamefont
  {A.}~\bibnamefont {Bombardi}}, \bibinfo {author} {\bibfnamefont
  {P.}~\bibnamefont {Manuel}}, \bibinfo {author} {\bibfnamefont
  {P.}~\bibnamefont {Gegenwart}}, \ and\ \bibinfo {author} {\bibfnamefont
  {R.}~\bibnamefont {Coldea}},\ }\href {\doibase 10.1103/PhysRevB.93.195158}
  {\bibfield  {journal} {\bibinfo  {journal} {Phys. Rev. B}\ }\textbf {\bibinfo
  {volume} {93}},\ \bibinfo {pages} {195158} (\bibinfo {year}
  {2016})}\BibitemShut {NoStop}%
\bibitem [{\citenamefont {Kitagawa}\ \emph {et~al.}(2018)\citenamefont
  {Kitagawa}, \citenamefont {Takayama}, \citenamefont {Matsumoto},
  \citenamefont {Kato}, \citenamefont {Takano}, \citenamefont {Kishimoto},
  \citenamefont {Bette}, \citenamefont {Dinnebier}, \citenamefont {Jackeli},\
  and\ \citenamefont {Takagi}}]{Kitagawa2018}%
  \BibitemOpen
  \bibfield  {author} {\bibinfo {author} {\bibfnamefont {K.}~\bibnamefont
  {Kitagawa}}, \bibinfo {author} {\bibfnamefont {T.}~\bibnamefont {Takayama}},
  \bibinfo {author} {\bibfnamefont {Y.}~\bibnamefont {Matsumoto}}, \bibinfo
  {author} {\bibfnamefont {A.}~\bibnamefont {Kato}}, \bibinfo {author}
  {\bibfnamefont {R.}~\bibnamefont {Takano}}, \bibinfo {author} {\bibfnamefont
  {Y.}~\bibnamefont {Kishimoto}}, \bibinfo {author} {\bibfnamefont
  {S.}~\bibnamefont {Bette}}, \bibinfo {author} {\bibfnamefont
  {R.}~\bibnamefont {Dinnebier}}, \bibinfo {author} {\bibfnamefont
  {G.}~\bibnamefont {Jackeli}}, \ and\ \bibinfo {author} {\bibfnamefont
  {H.}~\bibnamefont {Takagi}},\ }\href {\doibase 10.1038/nature25482}
  {\bibfield  {journal} {\bibinfo  {journal} {Nature}\ }\textbf {\bibinfo
  {volume} {554}},\ \bibinfo {pages} {341} (\bibinfo {year}
  {2018})}\BibitemShut {NoStop}%
\bibitem [{\citenamefont {Plumb}\ \emph {et~al.}(2014)\citenamefont {Plumb},
  \citenamefont {Clancy}, \citenamefont {Sandilands}, \citenamefont {Shankar},
  \citenamefont {Hu}, \citenamefont {Burch}, \citenamefont {Kee},\ and\
  \citenamefont {Kim}}]{Plumb2014}%
  \BibitemOpen
  \bibfield  {author} {\bibinfo {author} {\bibfnamefont {K.~W.}\ \bibnamefont
  {Plumb}}, \bibinfo {author} {\bibfnamefont {J.~P.}\ \bibnamefont {Clancy}},
  \bibinfo {author} {\bibfnamefont {L.~J.}\ \bibnamefont {Sandilands}},
  \bibinfo {author} {\bibfnamefont {V.~V.}\ \bibnamefont {Shankar}}, \bibinfo
  {author} {\bibfnamefont {Y.~F.}\ \bibnamefont {Hu}}, \bibinfo {author}
  {\bibfnamefont {K.~S.}\ \bibnamefont {Burch}}, \bibinfo {author}
  {\bibfnamefont {H.-Y.}\ \bibnamefont {Kee}}, \ and\ \bibinfo {author}
  {\bibfnamefont {Y.-J.}\ \bibnamefont {Kim}},\ }\href {\doibase
  10.1103/PhysRevB.90.041112} {\bibfield  {journal} {\bibinfo  {journal} {Phys.
  Rev. B}\ }\textbf {\bibinfo {volume} {90}},\ \bibinfo {pages} {041112}
  (\bibinfo {year} {2014})}\BibitemShut {NoStop}%
\bibitem [{\citenamefont {Sandilands}\ \emph {et~al.}(2015)\citenamefont
  {Sandilands}, \citenamefont {Tian}, \citenamefont {Plumb}, \citenamefont
  {Kim},\ and\ \citenamefont {Burch}}]{Sandilands2015}%
  \BibitemOpen
  \bibfield  {author} {\bibinfo {author} {\bibfnamefont {L.~J.}\ \bibnamefont
  {Sandilands}}, \bibinfo {author} {\bibfnamefont {Y.}~\bibnamefont {Tian}},
  \bibinfo {author} {\bibfnamefont {K.~W.}\ \bibnamefont {Plumb}}, \bibinfo
  {author} {\bibfnamefont {Y.-J.}\ \bibnamefont {Kim}}, \ and\ \bibinfo
  {author} {\bibfnamefont {K.~S.}\ \bibnamefont {Burch}},\ }\href {\doibase
  10.1103/PhysRevLett.114.147201} {\bibfield  {journal} {\bibinfo  {journal}
  {Phys. Rev. Lett.}\ }\textbf {\bibinfo {volume} {114}},\ \bibinfo {pages}
  {147201} (\bibinfo {year} {2015})}\BibitemShut {NoStop}%
\bibitem [{\citenamefont {Sears}\ \emph {et~al.}(2015)\citenamefont {Sears},
  \citenamefont {Songvilay}, \citenamefont {Plumb}, \citenamefont {Clancy},
  \citenamefont {Qiu}, \citenamefont {Zhao}, \citenamefont {Parshall},\ and\
  \citenamefont {Kim}}]{Sears2015}%
  \BibitemOpen
  \bibfield  {author} {\bibinfo {author} {\bibfnamefont {J.~A.}\ \bibnamefont
  {Sears}}, \bibinfo {author} {\bibfnamefont {M.}~\bibnamefont {Songvilay}},
  \bibinfo {author} {\bibfnamefont {K.~W.}\ \bibnamefont {Plumb}}, \bibinfo
  {author} {\bibfnamefont {J.~P.}\ \bibnamefont {Clancy}}, \bibinfo {author}
  {\bibfnamefont {Y.}~\bibnamefont {Qiu}}, \bibinfo {author} {\bibfnamefont
  {Y.}~\bibnamefont {Zhao}}, \bibinfo {author} {\bibfnamefont {D.}~\bibnamefont
  {Parshall}}, \ and\ \bibinfo {author} {\bibfnamefont {Y.-J.}\ \bibnamefont
  {Kim}},\ }\href {\doibase 10.1103/PhysRevB.91.144420} {\bibfield  {journal}
  {\bibinfo  {journal} {Phys. Rev. B}\ }\textbf {\bibinfo {volume} {91}},\
  \bibinfo {pages} {144420} (\bibinfo {year} {2015})}\BibitemShut {NoStop}%
\bibitem [{\citenamefont {Majumder}\ \emph {et~al.}(2015)\citenamefont
  {Majumder}, \citenamefont {Schmidt}, \citenamefont {Rosner}, \citenamefont
  {Tsirlin}, \citenamefont {Yasuoka},\ and\ \citenamefont
  {Baenitz}}]{Majumder2015}%
  \BibitemOpen
  \bibfield  {author} {\bibinfo {author} {\bibfnamefont {M.}~\bibnamefont
  {Majumder}}, \bibinfo {author} {\bibfnamefont {M.}~\bibnamefont {Schmidt}},
  \bibinfo {author} {\bibfnamefont {H.}~\bibnamefont {Rosner}}, \bibinfo
  {author} {\bibfnamefont {A.~A.}\ \bibnamefont {Tsirlin}}, \bibinfo {author}
  {\bibfnamefont {H.}~\bibnamefont {Yasuoka}}, \ and\ \bibinfo {author}
  {\bibfnamefont {M.}~\bibnamefont {Baenitz}},\ }\href {\doibase
  10.1103/PhysRevB.91.180401} {\bibfield  {journal} {\bibinfo  {journal} {Phys.
  Rev. B}\ }\textbf {\bibinfo {volume} {91}},\ \bibinfo {pages} {180401}
  (\bibinfo {year} {2015})}\BibitemShut {NoStop}%
\bibitem [{\citenamefont {Johnson}\ \emph {et~al.}(2015)\citenamefont
  {Johnson}, \citenamefont {Williams}, \citenamefont {Haghighirad},
  \citenamefont {Singleton}, \citenamefont {Zapf}, \citenamefont {Manuel},
  \citenamefont {Mazin}, \citenamefont {Li}, \citenamefont {Jeschke},
  \citenamefont {Valent\'{\i}},\ and\ \citenamefont {Coldea}}]{Johnson2015}%
  \BibitemOpen
  \bibfield  {author} {\bibinfo {author} {\bibfnamefont {R.~D.}\ \bibnamefont
  {Johnson}}, \bibinfo {author} {\bibfnamefont {S.~C.}\ \bibnamefont
  {Williams}}, \bibinfo {author} {\bibfnamefont {A.~A.}\ \bibnamefont
  {Haghighirad}}, \bibinfo {author} {\bibfnamefont {J.}~\bibnamefont
  {Singleton}}, \bibinfo {author} {\bibfnamefont {V.}~\bibnamefont {Zapf}},
  \bibinfo {author} {\bibfnamefont {P.}~\bibnamefont {Manuel}}, \bibinfo
  {author} {\bibfnamefont {I.~I.}\ \bibnamefont {Mazin}}, \bibinfo {author}
  {\bibfnamefont {Y.}~\bibnamefont {Li}}, \bibinfo {author} {\bibfnamefont
  {H.~O.}\ \bibnamefont {Jeschke}}, \bibinfo {author} {\bibfnamefont
  {R.}~\bibnamefont {Valent\'{\i}}}, \ and\ \bibinfo {author} {\bibfnamefont
  {R.}~\bibnamefont {Coldea}},\ }\href {\doibase 10.1103/PhysRevB.92.235119}
  {\bibfield  {journal} {\bibinfo  {journal} {Phys. Rev. B}\ }\textbf {\bibinfo
  {volume} {92}},\ \bibinfo {pages} {235119} (\bibinfo {year}
  {2015})}\BibitemShut {NoStop}%
\bibitem [{\citenamefont {Sandilands}\ \emph {et~al.}(2016)\citenamefont
  {Sandilands}, \citenamefont {Tian}, \citenamefont {Reijnders}, \citenamefont
  {Kim}, \citenamefont {Plumb}, \citenamefont {Kim}, \citenamefont {Kee},\ and\
  \citenamefont {Burch}}]{Sandilands2016}%
  \BibitemOpen
  \bibfield  {author} {\bibinfo {author} {\bibfnamefont {L.~J.}\ \bibnamefont
  {Sandilands}}, \bibinfo {author} {\bibfnamefont {Y.}~\bibnamefont {Tian}},
  \bibinfo {author} {\bibfnamefont {A.~A.}\ \bibnamefont {Reijnders}}, \bibinfo
  {author} {\bibfnamefont {H.-S.}\ \bibnamefont {Kim}}, \bibinfo {author}
  {\bibfnamefont {K.~W.}\ \bibnamefont {Plumb}}, \bibinfo {author}
  {\bibfnamefont {Y.-J.}\ \bibnamefont {Kim}}, \bibinfo {author} {\bibfnamefont
  {H.-Y.}\ \bibnamefont {Kee}}, \ and\ \bibinfo {author} {\bibfnamefont
  {K.~S.}\ \bibnamefont {Burch}},\ }\href {\doibase 10.1103/PhysRevB.93.075144}
  {\bibfield  {journal} {\bibinfo  {journal} {Phys. Rev. B}\ }\textbf {\bibinfo
  {volume} {93}},\ \bibinfo {pages} {075144} (\bibinfo {year}
  {2016})}\BibitemShut {NoStop}%
\bibitem [{\citenamefont {Banerjee}\ \emph {et~al.}(2016)\citenamefont
  {Banerjee}, \citenamefont {Bridges}, \citenamefont {Yan}, \citenamefont
  {Aczel}, \citenamefont {Li}, \citenamefont {Stone}, \citenamefont {Granroth},
  \citenamefont {Lumsden}, \citenamefont {Yiu}, \citenamefont {Knolle},
  \citenamefont {Bhattacharjee}, \citenamefont {Kovrizhin}, \citenamefont
  {Moessner}, \citenamefont {Tennant}, \citenamefont {G.},\ and\ \citenamefont
  {Nagler}}]{Banerjee2016}%
  \BibitemOpen
  \bibfield  {author} {\bibinfo {author} {\bibfnamefont {A.}~\bibnamefont
  {Banerjee}}, \bibinfo {author} {\bibfnamefont {C.~A.}\ \bibnamefont
  {Bridges}}, \bibinfo {author} {\bibfnamefont {J.-Q.}\ \bibnamefont {Yan}},
  \bibinfo {author} {\bibfnamefont {A.~A.}\ \bibnamefont {Aczel}}, \bibinfo
  {author} {\bibfnamefont {L.}~\bibnamefont {Li}}, \bibinfo {author}
  {\bibfnamefont {M.~B.}\ \bibnamefont {Stone}}, \bibinfo {author}
  {\bibfnamefont {G.~E.}\ \bibnamefont {Granroth}}, \bibinfo {author}
  {\bibfnamefont {M.~D.}\ \bibnamefont {Lumsden}}, \bibinfo {author}
  {\bibfnamefont {Y.}~\bibnamefont {Yiu}}, \bibinfo {author} {\bibfnamefont
  {J.}~\bibnamefont {Knolle}}, \bibinfo {author} {\bibfnamefont
  {S.}~\bibnamefont {Bhattacharjee}}, \bibinfo {author} {\bibfnamefont {D.~L.}\
  \bibnamefont {Kovrizhin}}, \bibinfo {author} {\bibfnamefont {R.}~\bibnamefont
  {Moessner}}, \bibinfo {author} {\bibfnamefont {D.~A.}\ \bibnamefont
  {Tennant}}, \bibinfo {author} {\bibfnamefont {M.~D.}\ \bibnamefont {G.}}, \
  and\ \bibinfo {author} {\bibfnamefont {S.~E.}\ \bibnamefont {Nagler}},\
  }\href {\doibase 10.1038/nmat4604} {\bibfield  {journal} {\bibinfo  {journal}
  {Nature materials}\ } (\bibinfo {year} {2016}),\
  10.1038/nmat4604}\BibitemShut {NoStop}%
\bibitem [{\citenamefont {Banerjee}\ \emph {et~al.}(2017)\citenamefont
  {Banerjee}, \citenamefont {Yan}, \citenamefont {Knolle}, \citenamefont
  {Bridges}, \citenamefont {Stone}, \citenamefont {Lumsden}, \citenamefont
  {Mandrus}, \citenamefont {Tennant}, \citenamefont {Moessner},\ and\
  \citenamefont {Nagler}}]{Banerjee2017}%
  \BibitemOpen
  \bibfield  {author} {\bibinfo {author} {\bibfnamefont {A.}~\bibnamefont
  {Banerjee}}, \bibinfo {author} {\bibfnamefont {J.}~\bibnamefont {Yan}},
  \bibinfo {author} {\bibfnamefont {J.}~\bibnamefont {Knolle}}, \bibinfo
  {author} {\bibfnamefont {C.~A.}\ \bibnamefont {Bridges}}, \bibinfo {author}
  {\bibfnamefont {M.~B.}\ \bibnamefont {Stone}}, \bibinfo {author}
  {\bibfnamefont {M.~D.}\ \bibnamefont {Lumsden}}, \bibinfo {author}
  {\bibfnamefont {D.~G.}\ \bibnamefont {Mandrus}}, \bibinfo {author}
  {\bibfnamefont {D.~A.}\ \bibnamefont {Tennant}}, \bibinfo {author}
  {\bibfnamefont {R.}~\bibnamefont {Moessner}}, \ and\ \bibinfo {author}
  {\bibfnamefont {S.~E.}\ \bibnamefont {Nagler}},\ }\href {\doibase
  10.1126/science.aah6015} {\bibfield  {journal} {\bibinfo  {journal}
  {Science}\ }\textbf {\bibinfo {volume} {356}},\ \bibinfo {pages} {1055}
  (\bibinfo {year} {2017})}\BibitemShut {NoStop}%
\bibitem [{\citenamefont {Do}\ \emph {et~al.}(2017)\citenamefont {Do},
  \citenamefont {Park}, \citenamefont {Yoshitake}, \citenamefont {Nasu},
  \citenamefont {Motome}, \citenamefont {Kwon}, \citenamefont {Adroja},
  \citenamefont {Voneshen}, \citenamefont {Kim}, \citenamefont {Jang},
  \citenamefont {Park}, \citenamefont {Choi},\ and\ \citenamefont
  {Ji}}]{Do2017}%
  \BibitemOpen
  \bibfield  {author} {\bibinfo {author} {\bibfnamefont {S.-H.}\ \bibnamefont
  {Do}}, \bibinfo {author} {\bibfnamefont {S.-Y.}\ \bibnamefont {Park}},
  \bibinfo {author} {\bibfnamefont {J.}~\bibnamefont {Yoshitake}}, \bibinfo
  {author} {\bibfnamefont {J.}~\bibnamefont {Nasu}}, \bibinfo {author}
  {\bibfnamefont {Y.}~\bibnamefont {Motome}}, \bibinfo {author} {\bibfnamefont
  {Y.~S.}\ \bibnamefont {Kwon}}, \bibinfo {author} {\bibfnamefont {D.~T.}\
  \bibnamefont {Adroja}}, \bibinfo {author} {\bibfnamefont {D.~J.}\
  \bibnamefont {Voneshen}}, \bibinfo {author} {\bibfnamefont {K.}~\bibnamefont
  {Kim}}, \bibinfo {author} {\bibfnamefont {T.~H.}\ \bibnamefont {Jang}},
  \bibinfo {author} {\bibfnamefont {J.~H.}\ \bibnamefont {Park}}, \bibinfo
  {author} {\bibfnamefont {K.-Y.}\ \bibnamefont {Choi}}, \ and\ \bibinfo
  {author} {\bibfnamefont {S.}~\bibnamefont {Ji}},\ }\href {\doibase
  10.1038/nphys4264} {\bibfield  {journal} {\bibinfo  {journal} {Nature
  Physics}\ }\textbf {\bibinfo {volume} {13}},\ \bibinfo {pages} {1079}
  (\bibinfo {year} {2017})}\BibitemShut {NoStop}%
\bibitem [{\citenamefont {Wu}\ \emph {et~al.}(2018)\citenamefont {Wu},
  \citenamefont {Little}, \citenamefont {Aldape}, \citenamefont {Rees},
  \citenamefont {Thewalt}, \citenamefont {Lampen-Kelley}, \citenamefont
  {Banerjee}, \citenamefont {Bridges}, \citenamefont {Yan}, \citenamefont
  {Boone}, \citenamefont {Patankar}, \citenamefont {Goldhaber-Gordon},
  \citenamefont {Mandrus}, \citenamefont {Nagler}, \citenamefont {Altman},\
  and\ \citenamefont {Orenstein}}]{Wu2018}%
  \BibitemOpen
  \bibfield  {author} {\bibinfo {author} {\bibfnamefont {L.}~\bibnamefont
  {Wu}}, \bibinfo {author} {\bibfnamefont {A.}~\bibnamefont {Little}}, \bibinfo
  {author} {\bibfnamefont {E.~E.}\ \bibnamefont {Aldape}}, \bibinfo {author}
  {\bibfnamefont {D.}~\bibnamefont {Rees}}, \bibinfo {author} {\bibfnamefont
  {E.}~\bibnamefont {Thewalt}}, \bibinfo {author} {\bibfnamefont
  {P.}~\bibnamefont {Lampen-Kelley}}, \bibinfo {author} {\bibfnamefont
  {A.}~\bibnamefont {Banerjee}}, \bibinfo {author} {\bibfnamefont {C.~A.}\
  \bibnamefont {Bridges}}, \bibinfo {author} {\bibfnamefont {J.-Q.}\
  \bibnamefont {Yan}}, \bibinfo {author} {\bibfnamefont {D.}~\bibnamefont
  {Boone}}, \bibinfo {author} {\bibfnamefont {S.}~\bibnamefont {Patankar}},
  \bibinfo {author} {\bibfnamefont {D.}~\bibnamefont {Goldhaber-Gordon}},
  \bibinfo {author} {\bibfnamefont {D.}~\bibnamefont {Mandrus}}, \bibinfo
  {author} {\bibfnamefont {S.~E.}\ \bibnamefont {Nagler}}, \bibinfo {author}
  {\bibfnamefont {E.}~\bibnamefont {Altman}}, \ and\ \bibinfo {author}
  {\bibfnamefont {J.}~\bibnamefont {Orenstein}},\ }\href {\doibase
  10.1103/PhysRevB.98.094425} {\bibfield  {journal} {\bibinfo  {journal} {Phys.
  Rev. B}\ }\textbf {\bibinfo {volume} {98}},\ \bibinfo {pages} {094425}
  (\bibinfo {year} {2018})}\BibitemShut {NoStop}%
\bibitem [{\citenamefont {Wulferding}\ \emph {et~al.}(2020)\citenamefont
  {Wulferding}, \citenamefont {Choi}, \citenamefont {Do}, \citenamefont {Lee},
  \citenamefont {Lemmens}, \citenamefont {Faugeras},\ and\ \citenamefont
  {Gallais}}]{Dirk2020}%
  \BibitemOpen
  \bibfield  {author} {\bibinfo {author} {\bibfnamefont {D.}~\bibnamefont
  {Wulferding}}, \bibinfo {author} {\bibfnamefont {Y.}~\bibnamefont {Choi}},
  \bibinfo {author} {\bibfnamefont {S.-H.}\ \bibnamefont {Do}}, \bibinfo
  {author} {\bibfnamefont {C.~H.}\ \bibnamefont {Lee}}, \bibinfo {author}
  {\bibfnamefont {P.}~\bibnamefont {Lemmens}}, \bibinfo {author} {\bibfnamefont
  {C.}~\bibnamefont {Faugeras}}, \ and\ \bibinfo {author} {\bibfnamefont
  {K.-Y.}\ \bibnamefont {Gallais}, \bibfnamefont {Yann~andChoi}},\ }\href
  {\doibase 10.1038/s41467-020-15370-1} {\bibfield  {journal} {\bibinfo
  {journal} {Nat. Commun.}\ }\textbf {\bibinfo {volume} {11}},\ \bibinfo
  {pages} {1603} (\bibinfo {year} {2020})}\BibitemShut {NoStop}%
\bibitem [{\citenamefont {Ruiz}\ \emph {et~al.}(2021)\citenamefont {Ruiz},
  \citenamefont {Breznay}, \citenamefont {Li}, \citenamefont {Rousochatzakis},
  \citenamefont {Allen}, \citenamefont {Zinda}, \citenamefont {Nagarajan},
  \citenamefont {Lopez}, \citenamefont {Islam}, \citenamefont {Upton},
  \citenamefont {Kim}, \citenamefont {Said}, \citenamefont {Huang},
  \citenamefont {Gog}, \citenamefont {Casa}, \citenamefont {Birgeneau},
  \citenamefont {Koralek}, \citenamefont {Analytis}, \citenamefont {Perkins},\
  and\ \citenamefont {Frano}}]{Ruiz2021}%
  \BibitemOpen
  \bibfield  {author} {\bibinfo {author} {\bibfnamefont {A.}~\bibnamefont
  {Ruiz}}, \bibinfo {author} {\bibfnamefont {N.~P.}\ \bibnamefont {Breznay}},
  \bibinfo {author} {\bibfnamefont {M.}~\bibnamefont {Li}}, \bibinfo {author}
  {\bibfnamefont {I.}~\bibnamefont {Rousochatzakis}}, \bibinfo {author}
  {\bibfnamefont {A.}~\bibnamefont {Allen}}, \bibinfo {author} {\bibfnamefont
  {I.}~\bibnamefont {Zinda}}, \bibinfo {author} {\bibfnamefont
  {V.}~\bibnamefont {Nagarajan}}, \bibinfo {author} {\bibfnamefont
  {G.}~\bibnamefont {Lopez}}, \bibinfo {author} {\bibfnamefont
  {Z.}~\bibnamefont {Islam}}, \bibinfo {author} {\bibfnamefont {M.~H.}\
  \bibnamefont {Upton}}, \bibinfo {author} {\bibfnamefont {J.}~\bibnamefont
  {Kim}}, \bibinfo {author} {\bibfnamefont {A.~H.}\ \bibnamefont {Said}},
  \bibinfo {author} {\bibfnamefont {X.-R.}\ \bibnamefont {Huang}}, \bibinfo
  {author} {\bibfnamefont {T.}~\bibnamefont {Gog}}, \bibinfo {author}
  {\bibfnamefont {D.}~\bibnamefont {Casa}}, \bibinfo {author} {\bibfnamefont
  {R.~J.}\ \bibnamefont {Birgeneau}}, \bibinfo {author} {\bibfnamefont {J.~D.}\
  \bibnamefont {Koralek}}, \bibinfo {author} {\bibfnamefont {J.~G.}\
  \bibnamefont {Analytis}}, \bibinfo {author} {\bibfnamefont {N.~B.}\
  \bibnamefont {Perkins}}, \ and\ \bibinfo {author} {\bibfnamefont
  {A.}~\bibnamefont {Frano}},\ }\href {\doibase 10.1103/PhysRevB.103.184404}
  {\bibfield  {journal} {\bibinfo  {journal} {Phys. Rev. B}\ }\textbf {\bibinfo
  {volume} {103}},\ \bibinfo {pages} {184404} (\bibinfo {year}
  {2021})}\BibitemShut {NoStop}%
\bibitem [{\citenamefont {Halloran}\ \emph {et~al.}(2022)\citenamefont
  {Halloran}, \citenamefont {Wang}, \citenamefont {Li}, \citenamefont
  {Rousochatzakis}, \citenamefont {Chauhan}, \citenamefont {Stone},
  \citenamefont {Takayama}, \citenamefont {Takagi}, \citenamefont {Armitage},
  \citenamefont {Perkins},\ and\ \citenamefont {Broholm}}]{Halloran2022}%
  \BibitemOpen
  \bibfield  {author} {\bibinfo {author} {\bibfnamefont {T.}~\bibnamefont
  {Halloran}}, \bibinfo {author} {\bibfnamefont {Y.}~\bibnamefont {Wang}},
  \bibinfo {author} {\bibfnamefont {M.}~\bibnamefont {Li}}, \bibinfo {author}
  {\bibfnamefont {I.}~\bibnamefont {Rousochatzakis}}, \bibinfo {author}
  {\bibfnamefont {P.}~\bibnamefont {Chauhan}}, \bibinfo {author} {\bibfnamefont
  {M.~B.}\ \bibnamefont {Stone}}, \bibinfo {author} {\bibfnamefont
  {T.}~\bibnamefont {Takayama}}, \bibinfo {author} {\bibfnamefont
  {H.}~\bibnamefont {Takagi}}, \bibinfo {author} {\bibfnamefont {N.~P.}\
  \bibnamefont {Armitage}}, \bibinfo {author} {\bibfnamefont {N.~B.}\
  \bibnamefont {Perkins}}, \ and\ \bibinfo {author} {\bibfnamefont
  {C.}~\bibnamefont {Broholm}},\ }\href {\doibase 10.1103/PhysRevB.106.064423}
  {\bibfield  {journal} {\bibinfo  {journal} {Phys. Rev. B}\ }\textbf {\bibinfo
  {volume} {106}},\ \bibinfo {pages} {064423} (\bibinfo {year}
  {2022})}\BibitemShut {NoStop}%
\bibitem [{\citenamefont {Yang}\ \emph {et~al.}(2022)\citenamefont {Yang},
  \citenamefont {Wang}, \citenamefont {Rousochatzakis}, \citenamefont {Ruiz},
  \citenamefont {Analytis}, \citenamefont {Burch},\ and\ \citenamefont
  {Perkins}}]{Yang2022}%
  \BibitemOpen
  \bibfield  {author} {\bibinfo {author} {\bibfnamefont {Y.}~\bibnamefont
  {Yang}}, \bibinfo {author} {\bibfnamefont {Y.}~\bibnamefont {Wang}}, \bibinfo
  {author} {\bibfnamefont {I.}~\bibnamefont {Rousochatzakis}}, \bibinfo
  {author} {\bibfnamefont {A.}~\bibnamefont {Ruiz}}, \bibinfo {author}
  {\bibfnamefont {J.~G.}\ \bibnamefont {Analytis}}, \bibinfo {author}
  {\bibfnamefont {K.~S.}\ \bibnamefont {Burch}}, \ and\ \bibinfo {author}
  {\bibfnamefont {N.~B.}\ \bibnamefont {Perkins}},\ }\href {\doibase
  10.1103/PhysRevB.105.L241101} {\bibfield  {journal} {\bibinfo  {journal}
  {Phys. Rev. B}\ }\textbf {\bibinfo {volume} {105}},\ \bibinfo {pages}
  {L241101} (\bibinfo {year} {2022})}\BibitemShut {NoStop}%
\bibitem [{\citenamefont {Li}\ \emph {et~al.}(2021{\natexlab{a}})\citenamefont
  {Li}, \citenamefont {Zhang}, \citenamefont {Said}, \citenamefont {Fabbris},
  \citenamefont {Mazzone}, \citenamefont {Yan}, \citenamefont {Mandrus},
  \citenamefont {Halasz}, \citenamefont {Okamoto}, \citenamefont {Murakami},
  \citenamefont {Dean}, \citenamefont {Lee},\ and\ \citenamefont
  {Miao}}]{Miao2020}%
  \BibitemOpen
  \bibfield  {author} {\bibinfo {author} {\bibfnamefont {H.}~\bibnamefont
  {Li}}, \bibinfo {author} {\bibfnamefont {T.~T.}\ \bibnamefont {Zhang}},
  \bibinfo {author} {\bibfnamefont {A.}~\bibnamefont {Said}}, \bibinfo {author}
  {\bibfnamefont {G.}~\bibnamefont {Fabbris}}, \bibinfo {author} {\bibfnamefont
  {D.~G.}\ \bibnamefont {Mazzone}}, \bibinfo {author} {\bibfnamefont {J.~Q.}\
  \bibnamefont {Yan}}, \bibinfo {author} {\bibfnamefont {D.}~\bibnamefont
  {Mandrus}}, \bibinfo {author} {\bibfnamefont {G.~B.}\ \bibnamefont {Halasz}},
  \bibinfo {author} {\bibfnamefont {S.}~\bibnamefont {Okamoto}}, \bibinfo
  {author} {\bibfnamefont {S.}~\bibnamefont {Murakami}}, \bibinfo {author}
  {\bibfnamefont {M.~P.~M.}\ \bibnamefont {Dean}}, \bibinfo {author}
  {\bibfnamefont {H.~N.}\ \bibnamefont {Lee}}, \ and\ \bibinfo {author}
  {\bibfnamefont {H.}~\bibnamefont {Miao}},\ }\href {\doibase
  10.1038/s41467-021-23826-1} {\bibfield  {journal} {\bibinfo  {journal}
  {Nature Communications}\ }\textbf {\bibinfo {volume} {12}},\ \bibinfo {pages}
  {3513} (\bibinfo {year} {2021}{\natexlab{a}})}\BibitemShut {NoStop}%
\bibitem [{\citenamefont {Mu}\ \emph {et~al.}(2022)\citenamefont {Mu},
  \citenamefont {Dixit}, \citenamefont {Wang}, \citenamefont {Abernathy},
  \citenamefont {Cao}, \citenamefont {Nagler}, \citenamefont {Yan},
  \citenamefont {Lampen-Kelley}, \citenamefont {Mandrus}, \citenamefont
  {Polanco}, \citenamefont {Liang}, \citenamefont {Hal\'asz}, \citenamefont
  {Cheng}, \citenamefont {Banerjee},\ and\ \citenamefont {Berlijn}}]{Mu2022}%
  \BibitemOpen
  \bibfield  {author} {\bibinfo {author} {\bibfnamefont {S.}~\bibnamefont
  {Mu}}, \bibinfo {author} {\bibfnamefont {K.~D.}\ \bibnamefont {Dixit}},
  \bibinfo {author} {\bibfnamefont {X.}~\bibnamefont {Wang}}, \bibinfo {author}
  {\bibfnamefont {D.~L.}\ \bibnamefont {Abernathy}}, \bibinfo {author}
  {\bibfnamefont {H.}~\bibnamefont {Cao}}, \bibinfo {author} {\bibfnamefont
  {S.~E.}\ \bibnamefont {Nagler}}, \bibinfo {author} {\bibfnamefont
  {J.}~\bibnamefont {Yan}}, \bibinfo {author} {\bibfnamefont {P.}~\bibnamefont
  {Lampen-Kelley}}, \bibinfo {author} {\bibfnamefont {D.}~\bibnamefont
  {Mandrus}}, \bibinfo {author} {\bibfnamefont {C.~A.}\ \bibnamefont
  {Polanco}}, \bibinfo {author} {\bibfnamefont {L.}~\bibnamefont {Liang}},
  \bibinfo {author} {\bibfnamefont {G.~B.}\ \bibnamefont {Hal\'asz}}, \bibinfo
  {author} {\bibfnamefont {Y.}~\bibnamefont {Cheng}}, \bibinfo {author}
  {\bibfnamefont {A.}~\bibnamefont {Banerjee}}, \ and\ \bibinfo {author}
  {\bibfnamefont {T.}~\bibnamefont {Berlijn}},\ }\href {\doibase
  10.1103/PhysRevResearch.4.013067} {\bibfield  {journal} {\bibinfo  {journal}
  {Phys. Rev. Res.}\ }\textbf {\bibinfo {volume} {4}},\ \bibinfo {pages}
  {013067} (\bibinfo {year} {2022})}\BibitemShut {NoStop}%
\bibitem [{\citenamefont {Knolle}\ \emph
  {et~al.}(2014{\natexlab{a}})\citenamefont {Knolle}, \citenamefont
  {Kovrizhin}, \citenamefont {Chalker},\ and\ \citenamefont
  {Moessner}}]{Knolle2014a}%
  \BibitemOpen
  \bibfield  {author} {\bibinfo {author} {\bibfnamefont {J.}~\bibnamefont
  {Knolle}}, \bibinfo {author} {\bibfnamefont {D.~L.}\ \bibnamefont
  {Kovrizhin}}, \bibinfo {author} {\bibfnamefont {J.~T.}\ \bibnamefont
  {Chalker}}, \ and\ \bibinfo {author} {\bibfnamefont {R.}~\bibnamefont
  {Moessner}},\ }\href {\doibase 10.1103/PhysRevLett.112.207203} {\bibfield
  {journal} {\bibinfo  {journal} {Phys. Rev. Lett.}\ }\textbf {\bibinfo
  {volume} {112}},\ \bibinfo {pages} {207203} (\bibinfo {year}
  {2014}{\natexlab{a}})}\BibitemShut {NoStop}%
\bibitem [{\citenamefont {Knolle}\ \emph
  {et~al.}(2014{\natexlab{b}})\citenamefont {Knolle}, \citenamefont {Chern},
  \citenamefont {Kovrizhin}, \citenamefont {Moessner},\ and\ \citenamefont
  {Perkins}}]{Knolle2014b}%
  \BibitemOpen
  \bibfield  {author} {\bibinfo {author} {\bibfnamefont {J.}~\bibnamefont
  {Knolle}}, \bibinfo {author} {\bibfnamefont {G.-W.}\ \bibnamefont {Chern}},
  \bibinfo {author} {\bibfnamefont {D.~L.}\ \bibnamefont {Kovrizhin}}, \bibinfo
  {author} {\bibfnamefont {R.}~\bibnamefont {Moessner}}, \ and\ \bibinfo
  {author} {\bibfnamefont {N.~B.}\ \bibnamefont {Perkins}},\ }\href {\doibase
  10.1103/PhysRevLett.113.187201} {\bibfield  {journal} {\bibinfo  {journal}
  {Phys. Rev. Lett.}\ }\textbf {\bibinfo {volume} {113}},\ \bibinfo {pages}
  {187201} (\bibinfo {year} {2014}{\natexlab{b}})}\BibitemShut {NoStop}%
\bibitem [{\citenamefont {Knolle}\ \emph {et~al.}(2015)\citenamefont {Knolle},
  \citenamefont {Kovrizhin}, \citenamefont {Chalker},\ and\ \citenamefont
  {Moessner}}]{Knolle2015}%
  \BibitemOpen
  \bibfield  {author} {\bibinfo {author} {\bibfnamefont {J.}~\bibnamefont
  {Knolle}}, \bibinfo {author} {\bibfnamefont {D.~L.}\ \bibnamefont
  {Kovrizhin}}, \bibinfo {author} {\bibfnamefont {J.~T.}\ \bibnamefont
  {Chalker}}, \ and\ \bibinfo {author} {\bibfnamefont {R.}~\bibnamefont
  {Moessner}},\ }\href {\doibase 10.1103/PhysRevB.92.115127} {\bibfield
  {journal} {\bibinfo  {journal} {Phys. Rev. B}\ }\textbf {\bibinfo {volume}
  {92}},\ \bibinfo {pages} {115127} (\bibinfo {year} {2015})}\BibitemShut
  {NoStop}%
\bibitem [{\citenamefont {Nasu}\ \emph {et~al.}(2014)\citenamefont {Nasu},
  \citenamefont {Udagawa},\ and\ \citenamefont
  {Motome}}]{nasu2014vaporization}%
  \BibitemOpen
  \bibfield  {author} {\bibinfo {author} {\bibfnamefont {J.}~\bibnamefont
  {Nasu}}, \bibinfo {author} {\bibfnamefont {M.}~\bibnamefont {Udagawa}}, \
  and\ \bibinfo {author} {\bibfnamefont {Y.}~\bibnamefont {Motome}},\ }\href
  {https://journals.aps.org/prl/abstract/10.1103/PhysRevLett.113.197205}
  {\bibfield  {journal} {\bibinfo  {journal} {Physical review letters}\
  }\textbf {\bibinfo {volume} {113}},\ \bibinfo {pages} {197205} (\bibinfo
  {year} {2014})}\BibitemShut {NoStop}%
\bibitem [{\citenamefont {Nasu}\ \emph {et~al.}(2016)\citenamefont {Nasu},
  \citenamefont {Knolle}, \citenamefont {Kovrizhin}, \citenamefont {Motome},\
  and\ \citenamefont {Moessner}}]{nasu2016fermionic}%
  \BibitemOpen
  \bibfield  {author} {\bibinfo {author} {\bibfnamefont {J.}~\bibnamefont
  {Nasu}}, \bibinfo {author} {\bibfnamefont {J.}~\bibnamefont {Knolle}},
  \bibinfo {author} {\bibfnamefont {D.~L.}\ \bibnamefont {Kovrizhin}}, \bibinfo
  {author} {\bibfnamefont {Y.}~\bibnamefont {Motome}}, \ and\ \bibinfo {author}
  {\bibfnamefont {R.}~\bibnamefont {Moessner}},\ }\href@noop {} {\bibfield
  {journal} {\bibinfo  {journal} {Nature Physics}\ }\textbf {\bibinfo {volume}
  {12}},\ \bibinfo {pages} {912} (\bibinfo {year} {2016})}\BibitemShut
  {NoStop}%
\bibitem [{\citenamefont {Perreault}\ \emph {et~al.}(2015)\citenamefont
  {Perreault}, \citenamefont {Knolle}, \citenamefont {Perkins},\ and\
  \citenamefont {Burnell}}]{Perreault2015}%
  \BibitemOpen
  \bibfield  {author} {\bibinfo {author} {\bibfnamefont {B.}~\bibnamefont
  {Perreault}}, \bibinfo {author} {\bibfnamefont {J.}~\bibnamefont {Knolle}},
  \bibinfo {author} {\bibfnamefont {N.~B.}\ \bibnamefont {Perkins}}, \ and\
  \bibinfo {author} {\bibfnamefont {F.~J.}\ \bibnamefont {Burnell}},\ }\href
  {\doibase 10.1103/PhysRevB.92.094439} {\bibfield  {journal} {\bibinfo
  {journal} {Phys. Rev. B}\ }\textbf {\bibinfo {volume} {92}},\ \bibinfo
  {pages} {094439} (\bibinfo {year} {2015})}\BibitemShut {NoStop}%
\bibitem [{\citenamefont {Perreault}\ \emph {et~al.}(2016)\citenamefont
  {Perreault}, \citenamefont {Knolle}, \citenamefont {Perkins},\ and\
  \citenamefont {Burnell}}]{Perreault2016}%
  \BibitemOpen
  \bibfield  {author} {\bibinfo {author} {\bibfnamefont {B.}~\bibnamefont
  {Perreault}}, \bibinfo {author} {\bibfnamefont {J.}~\bibnamefont {Knolle}},
  \bibinfo {author} {\bibfnamefont {N.~B.}\ \bibnamefont {Perkins}}, \ and\
  \bibinfo {author} {\bibfnamefont {F.~J.}\ \bibnamefont {Burnell}},\ }\href
  {\doibase 10.1103/PhysRevB.94.104427} {\bibfield  {journal} {\bibinfo
  {journal} {Phys. Rev. B}\ }\textbf {\bibinfo {volume} {94}},\ \bibinfo
  {pages} {104427} (\bibinfo {year} {2016})}\BibitemShut {NoStop}%
\bibitem [{\citenamefont {Hal\'asz}\ \emph {et~al.}(2016)\citenamefont
  {Hal\'asz}, \citenamefont {Perkins},\ and\ \citenamefont {van~den
  Brink}}]{Gabor2016}%
  \BibitemOpen
  \bibfield  {author} {\bibinfo {author} {\bibfnamefont {G.~B.}\ \bibnamefont
  {Hal\'asz}}, \bibinfo {author} {\bibfnamefont {N.~B.}\ \bibnamefont
  {Perkins}}, \ and\ \bibinfo {author} {\bibfnamefont {J.}~\bibnamefont
  {van~den Brink}},\ }\href {\doibase 10.1103/PhysRevLett.117.127203}
  {\bibfield  {journal} {\bibinfo  {journal} {Phys. Rev. Lett.}\ }\textbf
  {\bibinfo {volume} {117}},\ \bibinfo {pages} {127203} (\bibinfo {year}
  {2016})}\BibitemShut {NoStop}%
\bibitem [{\citenamefont {Hal\'asz}\ \emph {et~al.}(2017)\citenamefont
  {Hal\'asz}, \citenamefont {Perreault},\ and\ \citenamefont
  {Perkins}}]{Gabor2017}%
  \BibitemOpen
  \bibfield  {author} {\bibinfo {author} {\bibfnamefont {G.~B.}\ \bibnamefont
  {Hal\'asz}}, \bibinfo {author} {\bibfnamefont {B.}~\bibnamefont {Perreault}},
  \ and\ \bibinfo {author} {\bibfnamefont {N.~B.}\ \bibnamefont {Perkins}},\
  }\href {\doibase 10.1103/PhysRevLett.119.097202} {\bibfield  {journal}
  {\bibinfo  {journal} {Phys. Rev. Lett.}\ }\textbf {\bibinfo {volume} {119}},\
  \bibinfo {pages} {097202} (\bibinfo {year} {2017})}\BibitemShut {NoStop}%
\bibitem [{\citenamefont {Hal\'asz}\ \emph {et~al.}(2019)\citenamefont
  {Hal\'asz}, \citenamefont {Kourtis}, \citenamefont {Knolle},\ and\
  \citenamefont {Perkins}}]{Gabor2019}%
  \BibitemOpen
  \bibfield  {author} {\bibinfo {author} {\bibfnamefont {G.~B.}\ \bibnamefont
  {Hal\'asz}}, \bibinfo {author} {\bibfnamefont {S.}~\bibnamefont {Kourtis}},
  \bibinfo {author} {\bibfnamefont {J.}~\bibnamefont {Knolle}}, \ and\ \bibinfo
  {author} {\bibfnamefont {N.~B.}\ \bibnamefont {Perkins}},\ }\href {\doibase
  10.1103/PhysRevB.99.184417} {\bibfield  {journal} {\bibinfo  {journal} {Phys.
  Rev. B}\ }\textbf {\bibinfo {volume} {99}},\ \bibinfo {pages} {184417}
  (\bibinfo {year} {2019})}\BibitemShut {NoStop}%
\bibitem [{\citenamefont {Rousochatzakis}\ \emph {et~al.}(2019)\citenamefont
  {Rousochatzakis}, \citenamefont {Kourtis}, \citenamefont {Knolle},
  \citenamefont {Moessner},\ and\ \citenamefont
  {Perkins}}]{Rousochatzakis2019}%
  \BibitemOpen
  \bibfield  {author} {\bibinfo {author} {\bibfnamefont {I.}~\bibnamefont
  {Rousochatzakis}}, \bibinfo {author} {\bibfnamefont {S.}~\bibnamefont
  {Kourtis}}, \bibinfo {author} {\bibfnamefont {J.}~\bibnamefont {Knolle}},
  \bibinfo {author} {\bibfnamefont {R.}~\bibnamefont {Moessner}}, \ and\
  \bibinfo {author} {\bibfnamefont {N.~B.}\ \bibnamefont {Perkins}},\ }\href
  {\doibase 10.1103/PhysRevB.100.045117} {\bibfield  {journal} {\bibinfo
  {journal} {Phys. Rev. B}\ }\textbf {\bibinfo {volume} {100}},\ \bibinfo
  {pages} {045117} (\bibinfo {year} {2019})}\BibitemShut {NoStop}%
\bibitem [{\citenamefont {Udagawa}\ \emph {et~al.}(2021)\citenamefont
  {Udagawa}, \citenamefont {Takayoshi},\ and\ \citenamefont
  {Oka}}]{udagawa2021}%
  \BibitemOpen
  \bibfield  {author} {\bibinfo {author} {\bibfnamefont {M.}~\bibnamefont
  {Udagawa}}, \bibinfo {author} {\bibfnamefont {S.}~\bibnamefont {Takayoshi}},
  \ and\ \bibinfo {author} {\bibfnamefont {T.}~\bibnamefont {Oka}},\ }\href
  {\doibase 10.1103/PhysRevLett.126.127201} {\bibfield  {journal} {\bibinfo
  {journal} {Phys. Rev. Lett.}\ }\textbf {\bibinfo {volume} {126}},\ \bibinfo
  {pages} {127201} (\bibinfo {year} {2021})}\BibitemShut {NoStop}%
\bibitem [{\citenamefont {Wan}\ and\ \citenamefont {Armitage}(2019)}]{Wan2019}%
  \BibitemOpen
  \bibfield  {author} {\bibinfo {author} {\bibfnamefont {Y.}~\bibnamefont
  {Wan}}\ and\ \bibinfo {author} {\bibfnamefont {N.~P.}\ \bibnamefont
  {Armitage}},\ }\href {\doibase 10.1103/PhysRevLett.122.257401} {\bibfield
  {journal} {\bibinfo  {journal} {Phys. Rev. Lett.}\ }\textbf {\bibinfo
  {volume} {122}},\ \bibinfo {pages} {257401} (\bibinfo {year}
  {2019})}\BibitemShut {NoStop}%
\bibitem [{\citenamefont {Choi}\ \emph {et~al.}(2020)\citenamefont {Choi},
  \citenamefont {Lee},\ and\ \citenamefont {Kim}}]{Choi2020}%
  \BibitemOpen
  \bibfield  {author} {\bibinfo {author} {\bibfnamefont {W.}~\bibnamefont
  {Choi}}, \bibinfo {author} {\bibfnamefont {K.~H.}\ \bibnamefont {Lee}}, \
  and\ \bibinfo {author} {\bibfnamefont {Y.~B.}\ \bibnamefont {Kim}},\ }\href
  {\doibase 10.1103/PhysRevLett.124.117205} {\bibfield  {journal} {\bibinfo
  {journal} {Phys. Rev. Lett.}\ }\textbf {\bibinfo {volume} {124}},\ \bibinfo
  {pages} {117205} (\bibinfo {year} {2020})}\BibitemShut {NoStop}%
\bibitem [{\citenamefont {Ye}\ \emph {et~al.}(2020)\citenamefont {Ye},
  \citenamefont {Fernandes},\ and\ \citenamefont {Perkins}}]{Mengxing2020}%
  \BibitemOpen
  \bibfield  {author} {\bibinfo {author} {\bibfnamefont {M.}~\bibnamefont
  {Ye}}, \bibinfo {author} {\bibfnamefont {R.~M.}\ \bibnamefont {Fernandes}}, \
  and\ \bibinfo {author} {\bibfnamefont {N.~B.}\ \bibnamefont {Perkins}},\
  }\href {\doibase 10.1103/PhysRevResearch.2.033180} {\bibfield  {journal}
  {\bibinfo  {journal} {Phys. Rev. Research}\ }\textbf {\bibinfo {volume}
  {2}},\ \bibinfo {pages} {033180} (\bibinfo {year} {2020})}\BibitemShut
  {NoStop}%
\bibitem [{\citenamefont {Metavitsiadis}\ and\ \citenamefont
  {Brenig}(2020)}]{Metavitsiadis2020}%
  \BibitemOpen
  \bibfield  {author} {\bibinfo {author} {\bibfnamefont {A.}~\bibnamefont
  {Metavitsiadis}}\ and\ \bibinfo {author} {\bibfnamefont {W.}~\bibnamefont
  {Brenig}},\ }\href {\doibase 10.1103/PhysRevB.101.035103} {\bibfield
  {journal} {\bibinfo  {journal} {Phys. Rev. B}\ }\textbf {\bibinfo {volume}
  {101}},\ \bibinfo {pages} {035103} (\bibinfo {year} {2020})}\BibitemShut
  {NoStop}%
\bibitem [{\citenamefont {Feng}\ \emph {et~al.}(2021)\citenamefont {Feng},
  \citenamefont {Ye},\ and\ \citenamefont {Perkins}}]{Kexin2021}%
  \BibitemOpen
  \bibfield  {author} {\bibinfo {author} {\bibfnamefont {K.}~\bibnamefont
  {Feng}}, \bibinfo {author} {\bibfnamefont {M.}~\bibnamefont {Ye}}, \ and\
  \bibinfo {author} {\bibfnamefont {N.~B.}\ \bibnamefont {Perkins}},\ }\href
  {\doibase 10.1103/PhysRevB.103.214416} {\bibfield  {journal} {\bibinfo
  {journal} {Phys. Rev. B}\ }\textbf {\bibinfo {volume} {103}},\ \bibinfo
  {pages} {214416} (\bibinfo {year} {2021})}\BibitemShut {NoStop}%
\bibitem [{\citenamefont {Feng}\ \emph
  {et~al.}(2022{\natexlab{a}})\citenamefont {Feng}, \citenamefont
  {Shiralieva},\ and\ \citenamefont {Perkins}}]{Feng2022}%
  \BibitemOpen
  \bibfield  {author} {\bibinfo {author} {\bibfnamefont {K.}~\bibnamefont
  {Feng}}, \bibinfo {author} {\bibfnamefont {A.}~\bibnamefont {Shiralieva}}, \
  and\ \bibinfo {author} {\bibfnamefont {N.~B.}\ \bibnamefont {Perkins}},\
  }\href {\doibase 10.1103/PhysRevB.106.144424} {\bibfield  {journal} {\bibinfo
   {journal} {Phys. Rev. B}\ }\textbf {\bibinfo {volume} {106}},\ \bibinfo
  {pages} {144424} (\bibinfo {year} {2022}{\natexlab{a}})}\BibitemShut
  {NoStop}%
\bibitem [{\citenamefont {Winter}\ \emph {et~al.}(2017)\citenamefont {Winter},
  \citenamefont {Tsirlin}, \citenamefont {Daghofer}, \citenamefont {van~den
  Brink}, \citenamefont {Singh}, \citenamefont {Gegenwart},\ and\ \citenamefont
  {ValentÃ­}}]{Winter2017}%
  \BibitemOpen
  \bibfield  {author} {\bibinfo {author} {\bibfnamefont {S.~M.}\ \bibnamefont
  {Winter}}, \bibinfo {author} {\bibfnamefont {A.~A.}\ \bibnamefont {Tsirlin}},
  \bibinfo {author} {\bibfnamefont {M.}~\bibnamefont {Daghofer}}, \bibinfo
  {author} {\bibfnamefont {J.}~\bibnamefont {van~den Brink}}, \bibinfo {author}
  {\bibfnamefont {Y.}~\bibnamefont {Singh}}, \bibinfo {author} {\bibfnamefont
  {P.}~\bibnamefont {Gegenwart}}, \ and\ \bibinfo {author} {\bibfnamefont
  {R.}~\bibnamefont {ValentÃ­}},\ }\href
  {http://stacks.iop.org/0953-8984/29/i=49/a=493002} {\bibfield  {journal}
  {\bibinfo  {journal} {J. Phys.: Condens. Matter}\ }\textbf {\bibinfo {volume}
  {29}},\ \bibinfo {pages} {493002} (\bibinfo {year} {2017})}\BibitemShut
  {NoStop}%
\bibitem [{\citenamefont {Rau}\ \emph {et~al.}(2014)\citenamefont {Rau},
  \citenamefont {Lee},\ and\ \citenamefont {Kee}}]{Rau2014}%
  \BibitemOpen
  \bibfield  {author} {\bibinfo {author} {\bibfnamefont {J.~G.}\ \bibnamefont
  {Rau}}, \bibinfo {author} {\bibfnamefont {E.~K.-H.}\ \bibnamefont {Lee}}, \
  and\ \bibinfo {author} {\bibfnamefont {H.-Y.}\ \bibnamefont {Kee}},\ }\href
  {\doibase 10.1103/PhysRevLett.112.077204} {\bibfield  {journal} {\bibinfo
  {journal} {Phys. Rev. Lett.}\ }\textbf {\bibinfo {volume} {112}},\ \bibinfo
  {pages} {077204} (\bibinfo {year} {2014})}\BibitemShut {NoStop}%
\bibitem [{\citenamefont {Sizyuk}\ \emph {et~al.}(2014)\citenamefont {Sizyuk},
  \citenamefont {Price}, \citenamefont {W\"olfle},\ and\ \citenamefont
  {Perkins}}]{Sizyuk2014}%
  \BibitemOpen
  \bibfield  {author} {\bibinfo {author} {\bibfnamefont {Y.}~\bibnamefont
  {Sizyuk}}, \bibinfo {author} {\bibfnamefont {C.}~\bibnamefont {Price}},
  \bibinfo {author} {\bibfnamefont {P.}~\bibnamefont {W\"olfle}}, \ and\
  \bibinfo {author} {\bibfnamefont {N.~B.}\ \bibnamefont {Perkins}},\ }\href
  {\doibase 10.1103/PhysRevB.90.155126} {\bibfield  {journal} {\bibinfo
  {journal} {Phys. Rev. B}\ }\textbf {\bibinfo {volume} {90}},\ \bibinfo
  {pages} {155126} (\bibinfo {year} {2014})}\BibitemShut {NoStop}%
\bibitem [{\citenamefont {Wang}\ \emph {et~al.}(2017)\citenamefont {Wang},
  \citenamefont {Dong}, \citenamefont {Yu},\ and\ \citenamefont
  {Li}}]{Wang2017}%
  \BibitemOpen
  \bibfield  {author} {\bibinfo {author} {\bibfnamefont {W.}~\bibnamefont
  {Wang}}, \bibinfo {author} {\bibfnamefont {Z.-Y.}\ \bibnamefont {Dong}},
  \bibinfo {author} {\bibfnamefont {S.-L.}\ \bibnamefont {Yu}}, \ and\ \bibinfo
  {author} {\bibfnamefont {J.-X.}\ \bibnamefont {Li}},\ }\href {\doibase
  10.1103/PhysRevB.96.115103} {\bibfield  {journal} {\bibinfo  {journal} {Phys.
  Rev. B}\ }\textbf {\bibinfo {volume} {96}},\ \bibinfo {pages} {115103}
  (\bibinfo {year} {2017})}\BibitemShut {NoStop}%
\bibitem [{\citenamefont {Ran}\ \emph {et~al.}(2017)\citenamefont {Ran},
  \citenamefont {Wang}, \citenamefont {Wang}, \citenamefont {Dong},
  \citenamefont {Ren}, \citenamefont {Bao}, \citenamefont {Li}, \citenamefont
  {Ma}, \citenamefont {Gan}, \citenamefont {Zhang}, \citenamefont {Park},
  \citenamefont {Deng}, \citenamefont {Danilkin}, \citenamefont {Yu},
  \citenamefont {Li},\ and\ \citenamefont {Wen}}]{Ran2017}%
  \BibitemOpen
  \bibfield  {author} {\bibinfo {author} {\bibfnamefont {K.}~\bibnamefont
  {Ran}}, \bibinfo {author} {\bibfnamefont {J.}~\bibnamefont {Wang}}, \bibinfo
  {author} {\bibfnamefont {W.}~\bibnamefont {Wang}}, \bibinfo {author}
  {\bibfnamefont {Z.-Y.}\ \bibnamefont {Dong}}, \bibinfo {author}
  {\bibfnamefont {X.}~\bibnamefont {Ren}}, \bibinfo {author} {\bibfnamefont
  {S.}~\bibnamefont {Bao}}, \bibinfo {author} {\bibfnamefont {S.}~\bibnamefont
  {Li}}, \bibinfo {author} {\bibfnamefont {Z.}~\bibnamefont {Ma}}, \bibinfo
  {author} {\bibfnamefont {Y.}~\bibnamefont {Gan}}, \bibinfo {author}
  {\bibfnamefont {Y.}~\bibnamefont {Zhang}}, \bibinfo {author} {\bibfnamefont
  {J.~T.}\ \bibnamefont {Park}}, \bibinfo {author} {\bibfnamefont
  {G.}~\bibnamefont {Deng}}, \bibinfo {author} {\bibfnamefont {S.}~\bibnamefont
  {Danilkin}}, \bibinfo {author} {\bibfnamefont {S.-L.}\ \bibnamefont {Yu}},
  \bibinfo {author} {\bibfnamefont {J.-X.}\ \bibnamefont {Li}}, \ and\ \bibinfo
  {author} {\bibfnamefont {J.}~\bibnamefont {Wen}},\ }\href {\doibase
  10.1103/PhysRevLett.118.107203} {\bibfield  {journal} {\bibinfo  {journal}
  {Phys. Rev. Lett.}\ }\textbf {\bibinfo {volume} {118}},\ \bibinfo {pages}
  {107203} (\bibinfo {year} {2017})}\BibitemShut {NoStop}%
\bibitem [{\citenamefont {Chaloupka}\ \emph {et~al.}(2013)\citenamefont
  {Chaloupka}, \citenamefont {Jackeli},\ and\ \citenamefont
  {Khaliullin}}]{Chaloupka2013}%
  \BibitemOpen
  \bibfield  {author} {\bibinfo {author} {\bibfnamefont {J.~c.~v.}\
  \bibnamefont {Chaloupka}}, \bibinfo {author} {\bibfnamefont {G.}~\bibnamefont
  {Jackeli}}, \ and\ \bibinfo {author} {\bibfnamefont {G.}~\bibnamefont
  {Khaliullin}},\ }\href {\doibase 10.1103/PhysRevLett.110.097204} {\bibfield
  {journal} {\bibinfo  {journal} {Phys. Rev. Lett.}\ }\textbf {\bibinfo
  {volume} {110}},\ \bibinfo {pages} {097204} (\bibinfo {year}
  {2013})}\BibitemShut {NoStop}%
\bibitem [{\citenamefont {Rau}\ and\ \citenamefont {Kee}(2014)}]{JRau2014}%
  \BibitemOpen
  \bibfield  {author} {\bibinfo {author} {\bibfnamefont {J.~G.}\ \bibnamefont
  {Rau}}\ and\ \bibinfo {author} {\bibfnamefont {H.-Y.}\ \bibnamefont {Kee}},\
  }\href@noop {} {\enquote {\bibinfo {title} {Trigonal distortion in the
  honeycomb iridates: Proximity of zigzag and spiral phases in na2iro3},}\ }
  (\bibinfo {year} {2014}),\ \Eprint {http://arxiv.org/abs/1408.4811}
  {arXiv:1408.4811 [cond-mat.str-el]} \BibitemShut {NoStop}%
\bibitem [{\citenamefont {Gotfryd}\ \emph {et~al.}(2017)\citenamefont
  {Gotfryd}, \citenamefont {Rusna\ifmmode~\check{c}\else \v{c}\fi{}ko},
  \citenamefont {Wohlfeld}, \citenamefont {Jackeli}, \citenamefont
  {Chaloupka},\ and\ \citenamefont {Ole\ifmmode~\acute{s}\else
  \'{s}\fi{}}}]{Gotfryd2017}%
  \BibitemOpen
  \bibfield  {author} {\bibinfo {author} {\bibfnamefont {D.}~\bibnamefont
  {Gotfryd}}, \bibinfo {author} {\bibfnamefont {J.}~\bibnamefont
  {Rusna\ifmmode~\check{c}\else \v{c}\fi{}ko}}, \bibinfo {author}
  {\bibfnamefont {K.}~\bibnamefont {Wohlfeld}}, \bibinfo {author}
  {\bibfnamefont {G.}~\bibnamefont {Jackeli}}, \bibinfo {author} {\bibfnamefont
  {J.~c.~v.}\ \bibnamefont {Chaloupka}}, \ and\ \bibinfo {author}
  {\bibfnamefont {A.~M.}\ \bibnamefont {Ole\ifmmode~\acute{s}\else
  \'{s}\fi{}}},\ }\href {\doibase 10.1103/PhysRevB.95.024426} {\bibfield
  {journal} {\bibinfo  {journal} {Phys. Rev. B}\ }\textbf {\bibinfo {volume}
  {95}},\ \bibinfo {pages} {024426} (\bibinfo {year} {2017})}\BibitemShut
  {NoStop}%
\bibitem [{\citenamefont {Gohlke}\ \emph {et~al.}(2017)\citenamefont {Gohlke},
  \citenamefont {Verresen}, \citenamefont {Moessner},\ and\ \citenamefont
  {Pollmann}}]{Gohlke2017}%
  \BibitemOpen
  \bibfield  {author} {\bibinfo {author} {\bibfnamefont {M.}~\bibnamefont
  {Gohlke}}, \bibinfo {author} {\bibfnamefont {R.}~\bibnamefont {Verresen}},
  \bibinfo {author} {\bibfnamefont {R.}~\bibnamefont {Moessner}}, \ and\
  \bibinfo {author} {\bibfnamefont {F.}~\bibnamefont {Pollmann}},\ }\href
  {\doibase 10.1103/PhysRevLett.119.157203} {\bibfield  {journal} {\bibinfo
  {journal} {Phys. Rev. Lett.}\ }\textbf {\bibinfo {volume} {119}},\ \bibinfo
  {pages} {157203} (\bibinfo {year} {2017})}\BibitemShut {NoStop}%
\bibitem [{\citenamefont {Gohlke}\ \emph {et~al.}(2018)\citenamefont {Gohlke},
  \citenamefont {Wachtel}, \citenamefont {Yamaji}, \citenamefont {Pollmann},\
  and\ \citenamefont {Kim}}]{Gohlke2018}%
  \BibitemOpen
  \bibfield  {author} {\bibinfo {author} {\bibfnamefont {M.}~\bibnamefont
  {Gohlke}}, \bibinfo {author} {\bibfnamefont {G.}~\bibnamefont {Wachtel}},
  \bibinfo {author} {\bibfnamefont {Y.}~\bibnamefont {Yamaji}}, \bibinfo
  {author} {\bibfnamefont {F.}~\bibnamefont {Pollmann}}, \ and\ \bibinfo
  {author} {\bibfnamefont {Y.~B.}\ \bibnamefont {Kim}},\ }\href {\doibase
  10.1103/PhysRevB.97.075126} {\bibfield  {journal} {\bibinfo  {journal} {Phys.
  Rev. B}\ }\textbf {\bibinfo {volume} {97}},\ \bibinfo {pages} {075126}
  (\bibinfo {year} {2018})}\BibitemShut {NoStop}%
\bibitem [{\citenamefont {Gordon}\ \emph {et~al.}(2019)\citenamefont {Gordon},
  \citenamefont {Catuneanu}, \citenamefont {Sørensen},\ and\ \citenamefont
  {Kee}}]{Gordon2019}%
  \BibitemOpen
  \bibfield  {author} {\bibinfo {author} {\bibfnamefont {J.~S.}\ \bibnamefont
  {Gordon}}, \bibinfo {author} {\bibfnamefont {A.}~\bibnamefont {Catuneanu}},
  \bibinfo {author} {\bibfnamefont {E.~S.}\ \bibnamefont {Sørensen}}, \ and\
  \bibinfo {author} {\bibfnamefont {H.-Y.}\ \bibnamefont {Kee}},\ }\href
  {\doibase 10.1038/s41467-019-10405-8} {\bibfield  {journal} {\bibinfo
  {journal} {Nature Communications}\ }\textbf {\bibinfo {volume} {10}},\
  \bibinfo {pages} {2470} (\bibinfo {year} {2019})}\BibitemShut {NoStop}%
\bibitem [{\citenamefont {Osorio~Iregui}\ \emph {et~al.}(2014)\citenamefont
  {Osorio~Iregui}, \citenamefont {Corboz},\ and\ \citenamefont
  {Troyer}}]{Osorio2014}%
  \BibitemOpen
  \bibfield  {author} {\bibinfo {author} {\bibfnamefont {J.}~\bibnamefont
  {Osorio~Iregui}}, \bibinfo {author} {\bibfnamefont {P.}~\bibnamefont
  {Corboz}}, \ and\ \bibinfo {author} {\bibfnamefont {M.}~\bibnamefont
  {Troyer}},\ }\href {\doibase 10.1103/PhysRevB.90.195102} {\bibfield
  {journal} {\bibinfo  {journal} {Phys. Rev. B}\ }\textbf {\bibinfo {volume}
  {90}},\ \bibinfo {pages} {195102} (\bibinfo {year} {2014})}\BibitemShut
  {NoStop}%
\bibitem [{\citenamefont {Lee}\ \emph {et~al.}(2020)\citenamefont {Lee},
  \citenamefont {Kaneko}, \citenamefont {Chern}, \citenamefont {Okubo},
  \citenamefont {Yamaji}, \citenamefont {Kawashima},\ and\ \citenamefont
  {Kim}}]{Lee2020}%
  \BibitemOpen
  \bibfield  {author} {\bibinfo {author} {\bibfnamefont {H.-Y.}\ \bibnamefont
  {Lee}}, \bibinfo {author} {\bibfnamefont {R.}~\bibnamefont {Kaneko}},
  \bibinfo {author} {\bibfnamefont {L.~E.}\ \bibnamefont {Chern}}, \bibinfo
  {author} {\bibfnamefont {T.}~\bibnamefont {Okubo}}, \bibinfo {author}
  {\bibfnamefont {Y.}~\bibnamefont {Yamaji}}, \bibinfo {author} {\bibfnamefont
  {N.}~\bibnamefont {Kawashima}}, \ and\ \bibinfo {author} {\bibfnamefont
  {Y.~B.}\ \bibnamefont {Kim}},\ }\href {\doibase 10.1038/s41467-020-15320-x}
  {\bibfield  {journal} {\bibinfo  {journal} {Nature Communications}\ }\textbf
  {\bibinfo {volume} {11}},\ \bibinfo {pages} {1639} (\bibinfo {year}
  {2020})}\BibitemShut {NoStop}%
\bibitem [{\citenamefont {Burnell}\ and\ \citenamefont
  {Nayak}(2011)}]{Burnell2011}%
  \BibitemOpen
  \bibfield  {author} {\bibinfo {author} {\bibfnamefont {F.~J.}\ \bibnamefont
  {Burnell}}\ and\ \bibinfo {author} {\bibfnamefont {C.}~\bibnamefont
  {Nayak}},\ }\href {\doibase 10.1103/PhysRevB.84.125125} {\bibfield  {journal}
  {\bibinfo  {journal} {Phys. Rev. B}\ }\textbf {\bibinfo {volume} {84}},\
  \bibinfo {pages} {125125} (\bibinfo {year} {2011})}\BibitemShut {NoStop}%
\bibitem [{\citenamefont {Schaffer}\ \emph {et~al.}(2012)\citenamefont
  {Schaffer}, \citenamefont {Bhattacharjee},\ and\ \citenamefont
  {Kim}}]{Schaffer2012}%
  \BibitemOpen
  \bibfield  {author} {\bibinfo {author} {\bibfnamefont {R.}~\bibnamefont
  {Schaffer}}, \bibinfo {author} {\bibfnamefont {S.}~\bibnamefont
  {Bhattacharjee}}, \ and\ \bibinfo {author} {\bibfnamefont {Y.~B.}\
  \bibnamefont {Kim}},\ }\href {\doibase 10.1103/PhysRevB.86.224417} {\bibfield
   {journal} {\bibinfo  {journal} {Phys. Rev. B}\ }\textbf {\bibinfo {volume}
  {86}},\ \bibinfo {pages} {224417} (\bibinfo {year} {2012})}\BibitemShut
  {NoStop}%
\bibitem [{\citenamefont {Knolle}\ \emph {et~al.}(2018)\citenamefont {Knolle},
  \citenamefont {Bhattacharjee},\ and\ \citenamefont {Moessner}}]{Knolle2018}%
  \BibitemOpen
  \bibfield  {author} {\bibinfo {author} {\bibfnamefont {J.}~\bibnamefont
  {Knolle}}, \bibinfo {author} {\bibfnamefont {S.}~\bibnamefont
  {Bhattacharjee}}, \ and\ \bibinfo {author} {\bibfnamefont {R.}~\bibnamefont
  {Moessner}},\ }\href {\doibase 10.1103/PhysRevB.97.134432} {\bibfield
  {journal} {\bibinfo  {journal} {Phys. Rev. B}\ }\textbf {\bibinfo {volume}
  {97}},\ \bibinfo {pages} {134432} (\bibinfo {year} {2018})}\BibitemShut
  {NoStop}%
\bibitem [{\citenamefont {Wang}\ \emph {et~al.}(2019)\citenamefont {Wang},
  \citenamefont {Normand},\ and\ \citenamefont {Liu}}]{Wang2019}%
  \BibitemOpen
  \bibfield  {author} {\bibinfo {author} {\bibfnamefont {J.}~\bibnamefont
  {Wang}}, \bibinfo {author} {\bibfnamefont {B.}~\bibnamefont {Normand}}, \
  and\ \bibinfo {author} {\bibfnamefont {Z.-X.}\ \bibnamefont {Liu}},\ }\href
  {\doibase 10.1103/PhysRevLett.123.197201} {\bibfield  {journal} {\bibinfo
  {journal} {Phys. Rev. Lett.}\ }\textbf {\bibinfo {volume} {123}},\ \bibinfo
  {pages} {197201} (\bibinfo {year} {2019})}\BibitemShut {NoStop}%
\bibitem [{\citenamefont {Zhang}\ \emph {et~al.}(2021)\citenamefont {Zhang},
  \citenamefont {Hal\'asz}, \citenamefont {Zhu},\ and\ \citenamefont
  {Batista}}]{Zhang2021}%
  \BibitemOpen
  \bibfield  {author} {\bibinfo {author} {\bibfnamefont {S.-S.}\ \bibnamefont
  {Zhang}}, \bibinfo {author} {\bibfnamefont {G.~B.}\ \bibnamefont {Hal\'asz}},
  \bibinfo {author} {\bibfnamefont {W.}~\bibnamefont {Zhu}}, \ and\ \bibinfo
  {author} {\bibfnamefont {C.~D.}\ \bibnamefont {Batista}},\ }\href {\doibase
  10.1103/PhysRevB.104.014411} {\bibfield  {journal} {\bibinfo  {journal}
  {Phys. Rev. B}\ }\textbf {\bibinfo {volume} {104}},\ \bibinfo {pages}
  {014411} (\bibinfo {year} {2021})}\BibitemShut {NoStop}%
\bibitem [{\citenamefont {Hentrich}\ \emph {et~al.}(2018)\citenamefont
  {Hentrich}, \citenamefont {Wolter}, \citenamefont {Zotos}, \citenamefont
  {Brenig}, \citenamefont {Nowak}, \citenamefont {Isaeva}, \citenamefont
  {Doert}, \citenamefont {Banerjee}, \citenamefont {Lampen-Kelley},
  \citenamefont {Mandrus}, \citenamefont {Nagler}, \citenamefont {Sears},
  \citenamefont {Kim}, \citenamefont {B\"uchner},\ and\ \citenamefont
  {Hess}}]{Hentrich2018}%
  \BibitemOpen
  \bibfield  {author} {\bibinfo {author} {\bibfnamefont {R.}~\bibnamefont
  {Hentrich}}, \bibinfo {author} {\bibfnamefont {A.~U.~B.}\ \bibnamefont
  {Wolter}}, \bibinfo {author} {\bibfnamefont {X.}~\bibnamefont {Zotos}},
  \bibinfo {author} {\bibfnamefont {W.}~\bibnamefont {Brenig}}, \bibinfo
  {author} {\bibfnamefont {D.}~\bibnamefont {Nowak}}, \bibinfo {author}
  {\bibfnamefont {A.}~\bibnamefont {Isaeva}}, \bibinfo {author} {\bibfnamefont
  {T.}~\bibnamefont {Doert}}, \bibinfo {author} {\bibfnamefont
  {A.}~\bibnamefont {Banerjee}}, \bibinfo {author} {\bibfnamefont
  {P.}~\bibnamefont {Lampen-Kelley}}, \bibinfo {author} {\bibfnamefont {D.~G.}\
  \bibnamefont {Mandrus}}, \bibinfo {author} {\bibfnamefont {S.~E.}\
  \bibnamefont {Nagler}}, \bibinfo {author} {\bibfnamefont {J.}~\bibnamefont
  {Sears}}, \bibinfo {author} {\bibfnamefont {Y.-J.}\ \bibnamefont {Kim}},
  \bibinfo {author} {\bibfnamefont {B.}~\bibnamefont {B\"uchner}}, \ and\
  \bibinfo {author} {\bibfnamefont {C.}~\bibnamefont {Hess}},\ }\href {\doibase
  10.1103/PhysRevLett.120.117204} {\bibfield  {journal} {\bibinfo  {journal}
  {Phys. Rev. Lett.}\ }\textbf {\bibinfo {volume} {120}},\ \bibinfo {pages}
  {117204} (\bibinfo {year} {2018})}\BibitemShut {NoStop}%
\bibitem [{\citenamefont {Kasahara}\ \emph {et~al.}(2018)\citenamefont
  {Kasahara}, \citenamefont {Ohnishi}, \citenamefont {Mizukami}, \citenamefont
  {Tanaka}, \citenamefont {Ma}, \citenamefont {Sugii}, \citenamefont {Kurita},
  \citenamefont {Tanaka}, \citenamefont {Nasu}, \citenamefont {Motome},
  \citenamefont {Shibauchi},\ and\ \citenamefont {Matsuda}}]{Kasahara2018}%
  \BibitemOpen
  \bibfield  {author} {\bibinfo {author} {\bibfnamefont {Y.}~\bibnamefont
  {Kasahara}}, \bibinfo {author} {\bibfnamefont {T.}~\bibnamefont {Ohnishi}},
  \bibinfo {author} {\bibfnamefont {Y.}~\bibnamefont {Mizukami}}, \bibinfo
  {author} {\bibfnamefont {O.}~\bibnamefont {Tanaka}}, \bibinfo {author}
  {\bibfnamefont {S.}~\bibnamefont {Ma}}, \bibinfo {author} {\bibfnamefont
  {K.}~\bibnamefont {Sugii}}, \bibinfo {author} {\bibfnamefont
  {N.}~\bibnamefont {Kurita}}, \bibinfo {author} {\bibfnamefont
  {H.}~\bibnamefont {Tanaka}}, \bibinfo {author} {\bibfnamefont
  {J.}~\bibnamefont {Nasu}}, \bibinfo {author} {\bibfnamefont {Y.}~\bibnamefont
  {Motome}}, \bibinfo {author} {\bibfnamefont {T.}~\bibnamefont {Shibauchi}}, \
  and\ \bibinfo {author} {\bibfnamefont {Y.}~\bibnamefont {Matsuda}},\ }\href
  {\doibase 10.1038/s41586-018-0274-0} {\bibfield  {journal} {\bibinfo
  {journal} {Nature}\ }\textbf {\bibinfo {volume} {559}},\ \bibinfo {pages}
  {227} (\bibinfo {year} {2018})}\BibitemShut {NoStop}%
\bibitem [{\citenamefont {Pal}\ \emph {et~al.}(2020)\citenamefont {Pal},
  \citenamefont {Seth}, \citenamefont {Sakrikar}, \citenamefont {Ali},
  \citenamefont {Bhattacharjee}, \citenamefont {Muthu}, \citenamefont {Singh},\
  and\ \citenamefont {Sood}}]{Pal2020}%
  \BibitemOpen
  \bibfield  {author} {\bibinfo {author} {\bibfnamefont {S.}~\bibnamefont
  {Pal}}, \bibinfo {author} {\bibfnamefont {A.}~\bibnamefont {Seth}}, \bibinfo
  {author} {\bibfnamefont {P.}~\bibnamefont {Sakrikar}}, \bibinfo {author}
  {\bibfnamefont {A.}~\bibnamefont {Ali}}, \bibinfo {author} {\bibfnamefont
  {S.}~\bibnamefont {Bhattacharjee}}, \bibinfo {author} {\bibfnamefont
  {D.~V.~S.}\ \bibnamefont {Muthu}}, \bibinfo {author} {\bibfnamefont
  {Y.}~\bibnamefont {Singh}}, \ and\ \bibinfo {author} {\bibfnamefont {A.~K.}\
  \bibnamefont {Sood}},\ }\href {https://arxiv.org/abs/2011.00606} {\bibfield
  {journal} {\bibinfo  {journal} {arXiv:2011.00606}\ } (\bibinfo {year}
  {2020})}\BibitemShut {NoStop}%
\bibitem [{\citenamefont {Li}\ \emph {et~al.}(2021{\natexlab{b}})\citenamefont
  {Li}, \citenamefont {Said}, \citenamefont {Yan}, \citenamefont {Mandrus},
  \citenamefont {Lee}, \citenamefont {Okamoto}, \citenamefont {Halász},\ and\
  \citenamefont {Miao}}]{Haoxiang2021}%
  \BibitemOpen
  \bibfield  {author} {\bibinfo {author} {\bibfnamefont {H.}~\bibnamefont
  {Li}}, \bibinfo {author} {\bibfnamefont {A.}~\bibnamefont {Said}}, \bibinfo
  {author} {\bibfnamefont {J.~Q.}\ \bibnamefont {Yan}}, \bibinfo {author}
  {\bibfnamefont {D.~M.}\ \bibnamefont {Mandrus}}, \bibinfo {author}
  {\bibfnamefont {H.~N.}\ \bibnamefont {Lee}}, \bibinfo {author} {\bibfnamefont
  {S.}~\bibnamefont {Okamoto}}, \bibinfo {author} {\bibfnamefont {G.~B.}\
  \bibnamefont {Halász}}, \ and\ \bibinfo {author} {\bibfnamefont
  {H.}~\bibnamefont {Miao}},\ }\href@noop {} {} (\bibinfo {year}
  {2021}{\natexlab{b}}),\ \Eprint {http://arxiv.org/abs/2112.02015}
  {arXiv:2112.02015 [cond-mat.str-el]} \BibitemShut {NoStop}%
\bibitem [{\citenamefont {Hauspurg}\ \emph {et~al.}(2023)\citenamefont
  {Hauspurg}, \citenamefont {Zherlitsyn}, \citenamefont {Helm}, \citenamefont
  {Felea}, \citenamefont {Wosnitza}, \citenamefont {Tsurkan}, \citenamefont
  {Choi}, \citenamefont {Do}, \citenamefont {Ye}, \citenamefont {Brenig},\ and\
  \citenamefont {Perkins}}]{Andreas2023}%
  \BibitemOpen
  \bibfield  {author} {\bibinfo {author} {\bibfnamefont {A.}~\bibnamefont
  {Hauspurg}}, \bibinfo {author} {\bibfnamefont {S.}~\bibnamefont
  {Zherlitsyn}}, \bibinfo {author} {\bibfnamefont {T.}~\bibnamefont {Helm}},
  \bibinfo {author} {\bibfnamefont {V.}~\bibnamefont {Felea}}, \bibinfo
  {author} {\bibfnamefont {J.}~\bibnamefont {Wosnitza}}, \bibinfo {author}
  {\bibfnamefont {V.}~\bibnamefont {Tsurkan}}, \bibinfo {author} {\bibfnamefont
  {K.~Y.}\ \bibnamefont {Choi}}, \bibinfo {author} {\bibfnamefont {S.~H.}\
  \bibnamefont {Do}}, \bibinfo {author} {\bibfnamefont {M.}~\bibnamefont {Ye}},
  \bibinfo {author} {\bibfnamefont {W.}~\bibnamefont {Brenig}}, \ and\ \bibinfo
  {author} {\bibfnamefont {N.~B.}\ \bibnamefont {Perkins}},\ }\href@noop {} {}
  (\bibinfo {year} {2023}),\ \Eprint {http://arxiv.org/abs/2303.09288}
  {arXiv:2303.09288 [cond-mat.str-el]} \BibitemShut {NoStop}%
\bibitem [{\citenamefont {Feng}\ \emph
  {et~al.}(2022{\natexlab{b}})\citenamefont {Feng}, \citenamefont {Swarup},\
  and\ \citenamefont {Perkins}}]{Kexin2021-s}%
  \BibitemOpen
  \bibfield  {author} {\bibinfo {author} {\bibfnamefont {K.}~\bibnamefont
  {Feng}}, \bibinfo {author} {\bibfnamefont {S.}~\bibnamefont {Swarup}}, \ and\
  \bibinfo {author} {\bibfnamefont {N.~B.}\ \bibnamefont {Perkins}},\ }\href
  {\doibase 10.1103/PhysRevB.105.L121108} {\bibfield  {journal} {\bibinfo
  {journal} {Phys. Rev. B}\ }\textbf {\bibinfo {volume} {105}},\ \bibinfo
  {pages} {L121108} (\bibinfo {year} {2022}{\natexlab{b}})}\BibitemShut
  {NoStop}%
\bibitem [{\citenamefont {Metavitsiadis}\ \emph {et~al.}(2022)\citenamefont
  {Metavitsiadis}, \citenamefont {Natori}, \citenamefont {Knolle},\ and\
  \citenamefont {Brenig}}]{Metavitsiadis2022}%
  \BibitemOpen
  \bibfield  {author} {\bibinfo {author} {\bibfnamefont {A.}~\bibnamefont
  {Metavitsiadis}}, \bibinfo {author} {\bibfnamefont {W.}~\bibnamefont
  {Natori}}, \bibinfo {author} {\bibfnamefont {J.}~\bibnamefont {Knolle}}, \
  and\ \bibinfo {author} {\bibfnamefont {W.}~\bibnamefont {Brenig}},\ }\href
  {\doibase 10.1103/PhysRevB.105.165151} {\bibfield  {journal} {\bibinfo
  {journal} {Phys. Rev. B}\ }\textbf {\bibinfo {volume} {105}},\ \bibinfo
  {pages} {165151} (\bibinfo {year} {2022})}\BibitemShut {NoStop}%
\bibitem [{\citenamefont {Pippard}(1955)}]{Pippard1955}%
  \BibitemOpen
  \bibfield  {author} {\bibinfo {author} {\bibfnamefont {A.}~\bibnamefont
  {Pippard}},\ }\href {\doibase 10.1080/14786441008521122} {\bibfield
  {journal} {\bibinfo  {journal} {The London, Edinburgh, and Dublin
  Philosophical Magazine and Journal of Science}\ }\textbf {\bibinfo {volume}
  {46}},\ \bibinfo {pages} {1104} (\bibinfo {year} {1955})}\BibitemShut
  {NoStop}%
\bibitem [{\citenamefont {Akhiezer}\ \emph {et~al.}(1957)\citenamefont
  {Akhiezer}, \citenamefont {Kaganov},\ and\ \citenamefont
  {Lyubarskyi}}]{Akhiezer1957}%
  \BibitemOpen
  \bibfield  {author} {\bibinfo {author} {\bibfnamefont {A.~I.}\ \bibnamefont
  {Akhiezer}}, \bibinfo {author} {\bibfnamefont {M.~I.}\ \bibnamefont
  {Kaganov}}, \ and\ \bibinfo {author} {\bibfnamefont {G.~Y.}\ \bibnamefont
  {Lyubarskyi}},\ }\href {http://jetp.ras.ru/cgi-bin/e/index/e/5/4/p685?a=list}
  {\bibfield  {journal} {\bibinfo  {journal} {Sov. Phys. JETP}\ }\textbf
  {\bibinfo {volume} {5}},\ \bibinfo {pages} {685} (\bibinfo {year}
  {1957})}\BibitemShut {NoStop}%
\bibitem [{\citenamefont {Blount}(1959)}]{Blount1959}%
  \BibitemOpen
  \bibfield  {author} {\bibinfo {author} {\bibfnamefont {E.~I.}\ \bibnamefont
  {Blount}},\ }\href {\doibase 10.1103/PhysRev.114.418} {\bibfield  {journal}
  {\bibinfo  {journal} {Phys. Rev.}\ }\textbf {\bibinfo {volume} {114}},\
  \bibinfo {pages} {418} (\bibinfo {year} {1959})}\BibitemShut {NoStop}%
\bibitem [{\citenamefont {Tsuneto}(1961)}]{Tsuneto1961}%
  \BibitemOpen
  \bibfield  {author} {\bibinfo {author} {\bibfnamefont {T.}~\bibnamefont
  {Tsuneto}},\ }\href {\doibase 10.1103/PhysRev.121.402} {\bibfield  {journal}
  {\bibinfo  {journal} {Phys. Rev.}\ }\textbf {\bibinfo {volume} {121}},\
  \bibinfo {pages} {402} (\bibinfo {year} {1961})}\BibitemShut {NoStop}%
\bibitem [{\citenamefont {Batlogg}\ \emph {et~al.}(1985)\citenamefont
  {Batlogg}, \citenamefont {Bishop}, \citenamefont {Golding}, \citenamefont
  {Varma}, \citenamefont {Fisk}, \citenamefont {Smith},\ and\ \citenamefont
  {Ott}}]{Ott1985}%
  \BibitemOpen
  \bibfield  {author} {\bibinfo {author} {\bibfnamefont {B.}~\bibnamefont
  {Batlogg}}, \bibinfo {author} {\bibfnamefont {D.}~\bibnamefont {Bishop}},
  \bibinfo {author} {\bibfnamefont {B.}~\bibnamefont {Golding}}, \bibinfo
  {author} {\bibfnamefont {C.~M.}\ \bibnamefont {Varma}}, \bibinfo {author}
  {\bibfnamefont {Z.}~\bibnamefont {Fisk}}, \bibinfo {author} {\bibfnamefont
  {J.~L.}\ \bibnamefont {Smith}}, \ and\ \bibinfo {author} {\bibfnamefont
  {H.~R.}\ \bibnamefont {Ott}},\ }\href {\doibase 10.1103/PhysRevLett.55.1319}
  {\bibfield  {journal} {\bibinfo  {journal} {Phys. Rev. Lett.}\ }\textbf
  {\bibinfo {volume} {55}},\ \bibinfo {pages} {1319} (\bibinfo {year}
  {1985})}\BibitemShut {NoStop}%
\bibitem [{\citenamefont {Won}\ and\ \citenamefont {Maki}(1994)}]{Kazumi1994}%
  \BibitemOpen
  \bibfield  {author} {\bibinfo {author} {\bibfnamefont {H.}~\bibnamefont
  {Won}}\ and\ \bibinfo {author} {\bibfnamefont {K.}~\bibnamefont {Maki}},\
  }\href {\doibase 10.1103/PhysRevB.49.1397} {\bibfield  {journal} {\bibinfo
  {journal} {Phys. Rev. B}\ }\textbf {\bibinfo {volume} {49}},\ \bibinfo
  {pages} {1397} (\bibinfo {year} {1994})}\BibitemShut {NoStop}%
\bibitem [{\citenamefont {Lebert}\ \emph {et~al.}(2022)\citenamefont {Lebert},
  \citenamefont {Kim}, \citenamefont {Prishchenko}, \citenamefont {Tsirlin},
  \citenamefont {Said}, \citenamefont {Alatas},\ and\ \citenamefont
  {Kim}}]{LebertPRB2022}%
  \BibitemOpen
  \bibfield  {author} {\bibinfo {author} {\bibfnamefont {B.~W.}\ \bibnamefont
  {Lebert}}, \bibinfo {author} {\bibfnamefont {S.}~\bibnamefont {Kim}},
  \bibinfo {author} {\bibfnamefont {D.~A.}\ \bibnamefont {Prishchenko}},
  \bibinfo {author} {\bibfnamefont {A.~A.}\ \bibnamefont {Tsirlin}}, \bibinfo
  {author} {\bibfnamefont {A.~H.}\ \bibnamefont {Said}}, \bibinfo {author}
  {\bibfnamefont {A.}~\bibnamefont {Alatas}}, \ and\ \bibinfo {author}
  {\bibfnamefont {Y.-J.}\ \bibnamefont {Kim}},\ }\href {\doibase
  10.1103/PhysRevB.106.L041102} {\bibfield  {journal} {\bibinfo  {journal}
  {Phys. Rev. B}\ }\textbf {\bibinfo {volume} {106}},\ \bibinfo {pages}
  {L041102} (\bibinfo {year} {2022})}\BibitemShut {NoStop}%
\bibitem [{\citenamefont {Maksimov}\ and\ \citenamefont
  {Chernyshev}(2020)}]{MaksimovPRR2020}%
  \BibitemOpen
  \bibfield  {author} {\bibinfo {author} {\bibfnamefont {P.~A.}\ \bibnamefont
  {Maksimov}}\ and\ \bibinfo {author} {\bibfnamefont {A.~L.}\ \bibnamefont
  {Chernyshev}},\ }\href {\doibase 10.1103/PhysRevResearch.2.033011} {\bibfield
   {journal} {\bibinfo  {journal} {Phys. Rev. Res.}\ }\textbf {\bibinfo
  {volume} {2}},\ \bibinfo {pages} {033011} (\bibinfo {year}
  {2020})}\BibitemShut {NoStop}%
\bibitem [{\citenamefont {{Altland, Alexander and Simons, Ben
  D.}}(2010)}]{AltlandBook}%
  \BibitemOpen
  \bibfield  {author} {\bibinfo {author} {\bibnamefont {{Altland, Alexander and
  Simons, Ben D.}}},\ }\href@noop {} {\emph {\bibinfo {title} {Condensed Matter
  Field Theory}}},\ \bibinfo {edition} {2nd}\ ed.\ (\bibinfo  {publisher}
  {Cambridge University Press},\ \bibinfo {year} {2010})\BibitemShut {NoStop}%
\end{thebibliography}%

\end{document}